\documentclass[graybox,natbib]{svmult}

\setcitestyle{numbers,square}

\usepackage{multicol}        
\usepackage[bottom]{footmisc}

\usepackage{mathtools}
\usepackage{amssymb}

\usepackage{tikz}
\usetikzlibrary{decorations.markings,shadings}
\tikzset{middlearrow/.style={
        decoration={markings,
            mark= at position 0.55 with {\arrow{#1}} ,
        },
        postaction={decorate}
    }
}

\usepackage[varg]{txfonts}

\newcommand{\igc}[4]{\begin{figure}[htb]\centering\includegraphics[width=#1\linewidth]{#2}\caption{#4}\label{fig:#3}\end{figure}}

\DeclareMathOperator{\Tr}{Tr}
\DeclareMathOperator{\tr}{tr}
\DeclareMathOperator{\Det}{Det}
\newcommand*{\vx}{\vec{x}}
\newcommand*{\vy}{\vec{y}}
\newcommand*{\vp}{\vec{p}}
\newcommand*{\vA}{\vec{A}}
\newcommand*{\vD}{\vec{D}}

\newcommand{\va}{\vec{a}}
\newcommand{\ve}{\vec{e}}

\newcommand*{\vR}{\vec{R}}
\newcommand*{\vf}{\vec{f}}
\newcommand*{\vB}{\vec{B}}
\newcommand*{\cA}{\mathcal{A}}
\newcommand*{\cB}{\mathcal{B}}
\newcommand*{\cP}{\mathcal{P}}
\newcommand*{\cF}{\mathcal{F}}
\newcommand*{\ee}{\mathrm{e}}
\def\R{{\mathbb{R}}}
\newcommand{\RR}{\mathbb{R}}
\newcommand{\ZZ}{\mathbb{Z}}
\newcommand{\ii}{\mathrm{i}}
\newcommand{\dd}{\mathrm{d}}
\newcommand{\pli}{\prod\limits}
\newcommand{\il}{\int\limits}
\newcommand{\sli}{\sum\limits}
\newcommand{\lk}{\left(}
\newcommand{\rkx}{\right)}

\newcommand{\Dl}{D \hspace{-0.55em}/\hspace{.3em}}
\newcommand{\beq}{\begin{equation}}
\newcommand{\eeq}{\end{equation}}
\newcommand*{\dbar}[1][]{\mathop{\mathrm{d}\mkern-7mu\mathchar'26\mkern-1mu^{#1}}\mkern-4mu} 
\begin{document}
	
\title*{Effective Approaches to QCD\protect\footnote{Invited lecture given at the ``53rd Karpacz Winter School of Theoretical Physics'', 26 February to 4 March 2017.}}
\author{H. Reinhardt}
\institute{Universit\"at T\"ubingen,
Institut f\"ur Theoretische Physik,  
72076 T\"ubingen, Germany\\ 
\email{hugo.reinhardt@uni-tuebingen.de}}
%
%
\maketitle

\abstract{In this lecture I will explain the established pictures of the QCD vacuum and, in particular, the underlying confinement 
	mechanism. These are: the magnetic monopole condensation (dual Mei\ss ner effect), the center vortex picture and the Gribov--Zwanziger
	picture. I will start by giving a survey of the common order and disorder parameters of confinement: the temporal and spatial Wilson loop, 
	the Polyakov loop and the 't Hooft loop. Next
	the dual Mei\ss ner effect, which assumes a condensate of magnetic monopoles, will 
	be explained as a picture of confinement. I will also show how magnetic monopoles arises in QCD after the so-called Abelian 
	projection.\\
	The second lecture is devoted to the center vortex picture of confinement. Center vortices will be defined both 
	on the lattice and in the 
	continuum. Within the center vortex picture the emergence of the area law for the Wilson loop as well 
	as the deconfinement phase transition  at finite temperature will be explained. 
	Furthermore, lattice evidence for the center vortex picture will be provided. Finally, I will discuss the topological properties of center vortices and their relation to 
	magnetic monopoles. I will provide evidence from lattice calculations that center vortices are not only responsible for confinement but also for the spontaneous breaking of chiral symmetry.\\
	In the last lecture I will present the Hamiltonian approach to QCD in Coulomb gauge, which is then used to establish the Gribov--Zwanziger
	picture of confinement. Furthermore, I will also relate this scenario to the dual Mei\ss ner effect and the 
	center vortex picture. Finally, I will study QCD at finite temperature within the Hamiltonian approach in a novel way by 
	compactifying one spatial dimension. The Polyakov loop and the dual and chiral quark condensates will be evaluated as 
	function of the temperature.}

\section{QCD and phases of strongly interacting matter}
\begin{figure}[htb]
\centering
\includegraphics[width=.5\linewidth]{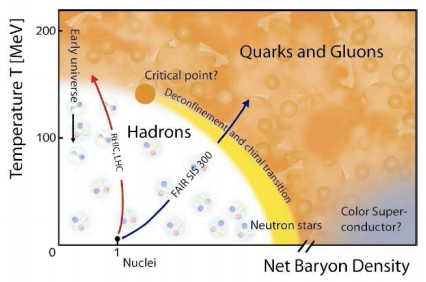}
\caption{Phase diagram of strongly interacting matter pictured in dependence on the temperature $T$ and baryon chemical potential $\mu_B$. 
Under extreme conditions hadronic matter undergoes a phase transition and becomes 
a hot soup of quarks and gluons --- \textit{the Quark-Gluon Plasma}}
\label{fig:part1:phase_diagram1}
\end{figure}

By the current state of knowledge quarks and gluons are considered as elementary particles. Under normal conditions they are 
{\em confined} inside hadrons and they acquire mass through the mechanism of spontaneous breaking of chiral symmetry (SBCS). 
On the other hand, under extreme conditions -- high temperature or high density -- quarks and gluons loose their confinement 
and form the so called {\em Quark-Gluon plasma}. The deconfinement transition is accompanied by a restoration of chiral symmetry. The phase diagram of QCD is schematically pictured in Fig.~\ref{fig:part1:phase_diagram1}. The detailed understanding of the phase diagram of QCD is one of the key challenges of particle and nuclear physics.

By means of ultra-relativistic heavy-ion collisions the properties of hadronic matter at high temperature and density are presently explored. From the theoretical point of view we have access to the finite temperature behavior of QCD by means of lattice calculations. Due to the notorious sign problem (the quark determinant 
becomes complex for gauge groups
$SU (N \geq 3)$ and finite chemical potential), this method fails, however, to describe baryonic matter at high density or, more technically, at large chemical potential. Various methods have been invented to overcome this problem like e.g. Taylor expansion in the chemical potential or analytic continuation of the chemical potential to imaginary values. However, these methods work only up to chemical potential of about twice the value of the temperature. Therefore, alternative, non-perturbative continuum approaches to QCD, which do not suffer from the sign problem, are needed. Furthermore, we have not yet fully understood the confinement mechanism itself. A thorough 
understanding of this mechanism certainly will not come from numerical simulations alone. 
(A strict analytic proof of confinement was formulated  as one of the millennium problems of the Clay Mathematics Institute.)
 However, during the last two decades a couple of consistent pictures of the QCD vacuum have emerged, which will be the subject of this lecture.

\subsection{Confinement}

All the presently known pictures of the QCD vacuum are formulated in or rely on a certain gauge. For example, the magnetic monopole condensation picture (\textit{dual Mei{\ss}ner effect}) \cite{tHooft:1981bkw,Mandelstam:1974pi} relies on the maximal Abelian gauge \cite{tHooft:1981bkw}. Similarly the center vortex condensation picture \cite{tHooft:1977nqb,Vinciarelli:1978kp,Yoneya:1978dt,Cornwall:1979hz,Mack:1978rq,Nielsen:1979xu} was established in the maximal center gauge \cite{DelDebbio:1998luz}. Finally there is the Gribov--Zwanziger mechanism \cite{Gribov:1977wm,Zwanziger:1988jt}, which relies on Coulomb gauge.

We may argue that confinement is a gauge invariant phenomenon, so we should also have a gauge invariant description of it.
However, this may be too ambitious a goal. 
Let me remind you that even the parton picture makes sense only for certain gauges, like the light-cone gauge (see the
lecture by J.~P.~Blaizot). 
So if the parton picture can be realized only for certain gauges we should not be surprised that in the non-perturbative regime the explanation of confinement requires fixing the gauge.

Before we develop the various pictures of confinement let us summarize some important phenomenological aspects of confinement 
and chiral symmetry breaking. In this context,
I will also 
summarize the various order parameters 
frequently used in QCD to distinguish its phases. They will show up when we explain the pictures of confinement.

\igc{0.45}{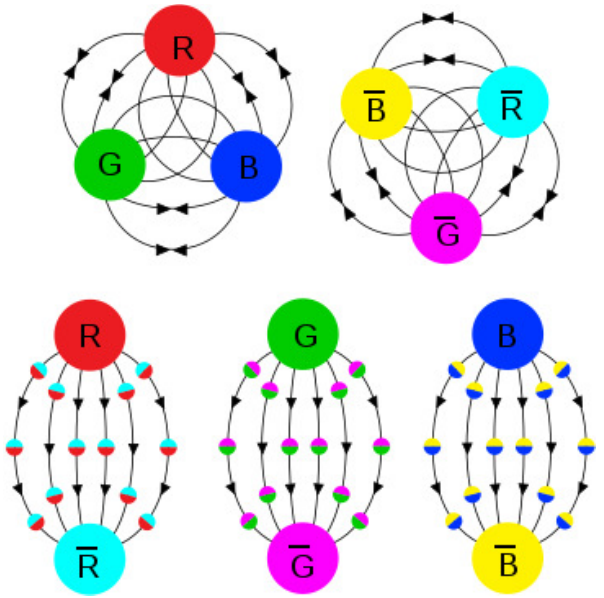}{part1:colors}{Illustration of the valence quark structure of hadrons: top left-baryon, top right-antibaryon, bottom-mesom.}

\begin{figure}
\centering
\includegraphics[width=.5\linewidth]{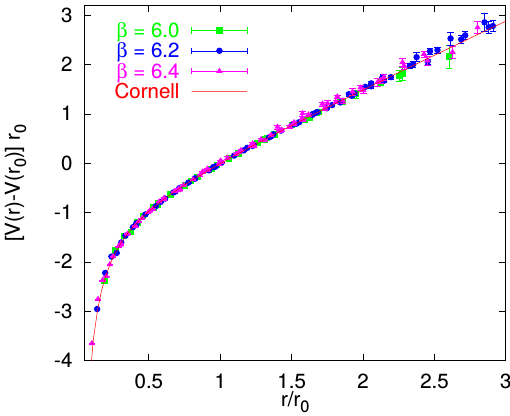}
\caption{Static quark--anti-quark potential obtained in a lattice calculation \cite{Bali:2000gf}.}
\label{fig:part1:static_quark_potential}
\end{figure}

The strongly interacting particles, i.e.~the hadrons, are built up from so-called valence quarks, 
which interact via the exchange of gluons. Mesons are built up from a quark and an anti-quark while baryons are formed by three quarks. Besides these valence quarks there are also ``sea'' quarks which fill the negative energy Dirac sea. Although the quarks 
carry, besides electric charge, also a ``color'' charge all
hadrons are colorless. Colored states do not exist in nature according to the confinement hypothesis. The simplest hadronic system is a heavy meson consisting of a heavy quark and a heavy anti-quark. 
These particles can be described by potential models and the corresponding 
potential can be measured on the lattice, see Fig.~\ref{fig:part1:static_quark_potential}.

\begin{figure}
\centering
\parbox{.45\linewidth}{\includegraphics[width=\linewidth]{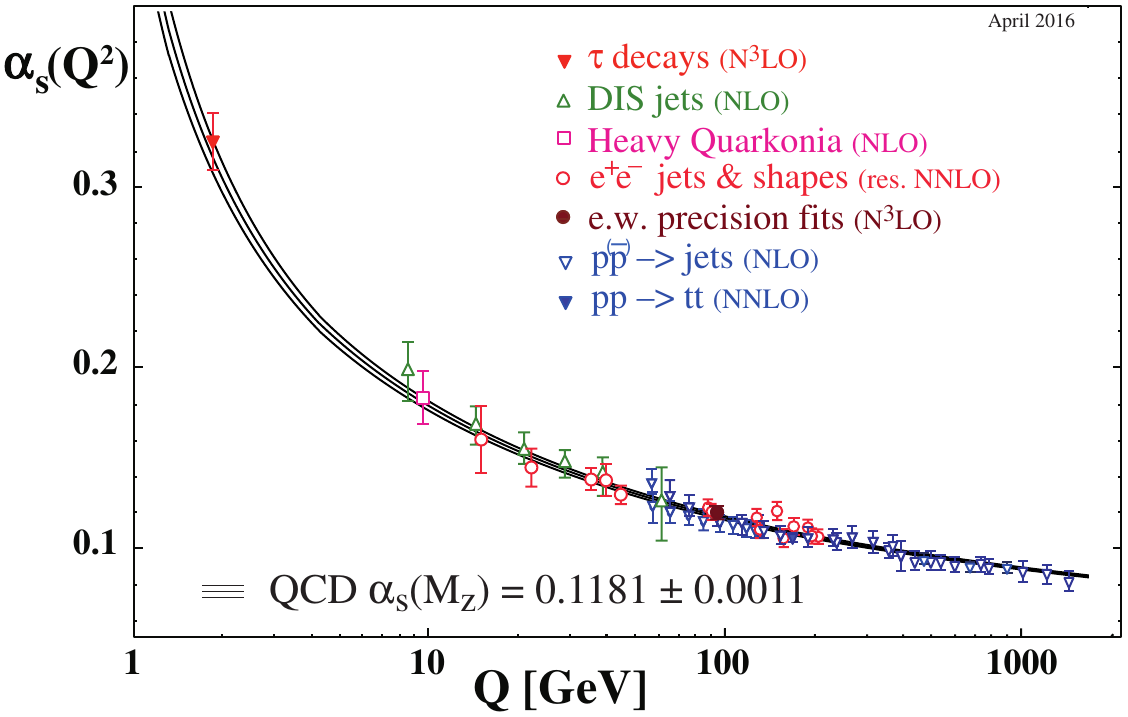}}\qquad
\parbox{.45\linewidth}{\includegraphics[width=\linewidth]{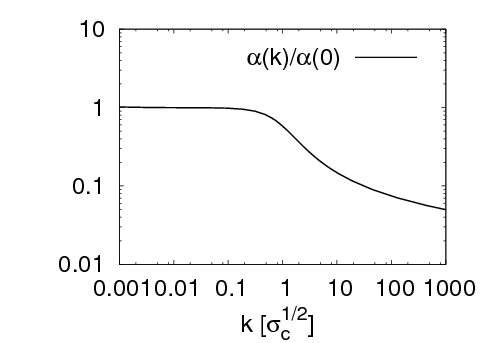}}
\caption{The running coupling $\alpha_s(Q)$. (left) Perturbative calculation. (right) Non-perturbative calculation from Ref.~\cite{Schleifenbaum:2006bq}.}
\label{fig:part1:coupling_nonperturbative}
\end{figure}

The lattice QCD simulations show that at large distances the potential grows linearly with the distance between the two static color charges, with a coefficient referred to as string tension (taken in units of MeV$^2\equiv$ energy per length)
\begin{equation}
V(r) = -\frac{\alpha_s}{r} + \sigma r,\quad\quad\sigma=(440\,\mathrm{MeV})^2 .
\label{eq:part1:static_quark_potential}
\end{equation}
Converting this quantity to more familiar units one obtains 
\begin{equation}
\label{120-5x}
\sigma = 157\,\mathrm{kN}  \, ,
\end{equation}
which corresponds to the weight of a mass of about 16 tons and 
which justifies calling this type of interactions `strong.' At short distance the potential behaves like an ordinary Coulomb potential, with a   coefficient 
given by the {\em running coupling} constant $\alpha_s$.  This quantity can be evaluated in perturbation theory in the high-momentum regime (Fig.~\ref{fig:part1:coupling_nonperturbative}, left), while non-perturbative 
methods are required in the
low-momentum regime, where it saturates (Fig.~\ref{fig:part1:coupling_nonperturbative}, right).

\subsection{Chiral symmetry}

The quarks receive a mass through the Higgs mechanism. This is the so-called ``current mass'', which enters
the QCD Lagrangian. This mass is small for the light
quark flavors $u$, $d$ (and $s$), see Table~\ref{tab:part1:quark_masses} (top). There is, however,
a large discrepancy between the sum of bare (current) masses of the valence quarks inside a hadron and the mass of the
hadron that they form (see Table.~\ref{tab:part1:quark_masses}, bottom). Ordinary hadrons,
like the proton or the neutron, receive most of their mass through the mechanism of
spontaneous breaking of chiral symmetry (SBCS), which converts the light bare quarks into massive constituent quarks (see the lecture by R.~Pisarski).

\begin{table}
\centering
\begin{tabular*}{.5\linewidth}{c@{\extracolsep\fill}c@{\extracolsep\fill}c}
$m_u \approx 2.3 \,\mathrm{MeV}$ & $m_d \approx 4.8 \,\mathrm{MeV}$ &$m_s \approx 95 \,\mathrm{MeV}$\\[2ex]
\textbf{Particle} & \textbf{Quark Content} & \textbf{Mass} (in MeV) \\
p & $uud$ & 938\\
n & $udd$ & 939\\
$\rho^+$ & $u\bar{d}$ & 775\\
$\omega$ & $\dfrac{u\bar{u} + d\bar{d}}{\sqrt{2}}$ & 782\\
$\phi$ & $s\bar{s}$ & 1020\\
\end{tabular*}
\caption{Current quark masses (top) and masses of baryons (bottom). One can see that the
small quark masses do not add up to the masses of the baryons that they form.}
\label{tab:part1:quark_masses}
\end{table}

Let us focus now on the mass term of the QCD Lagrangian:
\begin{equation}
\mathcal{L}_m = m_u\bar{u}u + m_d\bar{d}d + m_s\bar{s}s = 
\begin{pmatrix} \bar{u} & \bar{d} & \bar{s} \end{pmatrix} 
\begin{pmatrix} \,\,m_u \,\,& & \\ & m_d & \\ & & m_s\end{pmatrix} \begin{pmatrix} \,u\, \\ d \\ s \end{pmatrix}
\label{eq:part1:mass_term}
\end{equation}
Assuming the light quark masses to be equal $(m_n = m_d =  m_s)$, the QCD Lagrangian has the {\em flavor symmetry}, i.e. it is 
invariant under the transformation: 
\begin{equation}
\begin{pmatrix}  \,u \, \\ d \\ s \end{pmatrix} \rightarrow U \begin{pmatrix}  \,u \, \\ d \\ s  \end{pmatrix} ,
\label{eq:part1:flavor_symmetry}
\end{equation}
where $U$ is an arbitrary unitary $3 \times 3$-matrix. The flavor transformation (\ref{eq:part1:flavor_symmetry}) mixes the light quark 
flavors and the entirety of such transformations forms the flavor group $SU (N_F = 3$). 
If we neglect the small current masses $(m_n = m_d = m_s = 0)$ the QCD Lagrangian is also 
invariant under the {\em chiral transformation}:
\begin{equation}
\begin{pmatrix} \,\,u\,\, \\ d \\ s \end{pmatrix} \rightarrow U^{\gamma_5} 
\begin{pmatrix} \,\,u\,\, \\ d \\ s \end{pmatrix}
\label{eq:part1:chiral_symmetry}
\end{equation}
which differs from the flavor symmetry by the $\gamma_5$ 
in the exponent of the flavor matrix $U$. Expressing $\gamma_5$ in terms of the left and right projectors
\begin{equation}
P_{R/L} = \frac{1}{2} \left( 1 \pm \gamma_5 \right),
\end{equation}
we find 
\begin{equation}
U^{\gamma_5} = P_RU + P_LU^{-1} .
\label{eq:part1:projection_operators}
\end{equation}
The left and right handed quarks transform oppositely in flavour space under a chiral transformation, while
left and right quarks transform in the same way under a flavour transformation.

Obviously chiral symmetry makes no sense for the heavy quark flavors ($c$,$b$,$t$). It
is, however, a good approximation for the up and down quarks and a good starting point for the strange quark sector.

In QCD chiral symmetry is broken in three different ways:
\begin{itemize}
	\item Explicit breaking by the current quark masses, see eq.(\ref{eq:part1:mass_term}).
	\item Spontaneous symmetry breaking (most important in the context of this lecture).
	\item Anomalous breaking due to quantum effects (see R. Pisarski's lecture).
\end{itemize}
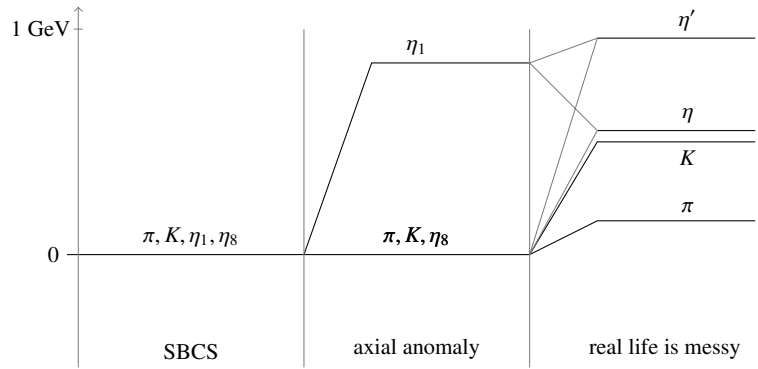
\begin{figure}[htb]
\centering
\begin{tikzpicture}[scale=3]
  \draw (-0.05,0) -- (1,0); 
  \node[left] at (-0.05,0) {0};
  \node[above] at (0.5,-0.5) {SBCS};
  \draw[->,gray] (0,-.5) -- (0,1.1);
  \draw[gray] (-0.02,1) -- (0.02,1);
  \node[left] at (0,1) {1 GeV};
  \node[above] at (0.5,0) {$\pi, K, \eta_1, \eta_8$};
  \draw (1,0) -- (2,0); 
  \node[above] at (1.5,0) {$\pi, K, \eta_8$};
  \node[above] at (1.5,-0.5) {axial anomaly};
  \draw (1,0) -- (1.3,0.85) -- (2,0.85); 
  \node[above] at (1.5,0) {$\pi, K, \eta_8$};
  \node[above] at (1.5,0.85) {$\eta_1$};
  \draw[gray] (1,-.5) -- (1,1);
  \draw (2,0) -- (2.3,0.15) -- (3,0.15); 
  \draw (2,0) -- (2.3,0.5) -- (3,0.5); 
  \draw (2.3,0.96) -- (3,0.96); 
  \draw (2.3,0.55) -- (3,0.55); 
  \draw[gray] (2,0) -- (2.3,0.55) -- (2,0.85) -- (2.3,0.96) -- (2,0); 
  \draw[gray] (2,-.5) -- (2,1);
  \node[above] at (2.6,-0.5) {real life is messy};
  \node[above] at (2.7,0.15) {$\pi$};
  \node[below] at (2.7,0.5) {$K$};
  \node[above] at (2.7,0.55) {$\eta$};
  \node[above] at (2.7,0.96) {$\eta'$};
\end{tikzpicture}
\caption{Consequences of the various forms of chiral symmetry breaking for the pseudoscalar meson masses.}
\label{fig:part1:CSSB_pattern}
\end{figure}
The consequences of the various forms of breaking the chiral symmetry for the masses of the pseudo-scalar mesons are the following (cf.~Fig.~\ref{fig:part1:CSSB_pattern}):
\begin{itemize}
	\item If one ignores the current quark masses the pseudoscalar mesons 
	(identified with Goldstone bosons of the SBCS) become exactly massless.
	\item Switching on the chiral (axial) 
	anomaly (the violation of the chiral symmetry by quantum effects) induces a mass for the $\eta_1$ meson of about 1GeV.
	\item Including also the current quark masses, the pseudoscalar mesons become massive and furthermore $\eta_1$ and $\eta_8$ mix, resulting in $\eta$ and $\eta'$.
\end{itemize}

\subsection{Order parameters in QCD}

As mentioned in the introduction QCD exists in several phases depending on the external conditions like temperature or matter density. The different phases are distinguished by order parameters, which I summarize below. 
\subsubsection{Quark condensate}
\begin{figure}[t]
\begin{tikzpicture}[scale=3]
\draw[fill=teal!30] (1.8,0) arc [start angle=0, end angle=90, x radius=1.8, y radius=1] ;
\draw[teal!30,fill=teal!30] (0,1) -- (0,0) -- (1.8,0) -- (0,1) ;
\draw[<->,thick] (0,2)node[left]{$T\;$} -- (0,0) -- (3,0) node[below]{$\rho$} ;
\draw (1.715,0.3) arc [start angle=170, end angle=90, x radius=1, y radius=0.5] ;
\node at (1,1.8) {quark-gluon plasma} ;
\node at (1,1.5) {$\langle\bar{q}q\rangle=0$} ;
\draw (.36,.48) circle [radius=.18] ;
\draw[red,fill=red] (.3,.55) circle [radius=.05] ;
\draw[olive,fill=olive] (.46,.49) circle [radius=.05] ;
\draw[blue,fill=blue] (.31,.4) circle [radius=.05] ;
\draw (.62,.72) circle [radius=.13] ;
\draw[red,fill=red] (.66,.76) circle [radius=.05] ;
\draw[red,very thick] (.58,.68) circle [radius=.05] ;
\node at (1,0.3) {$\langle\bar{q}q\rangle\neq0$} ;
\node at (2.4,0.3) [align=left] {color\\superconductivity} ;
\end{tikzpicture} 
\caption{The chiral phases of QCD distinguished by the quark condensate 
	$\langle \bar{q} q \rangle$.}
\label{part1:phase_diagram}
\end{figure}
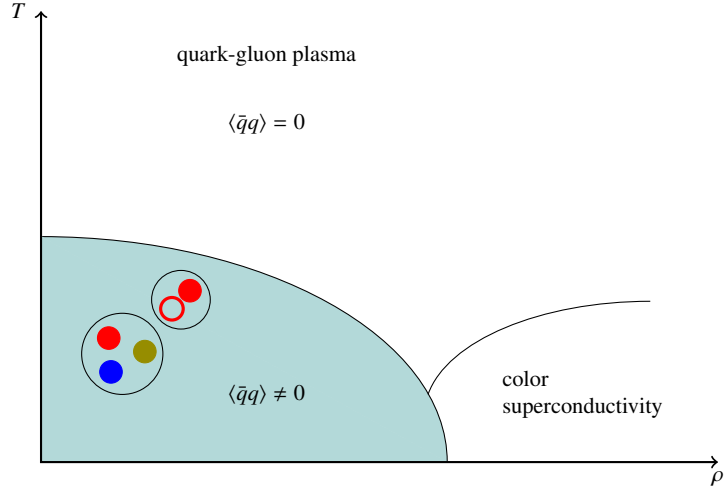

The order parameter of the chiral symmetry breaking is the \textbf{quark condensate} $\langle \bar{q} q \rangle$. It
 is non-zero in the hadronic phase and vanishes in the deconfined phase, see Fig.~\ref{part1:phase_diagram}. 
 From hadron phenomenolgy one extracts the value $\langle \bar{q} q \rangle = (- 230 \, MeV)^3$.  
 A non-zero quark condensate violates the chiral symmetry in analogy to what happens in superconductors --- the electron pair condensation violates the particle number conservation. We will see later that in many respects
  the QCD vacuum indeed resembles a superconductor:
   the quark sector behaves like an ordinary superconductor (the quark vacuum wave functional looks, in a certain approximation, like a BCS-state, see Sec.~\ref{sect4.4}), while the gluon sector can be 
   interpreted as a {\em dual} superconductor (see Sec.~\ref{sect2}).

\subsubsection{The Wilson loop}
At zero temperature the most familiar order parameter of confinement is the Wilson loop
which is defined by\footnote{This quantity was originally introduced as an order parameter by F.~J.~Wegner \cite{Wegner:1984qt}.}
\begin{equation}
W [A] (C) = \frac{1}{d_r} \tr~P~\exp\left(\ii g \oint_C \dd x^{\mu} \, A_{\mu}(x) \right) \, ,
\label{eq:part1:wilson_loop1}
\end{equation}
where $d_r$ is the dimension of the (group) representation of the gauge field $A_\mu$. Furthermore $C$ is an arbitrary closed loop in four-dimensional Euclidean space and $P$ denotes the path ordering along this loop.
For a non-Abelian gauge theory the gauge field\footnote{$t_a$ are the (hermitian) generators of the gauge group 
satisfying $[t_a, t_b] = i f_{abc} t_c$, where the $f_{abc}$ are the structure constants.}  $A_\mu = A^a_\mu t_a$ 
is matrix-valued and the path ordering complicates the evaluation of $W (C)$ considerably.

To exhibit the physical meaning of the Wilson loop let us rewrite it as
\begin{equation}
W [A] (C) = \tr~P~\exp \left(\ii\int \dd^4 x \, j^{\mu}(x;C) A_{\mu}(x) \right) \, ,
\label{eq:part1:wilson_loop2}
\end{equation}
where 
\begin{equation}
j^{\mu}(x;C) = g \oint_C \dd y^{\mu} \, \delta(x-y)
\label{eq:part1:wilson_loop_current}
\end{equation}
is the current generated by a point color charge on the trajectory $C$. The integrand
in Eq.~(\ref{eq:part1:wilson_loop2}) is 
the interaction Lagrangian of a unit  point charge on the trajectory $C$ with  a gauge field $A_\mu (x)$.

As we will see later, 
the expectation value of the Wilson loop obeys an area law in a confining theory and a perimeter law in a non-confining theory
\begin{equation}
\label{F10*}
\langle W  (C) \rangle =
\begin{cases}
\ee^{- \sigma \cA (C)} , & \text{confinement,} \\
\ee^{- \kappa \cP (C)} , & \text{deconfinement.}
\end{cases}
\end{equation}
Here $\cA (C)$ is the area included by the loop $C$ and $\cP (C)$ is its perimeter. The coefficient $\sigma$ is called {\em string tension}
for reasons which will be explained below. Due to the property (\ref{F10*}) the Wilson loop is an order parameter of confinement.

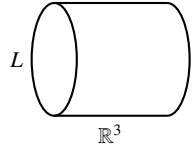
\begin{figure}[hbt]
\centering
\begin{tikzpicture}[scale=1.5,style=thick]
 \draw (0,0) node[left]{$L\quad$} ellipse (.2 and 0.5) ;
 \draw (1,-0.5) arc (-90:90:.2 and 0.5) ;
 \draw (0,-0.5) -- node[below]{$\mathbb{R}^3$}(1,-0.5) ;
 \draw (0,0.5) -- (1,0.5) ;
\end{tikzpicture}
\caption{Illustration of the Euclidean space-time manifold $S^1(L) \times \mathbb{R}^3$ at finite temperature $L^{-1}$.}
\label{fig:part1:finite_temperature_cylinder}
\end{figure}

At zero temperature the Euclidean space-time manifold is $\RR^4$, and the theory exhibits an
O(4)-symmetry, i.e.
the  position and orientation 
of the loop $C$ in Euclidean space is irrelevant for the behavior of the Wilson loop $\langle W (C) \rangle$. 
This changes at finite temperature $T = L^{- 1}$, where the fields in the functional integral of the grand canonical 
partition function $\Tr \exp (- L H)$
 have to satisfy periodic (for Bose fields $A$) and anti-periodic (for Fermi fields $\Psi$) boundary conditions in the Euclidean time 
 $x^4 =  \ii x^0, x^0 = ct$
\begin{eqnarray}
\label{235-2x}
A_\mu  (x^4 = L/2) & = & A_\mu (x^4 = - L/2) \, , \nonumber\\
\Psi (x^4 = L/2) & = & - \Psi (x^4 = - L/2) \, .
\end{eqnarray}
These boundary conditions compactify the Euclidean space-time manifold to $S^1 (L) \times \RR^3$ (see 
Fig.~\ref{fig:part1:finite_temperature_cylinder}). In this case it matters whether $C$ is a closed loop in $\RR^3$ (existing only during a single 
time instant) or whether $C$ evolves also along the (Euclidean) time axis. In the latter case $W (C)$
is called {\em temporal} Wilson loop, in the former case {\em spatial}
Wilson loop.

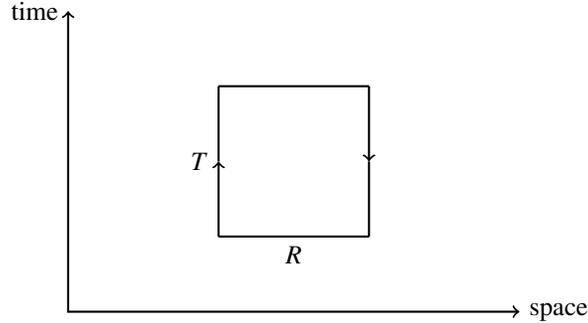
\begin{figure}
\centering\normalsize
\begin{tikzpicture}
\draw[thick,<->] (0,4) node[left]{time} -- (0,0) -- (6,0) node[right]{space};
\draw[thick] (2,1) -- (3,1) node[below]{$R$} -- (4,1) ;
\draw[thick] (2,3) -- (4,3) ;
\draw[thick,->] (2,1) -- (2,2) node[left]{$T$} ;
\draw[thick] (2,2) -- (2,3) ;
\draw[thick,->] (4,3) -- (4,2) ;
\draw[thick] (4,2) -- (4,1) ;
\end{tikzpicture}
\caption{The temporal Wilson loop.}
\label{fig:part1:temporal_wilson_loop}
\end{figure}

To exhibit the physical meaning of the
 temporal Wilson loop consider the case where $C$ is a rectangular loop with its
  edges running parallel or perpendicular to the time-axis, see Fig.~\ref{fig:part1:temporal_wilson_loop}. This loop can be interpreted as the creation of a particle and an 
 antiparticle 
 at some initial time, which are separated then by a spatial distance $R$, followed by the time-evolution of the particle and anti-particle over a time-period $T$ and the subsequent annihilation of the particle 
 and antiparticle pair. 
It can be shown that for $T \gg R$ the temporal Wilson loop is related to the particle-antiparticle potential $V(R)$ by
\begin{equation}
\label{243-F11}
\langle W (C) \rangle \sim \ee^{- V (R) T} , \qquad T \to \infty .
\end{equation}
In a confining theory with $\langle W (C) \rangle$ obeying the area law (\ref{F10*}) this potential obviously rises linearly at large distances
\begin{equation}
\label{248-F11-s}
V (R) \sim \sigma R , \qquad R \to \infty .
\end{equation}

To exhibit the physical meaning of a {\em spatial} Wilson loop let us consider QED where the path ordering is irrelevant and the trace does not occur since $d_r = 1$:
\begin{equation}
\label{253-F11-daq}
W (C)  = \exp \left(\ii g \oint\limits_C \dd x^\mu \, A_\mu (x) \right) .
\end{equation}
For a spatial Wilson loop the path $C$ is a closed loop in $\RR^3$. 
Using Stoke's theorem we obtain
\begin{equation}
\label{259-222}
W (C) = \exp \left(\ii g \oint\limits_{\Sigma (C)} \dd\vf \cdot \vB \right) ,
\end{equation}
where $\vB = \vec{\nabla} \times \vA$ is the magnetic field and $\Sigma (C)$ any area bounded by $C$. The quantity in the exponent 
\begin{equation}
\label{264-222-1}
\Phi   = \oint\limits_{\Sigma (C)} \dd\vf \cdot \vB 
\end{equation}
is the magnetic flux through the loop $C$. The spatial Wilson loop $\langle W (C) \rangle$ measures the magnetic flux through $C$ also in a non-Abelian gauge theory. However, in that case a non-Abelian version of Stokes' theorem is required to establish this connection. 

At zero temperature the spatial and temporal Wilson loops behave in exactly the same way due to the exact O(4) symmetry 
of the Euclidean space-time manifold $\RR^4$. 
This symmetry is spoiled at finite temperature $L^{- 1}$ due to the finite length $L$ of the 
Euclidean time axis and its compactification caused by the (anti-)periodic
boundary condition (\ref{235-2x}). For non-zero temperatures below the deconfinement transition the spatial and temporal 
string tension are approximately equal to their common zero temperature value. During the deconfinement transition the temporal string 
tension drops to zero,
while the spatial string tension slightly increases with the temperature. 

At finite temperatures $L^{- 1} \geq R$ the temporal Wilson loop can no longer serve as order parameter of confinement, since
the static quark--anti-quark potential can no  longer be  
extracted from it. In this case there exists a more efficient order parameter, which is the Polyakov loop. 

\subsubsection{The Polyakov loop}\label{sect133}

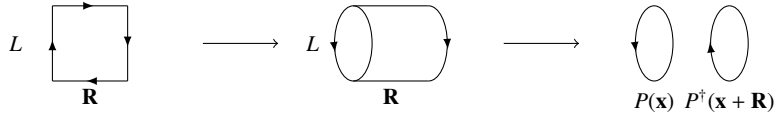
\begin{figure}[h]
	\centering
	\begin{tikzpicture}[scale=1.0]
	\draw[middlearrow=latex] (0,0) -- node[left]{$L\quad$} (0,1) ;
	\draw[middlearrow=latex] (0,1) -- (1,1) ;
	\draw[middlearrow=latex] (1,1) -- (1,0) ;
	\draw[middlearrow=latex] (1,0) -- node[below]{$\vec{R}$} (0,0) ;
	%
	\draw[->] (2,0.5) -- (3,0.5) ;
	\draw (4,1) -- (5,1) ;
	\draw (4,0) -- node[below]{$\vec{R}$}(5,0) ;
	\draw[middlearrow=latex] (4,0.5) node[left]{$L\quad\,$} ellipse (.25 and .5) ;
	\draw[middlearrow=latex reversed] (5,0) arc (-90:90:.25 and 0.5);
	\draw[->] (6,0.5) -- (7,0.5) ;
	\draw[middlearrow=latex] (8,0.5) ellipse (.25 and .5) ;
	\draw[middlearrow=latex reversed] (9,0.5) ellipse (.25 and 0.5) ;
	\node[below] at (8,0) {$\vphantom{P^\dag\vec{R}}P(\vec{x})$};
	\node[below] at (9,0) {$P^\dag(\vec{x}+\vec{R})$};
	\end{tikzpicture}
	\caption{Schematic illustration of the emergence of the Polyakov loop $P (\vx)$ and its
		complex conjugate $P^\dagger (\vx + \vR)$ from a temporal Wilson loop extending over
		the whole Euclidean time axis of length $L$.}
	\label{part1:wilson:polyakov_correlator}
\end{figure}

Consider a rectangular temporal Wilson loop in finite temperature QED
which extends over the whole Euclidean
time axis. Due to the compactification of the time axis the two spatial pieces (of length $R$) of the loop $C$ fall on top of each other, see Fig.~\ref{part1:wilson:polyakov_correlator}. Since they are traveled in opposite 
direction their contribution to the Wilson loop cancel. What is left from the original Wilson loop are two loops (both of length $L$) a spatial distance $\vR$ apart, running in opposite direction 
along the compactified time-axis. Wilson loops running along the whole compactified time axis
\begin{equation}
\label{334-13b}
P (\vx) = \frac{1}{d_r} \tr P \exp\biggl\{i g \int^\beta_0 \dd x^4 \, A_4 (x^4, \vx) \biggr\}
\end{equation}
are referred to as Polyakov loops. What we have shown above is that (in QED) a temporal rectangular Wilson loop of spatial extension $R$ running along the whole compactified time-axis is equivalent to the product ${P} (\vx) {P}^\dagger (\vx + \vR)$
of two Polyakov loops. With Eq.~(\ref{243-F11}) this implies that
\begin{equation}
\label{339-2-x}
\bigl< P (\vx) P^\dagger (\vx + \vR) \bigr> \sim \ee^{- L V (R)} .
\end{equation}
It can be shown that this equation remains valid in the non-Abelian case provided the r.h.s. is replaced by its color average \cite{Svetitsky:1985ye}
\begin{equation}
\label{344-dkfd}
\ee^{- L V (R)} \to \ee^{- L V_\mathrm{singlet}(R)} + (N^2 - 1) \ee^{- L V_\mathrm{adjoint}(R)} ,
\end{equation}
where ``singlet'' and ``adjoint'' refer to the color representations of the external color sources. 

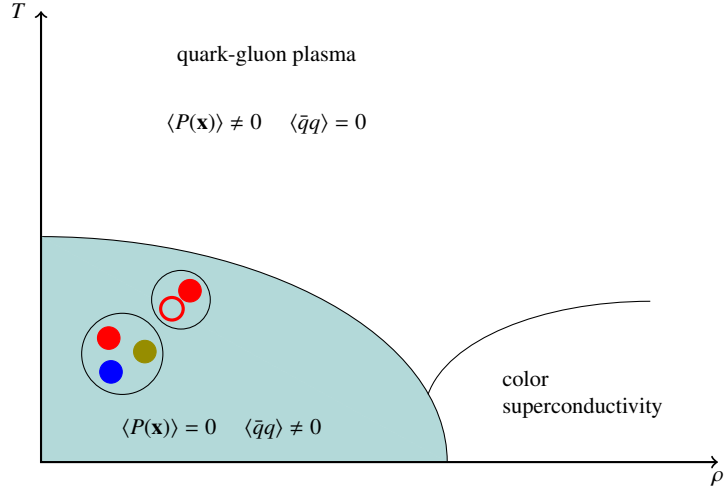
\begin{figure}
\begin{tikzpicture}[scale=3]
\draw[fill=teal!30] (1.8,0) arc [start angle=0, end angle=90, x radius=1.8, y radius=1] ;
\draw[teal!30,fill=teal!30] (0,1) -- (0,0) -- (1.8,0) -- (0,1) ;
\draw[<->,thick] (0,2)node[left]{$T\;$} -- (0,0) -- (3,0) node[below]{$\rho$} ;
\draw (1.715,0.3) arc [start angle=170, end angle=90, x radius=1, y radius=0.5] ;
\node at (1,1.8) {quark-gluon plasma} ;
\node at (1,1.5) {$\langle P(\vec{x})\rangle\neq0$ \quad $\langle\bar{q}q\rangle=0$} ;
\draw (.36,.48) circle [radius=.18] ;
\draw[red,fill=red] (.3,.55) circle [radius=.05] ;
\draw[olive,fill=olive] (.46,.49) circle [radius=.05] ;
\draw[blue,fill=blue] (.31,.4) circle [radius=.05] ;
\draw (.62,.72) circle [radius=.13] ;
\draw[red,fill=red] (.66,.76) circle [radius=.05] ;
\draw[red,very thick] (.58,.68) circle [radius=.05] ;
\node at (0.8,0.15) {$\langle P(\vec{x})\rangle=0$ \quad $\langle\bar{q}q\rangle\neq0$} ;
\node at (2.4,0.3) [align=left] {color\\superconductivity} ;
\end{tikzpicture} 
\caption{The phases of QCD distinguished by the Polyakov loop $\langle P \rangle$.}\label{phase_diagram_again}
\end{figure}

Instead of the Polyakov loop correlator \eqref{339-2-x} one can use the Polyakov loop itself as order parameter. This is because the Polyakov loop is related to the free energy of a static quark:
\begin{equation}
\bigl< P(\vec{x}) \bigr> = \ee^{-L(F_q-F_0)}.
\label{eq:part1:free_quark_energy}
\end{equation}
Here $F_q$ and $F_0$ are the free energies in the presence and absence, respectively, of an isolated quark.
For confined systems isolated quarks have infinite free energy and the Polyakov loop vanishes:
\begin{equation}
(F_q - F_0) \to\infty \qquad\Rightarrow\qquad \bigl< P(\vec{x}) \bigr>=0 .
\label{eq:part1:quark_energy_confined}
\end{equation}
In the deconfined phase, however, the free energy of a single quark is finite and the Polyakov loop is non-zero 
(see Fig.~\ref{phase_diagram_again})
\begin{equation}
(F_q - F_0)=\text{finite}\qquad\Rightarrow\qquad \bigl< P(\vec{x}) \bigr> \neq 0 .
\label{eq:part1:quark_energy_deconfined}
\end{equation}
Therefore, at finite  temperature the Polyakov loop can serve as order parameter for confinement. 

\subsubsection{The spatial 't Hooft loop}
There is yet another order parameter, or precisely a {\em disorder}
parameter, of confinement, which is the {\em spatial 't Hooft loop} \cite{tHooft:1977nqb}. The proper definitions will 
be given later in Eq.~(\ref{eq:part_iii:thooft_loop}), but it is worthwhile to introduce it already here in the context of the other order parameters.

While the spatial Wilson loop measures the magnetic flux,
the spatial 't Hooft loop $V (C)$
measures the electric flux through the loop $C$ (Fig.~\ref{fig:part1:wilson_thooft}).
One can show that its expectation value obeys a perimeter and an area law in the confined 
and deconfined, respectively, phase \cite{tHooft:1979rtg}
\begin{equation}
\label{34716}
\langle V (C) \rangle = 
\begin{cases}
\ee^{ - \tilde{\kappa} \cP(C)} , & \text{confinement,} \\
\ee^{ - \tilde{\sigma} \cA(C)} , & \text{deconfinement.}
\end{cases}
\end{equation}
This is opposite to the temporal Wilson loop, c.f.~Eq.~(\ref{F10*}). 

\igc{0.8}{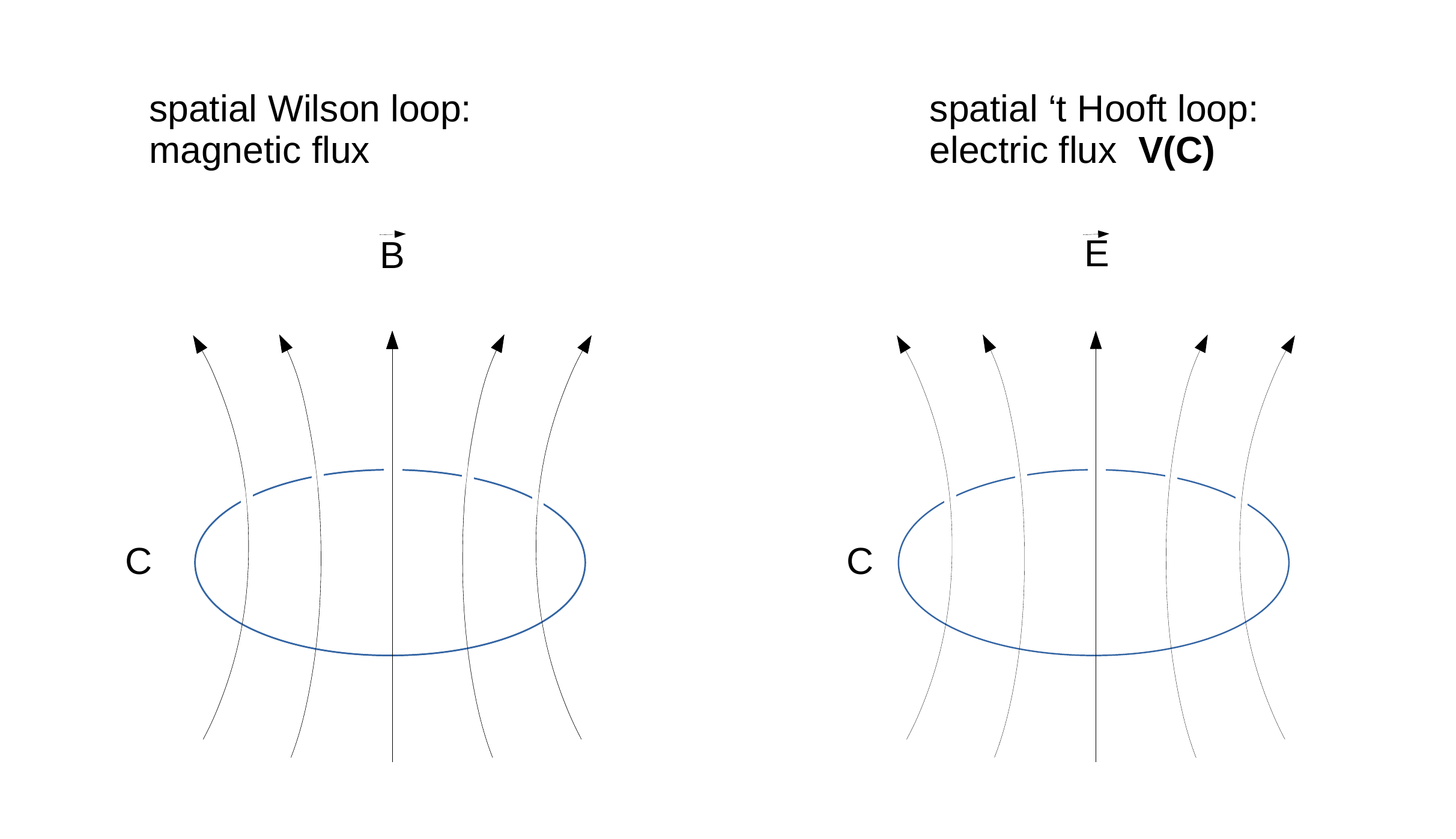}{part1:wilson_thooft}{Illustration of the spatial Wilson loop and the spatial 't Hooft loop. They measure the magnetic and electric flux, respectively.}


After having  summarized some basic features of QCD we will now attempt to explain how these features emerge from the underlying theory. We will focus our attention on 
the confinement phenomenon and the deconfinement phase transition at finite temperature. Further issues will be the spontaneous breaking of chiral symmetry and the topological  properties of gauge fields.
We begin with the dual Mei{\ss}ner effect as a possible explanation of confinement.


\section{The Magnetic monopole picture of confinement}\label{sect2}

\begin{figure}[tb]
\includegraphics[width=.45\linewidth]{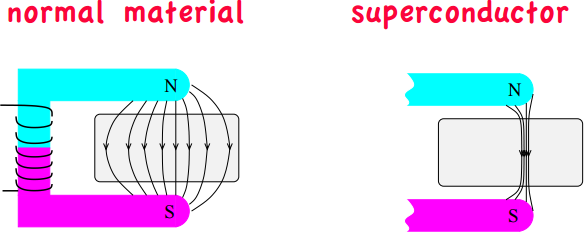}\hfill
\includegraphics[width=.35\linewidth]{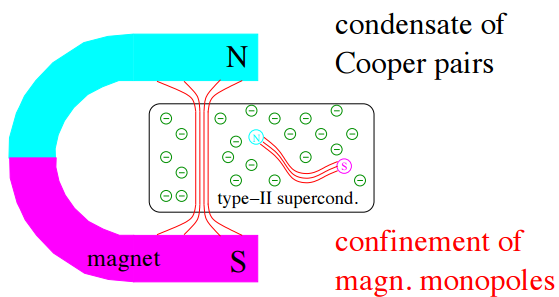}
\caption{ (left) The distribution of magnetic field lines in the presence of a ``normal'' material and of a type-II superconductor.\newline
(right) Type-II superconductor and confinement of magnetic monopoles.}
\label{fig:part1:normal_vs_superconductor}
\end{figure}

When a {\em normal}
material is brought into a magnetic field the field lines are somewhat disturbed but pass through the material, see Fig.~\ref{fig:part1:normal_vs_superconductor}. This is different for a superconductor. 
The magnetic field lines cannot penetrate into a type I-superconductor  but are expelled from 
it (Mei{\ss}ner effect) 
due to presence of the so-called {\em London currents} which are induced at its surface by the magnetic field. 
The situation is different  in a type II superconductor. If the magnetic field exceeds a critical strength it can pass through a type II superconductor in the form vortex lines, i.e.~small magnetic flux tubes. Inside these flux tubes the material is in its normal conducting phase.

\subsection{The dual Mei\ss ner effect}

In the vacuum 
the magnetic field of a (hypothetical) 
magnetic monopole-antimonopole pair is given by Coulomb's law like the electric field of two opposite (electric) 
point charges. When the monopole-antimonopole pair is brought into a type II superconductor the field lines are squeezed into magnetic flux tubes and as a consequence
the static potential between the magnetic point charges is no longer Coulombic but linearly rising.
This implies that magnetic monopoles are confined inside a type II superconductor.

\begin{figure}
\hfill
\includegraphics[width=.35\linewidth]{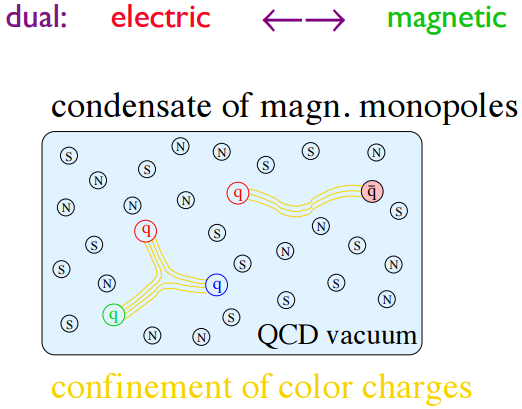}\hfill\hfill
\includegraphics[width=.45\linewidth]{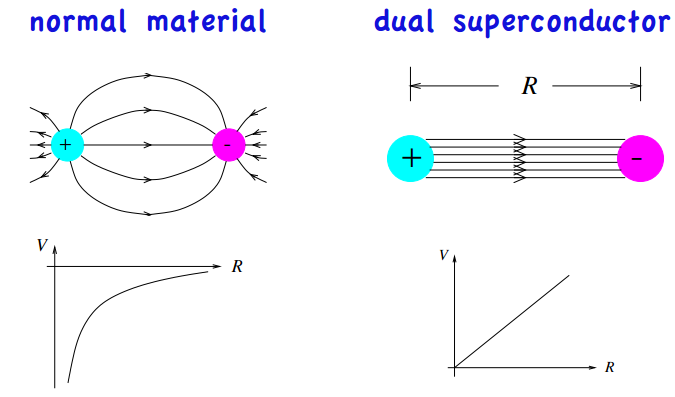}\hfill\null
\caption{(left) Confinement of color charges in a condensate of magnetic monopoles.
(right) (Electric) charge-anticharge potential in a {\em normal} material (or in the vacuum) and in a {\em dual superconductor}, 
respectively. }
\label{fig:part1:dual_meissner_superconductor}
\end{figure}

The QCD vacuum can be understood as a so-called {\em dual superconductor}. {\em Dual}
means that the roles of electric and magnetic fields and charges are interchanged. While the ordinary superconductor consists of a condensation of electron pairs, the dual superconductor consists of a condensation of magnetic monopoles.
 
In a dual superconductor the electric field lines 
emitted from a pair of opposite electric charges are squeezed into flux tubes resulting in a 
linearly rising potential and thus confinement of electric charges (dual Mei{\ss}ner effect). The quarks carry color electric charges and are 
then confined in a dual (color) superconductor. 
The dual superconductor picture of the QCD vacuum was suggested independently by Nambu \cite{Nambu:1974zg}, 
't Hooft \cite{tHooft:1981bkw} and
Mandelstam \cite{Mandelstam:1974pi}.
Dual superconductor models of the QCD vacuum analogous to the Ginzburg--Landau  type theories of the ordinary superconductor 
were developed and are reviewed in Ref.~\cite{Baker:1991bc}.

\subsection{Emergence of magnetic monopoles in QCD}
The dual superconductor picture of the QCD vacuum, giving rise to a linear quark--anti-quark potential, has been confronted with  lattice QCD calculations. Strong evidence for the realization of the dual Mei{\ss}ner effect in the QCD vacuum was found in the so-called maximal Abelian gauge 
(suggested by 't Hooft in Ref.~\cite{tHooft:1981bkw}). 
Let us consider as an example the color group  SU(2).
We choose the generator $T_3$ to generate the Abelian subgroup U(1) while the remaining generators $T_1$, $T_2$ belong to the coset SU(2)$/$U(1)
\begin{equation}
SU(2) =
{\underbrace{\mathrm{U}(1)}_{\displaystyle\underset{\text{Abelian (neutral)}}{A_\mu^3}}}
\times
{\underbrace{\mathrm{SU}(2)/\mathrm{U}(1)}_{\displaystyle\underset{\text{non Abelian (charged)}}{A_\mu^1, A_\mu^2}}}
\label{eq:part1:gauges_groups}
\end{equation}
The non-Abelian components of the gauge field $A^1$, $A^2$ are charged with respect to the Abelian U(1) subgroup while the Abelian field $A^3 (x)$ is, 
of course, (color) neutral, like the photon.

The maximal Abelian  gauge is defined by minimizing the  norm (module) of the non-Abelian components of the gauge field
\begin{equation}
\int \dd^3 x \left[ \left(A_{\mu}^1(x)\right)^2 + \left(A_{\mu}^2(x)\right)^2\right]\quad\longrightarrow\quad\textrm{min}
\label{eq:part1:gauge_fixing}
\end{equation}
Having implemented such a gauge, one performs an Abelian projection, putting the non-Abelian components of the gauge fields to zero
\begin{equation}
A_{\mu}^1(x) = A_{\mu}^2(x) = 0 .
\label{eq:part1:Abelian_projection}
\end{equation}
On the lattice one finds that after Abelian projection the (temporal) string tension is found to be about $95 \%$ of the string tension of the full (i.e.~not Abelian projected) theory
\begin{equation}
\label{426-19}
\sigma_{AP} \sim 0.95\sigma_\mathrm{exp}.
\end{equation}

Contrary to instantons, magnetic monopoles do not arise as stable classical solutions of Yang--Mills theory but are {\em artifacts}
of the Abelian projection, i.e.~they show up only in the  Abelian projected configuration but not in the full (unprojected) gauge field:

Let $V$ be the gauge transformation required to bring a given gauge field configuration\footnote{We use here 
anti-hermitian generators $T_a$ satisfying $[T_a, T_b] = f_{abc} T_c$. Furthermore, we have absorbed the 
coupling constant $g$ into the gauge field.\label{foot18} Sometimes we will also use hermitian ge\-nerators $t_a$
satisfying $[t_a, t_b] =  i f_{abc} t_c$, which are related to the anti-hermitian ones 
by $T_a = - i t_a$.} $A_\mu = g A^a_\mu T_a$ into the (maximal) Abelian gauge
\begin{equation}
\label{F20A-xx}
A^V_\mu = V \partial_\mu V^\dagger + V A_\mu V^\dagger .
\end{equation}
The magnetic monopoles show up in the Abelian projection of the induced gauge field $V \partial_\mu V^\dagger =\mathrel{\mathop:} a_\mu$, i.e.~the 
Abelian magnetic field
\begin{equation}
\label{kjdf-2}
\vec{b} = \vec{\nabla} \times \vec{a}^3 
\end{equation}
contains an ordinary Dirac monopole with a Dirac string. If this is the case then the Abelian component $\cB^3$
of the non-Abelian magnetic field
\begin{equation}
\label{kadkj-2-1}
\cB_k = \frac{1}{2} \varepsilon_{kij} [a_i, a_j]
\end{equation}
contains an anti-monopole without a Dirac string so that in the Abelian component of the total field strength 
\begin{equation}
 \label{proc-641}
 F_{\mu \nu} [A] = \partial_\mu A_\nu - \partial_\nu A_\mu + [A_\mu, A_\nu]
\end{equation}
of the 
induced gauge field $a_\mu$
\begin{equation}
\label{qqq-3}
B^3_k = \frac{1}{2} \varepsilon_{kij} F^3_{ij} [a] = b_k + \cB^3_k
\end{equation}
only the unobservable Dirac string survives but not a magnetic monopole, see ref. \cite{Reinhardt:1997rm} for more details.

An interesting remark is that the charge of the magnetic monopole arising after Abelian projection
is topologically quantized. To be more precise the magnetic charge of monopoles arising after Abelian projection is given by the winding number of the 
mapping of a two-sphere $S^2$ around the monopole onto the coset SU(2)$/$U(1) of the gauge group. These mappings fall into the second homotopy group
$\Pi_2 (SU(2)/U(1) = \ZZ)$. 

 Magnetic monopoles arise in the Abelian 
projected field $\vec{a}^3 = (V \vec{\nabla} V^\dagger)^3$ if the original gauge field configuration contains 
topological gauge fixing obstructions, which turn into localized singularities (magnetic monopoles) unter 
Abelian projection\footnote{This has been studied in detail in the Polyakov gauge (131) (which is another Abelian gauge),
see refs. \cite{Reinhardt:1997rm}, \cite{Quandt:1998ik}.}. 
Though the magnetic monopoles are ``artifacts'' of the Abelian projections the topological obstructions are  
gauge invariant features
of the original gauge field. It is therefore not surprising that the topological charge of a gauge field (see also Sec.~\ref{sect3.4})  
\begin{equation}
\label{ert-4}
\nu = \frac{1}{32} \int d^4 x \epsilon^{\mu \nu \kappa \lambda} F_{\mu \nu} [A] F_{\kappa \lambda} [A]
\end{equation}
can be entirely expressed by the magnetic monopoles contained in the Abelian projected gauge field configuration, for more details see ref. 
\cite{Reinhardt:1997rm}, \cite{Quandt:1998ik}.

The lattice allows one to calculate observables from only the magnetic monopole content of the gauge fields by taking into account 
only Abelian projected configurations $A^3$ which contain magnetic monopoles. One finds then still about $90 \%$ of the full string tension. This fact was interpreted as evidence that magnetic monopoles are the dominant IR degrees of freedom which are responsible
for confinement. 

Lattice calculation in the maximal Abelian gauge were initiated in Ref.~\cite{Kronfeld:1987ri} 
and later on performed on large scale in Refs.~\cite{Suzuki:1989gp,Chernodub:1994pw,Bali:1996dm}.

A final comment is in order: One may ask about the role of instantons in the confinement mechanism. 
After all instantons are 
classical solutions of Yang--Mills theory in Euclidean space. 
On the lattice the instanton content of a gauge field can be extracted by the so-called cooling method.
One finds that the instantons account only for $10 \%$ of the string tension. 
Hence these objects seem to be of minor relevance for confinement. 


\section{The Center vortex picture of confinement}
\subsection{Introduction}
In everyday life one can find vortices in fluids, like water or air. Intuitively, the vortex is the part of a fluid which rotates around some  line, either closed or open. Its topology 
depends on the dimensionality 
of the considered system. 
In two dimensions vortices are isolated points while
in three dimensions they form closed loops. Finally, in four dimensions vortices become closed surfaces, which may be
self-intersecting and non-oriented. In a fluid the characteristic quantity of a vortex is its vorticity $\vec{\omega}$, which is defined as the curl of the velocity 
field $\vec{v}$, 
$\vec{\omega}=\vec{\nabla}
 \times \vec{v}$.
 
There is a formal analogy between hydrodynamics and a gauge field theory: The gauge potential $\vec{A}$ plays the role of the velocity
field $\vec{v}$, while the magnetic field $\vB = \vec{\nabla} \times \vA$ corresponds to the vorticity $\vec{\omega}$. 
This analogy is presented in more details in 
the following table

\medskip
\begin{center}
\small
\begin{tabular}{l@{\qquad\qquad}l}
\hline
Fluid dynamics & Gauge theory  \\
\hline
Velocity field  $\vec{v}$ & Gauge potential $\vec{A}$\\
Vorticity   & Magnetic field  \\
$\vec{\omega}=\vec{\nabla}\times \vec{v}$, $\nabla \cdot \vec{\omega}=0$ & $\vec{B}=\vec{\nabla}\times \vec{A}$, $\nabla \cdot \vec{B}=0$\\
Circulation    & Magnetic flux   \\
$\oint\limits_C \D\vec{x}\vec{v}=\int\limits_{\Sigma(C)}\D\vec{\Sigma}\vec{\omega}$ & $\oint\limits_C \D\vec{x}\vec{A}=\int\limits_{\Sigma(C)}\D\vec{\Sigma}\vec{B}$ \\
Magnus force   & Lorentz force   \\
$\vec{\omega} \times \dot{\vec{x}}$& $ q\vec{B}\times \dot{\vec{x}}$\\
\hline
\end{tabular}
\end{center}

\medskip
This lecture is devoted to special vortices occurring in non-Abelian gauge theories: the center vortices. Lattice calculations give 
strong evidence that these objects are the dominant IR degrees of freedom of QCD, which are responsible for confinement and 
spontaneous breaking of chiral symmetry, as I will show later. Before defining the center vortices it is worthwhile to recall 
some facts from group theory.

The center $Z$ of a group $G$ contains all elements $z\in G$ which commute with every other element $g$ of this group,
\begin{equation}
Z\mathrel{\mathop:}=\{ \, z\in G \mid gz=zg \quad \forall g\in G \, \}.
\label{eq:part_ii:center_definition}
\end{equation}
The center of a group forms itself a group. 
The center of an Abelian group is the whole group, while 
some groups, for example G(2), have a trivial center containing only the neutral group 
element. Due to their physical significance we are particularly interested in the special unitary groups 
SU($N$). The center of the SU(2) group is the following set
\begin{equation}
Z(2)= \{-\hat{1},\hat{1}\}
= \left\{ \begin{pmatrix} -1 & 0 \\[-.5ex] 0 & -1 \end{pmatrix} , \begin{pmatrix} \,1\, & \,0\, \\[-.5ex] 0 & 1 \end{pmatrix} \right\} .
\label{eq:part_ii:su2_center}
\end{equation}
In general, the center of the SU(N) group contains $N$ elements,
\begin{equation}
Z(N)=\{\,z_k \mid k=0,1\ldots N-1\,\},
\label{eq:part_ii:suN_center}
\end{equation}
which are the $N$th roots of unity,
\begin{equation}
z_k=\E^{2\pi\I k/N}\hat{1},
\label{eq:part_ii:suN_center_elements}
\end{equation}
where $\hat{1}$ is the $N\times N$ unit matrix. Particularly, for the SU(3) group, which is the gauge group of QCD, the center is
\begin{equation}
Z(3) = \{ \, z_0=\hat{1}, z_1 = \E^{2\pi\I/3}\hat{1}, z_2 = \E^{4\pi\I/3}\hat{1} \, \}.
\label{eq:part_ii:su3_center}
\end{equation}
The center vortex is a specific configuration of a gauge field, which is defined as follows:

Consider a gauge field $A (C_1)$, whose (magnetic) flux is localized on a closed curve $C_1$ in three dimensions. 
This field forms a center vortex if a Wilson loop (\ref{eq:part1:wilson_loop1}) 
calculated along a curve $C_2$ picks up a non-trivial center element of the gauge group if
the curves $C_1$ and $C_2$ are linked in a non-trivial way. Mathematically speaking, the gauge field $A(C_1)$ is a center vortex if
\begin{equation}
W[A(C_1)](C_2)=z^{L(C_1,C_2)},
\label{eq:part_ii:center_vortex_definition}
\end{equation}
where 
\begin{equation} 
\label{proc_765_G21}
L(C_1,C_2)= \frac{1}{4 \pi} \oint\limits_{C_1} d x_i \oint\limits_{C_2} d x'_j \epsilon_{ijk} 
\frac{x_k  -  x'_k}{| \vx - \vx'|^3} 
 \end{equation}
is the Gauss' linking number of the closed 
curves $C_1$ and $C_2$ and $z$ is a non-trivial center element. Since the linking number is a topological invariant, 
it does not depend on details of the shape of $C_2$ or $C_1$. For example, the group SU(2) has only one non-trivial center 
element, $- \hat{1}$, and hence there is only one type of 
center vortex field, satisfying $W[A(C_1)](C_2)=(- \hat{1})^{L(C_1,C_2)}$.

dvantbove given definition can be generalized to arbitrary vortex surfaces in four dimensions. Let $A (\partial \Sigma)$
be a gauge field whose (magnetic or electric) flux is localized on a closed surface $\partial \Sigma$ in $\RR^4$. This field 
forms a center vortex if its Wilson loop is given by
\begin{equation}
 \label{840-22}
W [A (\partial \Sigma)] (C) = z^{L (\partial \Sigma, C)} \, , 
\end{equation}
where $L (\partial \Sigma, C)$ is the linking number between the closed surface $\partial \Sigma$ and the 
closed loop $C$. (Note in $\RR^4$ loops can be non-trivially linked only to two-dimensional surfaces.) 

Lattice data suggests that center vortices are the dominant field configurations in the infrared region and are responsible
for both confinement and the spontaneous breaking of chiral symmetry. Therefore it is worthwhile to discuss the properties 
of center vortices. These can best be studied on discretized space-time, i.e.~on the lattice. 

\subsection{Center vortices on the lattice}
In the lattice approach to gauge field theory one approximates the continuous space-time by a discrete lattice of spacing $a$. Gauge fields are represented by link variables, $U_\mu(x)=\exp[\I a A_\mu(x)]$. The calculation of the Wilson loop on the lattice
is greatly simplified in comparison to 
continuum calculations. 
On the lattice the Wilson loop $W (C)$ is given by the ordered product of link variables $U_l$ 
along the chosen curve $C$
\begin{equation}
W [U] (C) = P \pli_{\mu \in C} U_\mu .
\end{equation}
The partition function on the lattice is given by a sum over all link configurations 
\begin{equation}
Z[U]=\int\prod\limits_x\prod\limits_{\mu=0}^3\D U_\mu(x) \E^{-S[U]},
\label{eq:part_ii:lattice_partition_function}
\end{equation}
where $S[U]$ is a lattice action. Its simplest form, the Wilson action, is given by
\begin{equation}
S[U]=\beta \sum\limits_{\mu<\nu,x}\biggl[1-\frac{1}{2N}\bigl(\Tr P_{\mu\nu}(x)+\mathrm{c.c.}\bigr)\biggr],
\label{eq:part_ii:lattice_action_definition}
\end{equation}
where $P_{\mu\nu}(x)=U_\mu(x)U_\nu(x+\mu)U^\dag_\mu(x+\nu)U^\dag_\nu(x)$ is the product of link variables around 
an elementary lattice square called a {\em plaquette}. This action is invariant with respect to local gauge transformations:
\begin{equation}
U_\mu(x)\rightarrow \E^{\I\Theta(x)}U_\mu(x)\E^{\I\Theta(x+\mu)} ,
\label{eq:part_ii:lattice_gauge_transformation}
\end{equation}
where  $\Theta (x) = \Theta^a (x) t_a$ .
Lattice gauge theory is a powerful tool to study non-perturbative effects in gauge field theories. For more details, see the lecture by O. Kaczmarek.

By Eq.~(\ref{eq:part_ii:center_vortex_definition}) center vortices can in principle be defined in a gauge invariant way. For 
practical purposes, however, one has to fix the gauge to isolate the center vortex content of the gauge fields.
Then it turns out that the properties of the identified center vortices depend on the chosen gauge.
So far, one has found only one gauge in which the emerging center vortices are physical in the sense that their density scales properly in the continuum 
limit \cite{Langfeld:1997jx}. This is the 
so-called {\em maximal center gauge} \cite{DelDebbio:1998luz}.

\begin{figure}[t]
	\includegraphics[width=0.49\linewidth]{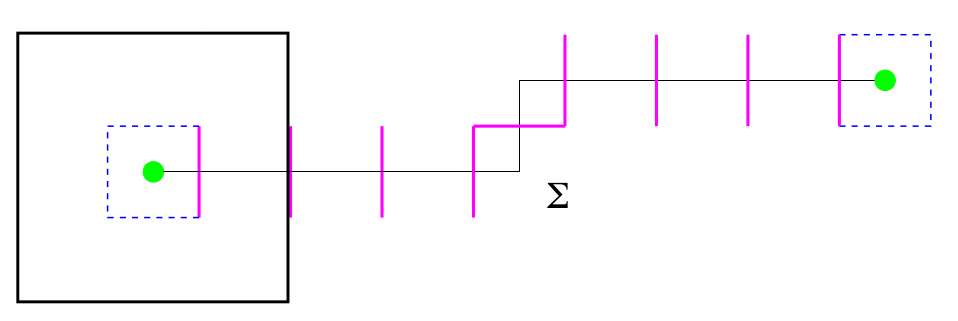}
	\includegraphics[width=0.49\linewidth]{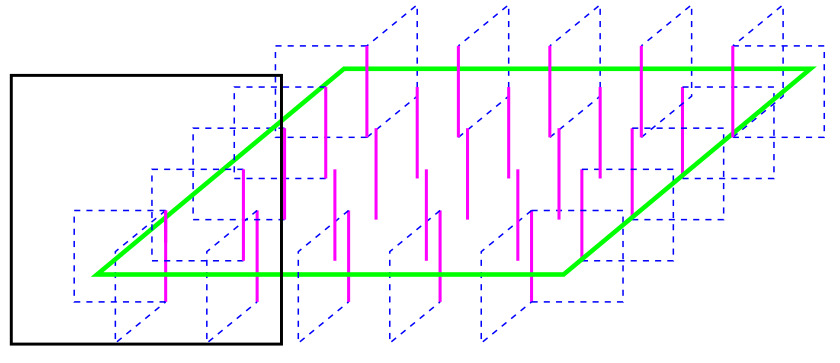}
	\caption{(left) Line of non-trivial center elements (purple lines) on a two-dimensional lattice. (right) Surface of non-trivial center elements on a three-dimensional lattice. In both pictures the black contour represents the Wilson loop. Trivial center elements are not shown.}
	\label{fig:part_ii:center_projected_vortices}
\end{figure}

Let the gauge group be SU(2), for simplicity. In this case, link variables can be expressed by the $S_4$ parametrization, 
\begin{equation}
\label{654-25}
U_\mu=\alpha_0+\ii\,\vec{\alpha}\cdot\vec{\tau} \, ,
\end{equation}
where $\vec\tau$ are the Pauli matrices and the real parameters $\alpha_\mu$, $\mu=0$, $1$, $2$, $3$ satisfy 
the constraint
\begin{equation}
\alpha^2_0 + \vec{\alpha}^2 = 1,
\label{511*}
\end{equation} 
which defines the unit sphere $S^3$ in the four-dimensional space
with coordinates $\alpha_\mu$.
The {\em maximal center gauge}
is defined by bringing each link as close as possible to its
nearest center element, which in this case is either $1$ or $-1$. In other words, one exploits the gauge freedom to maximize $(\alpha_0)^2$. By the constraint (\ref{511*}) this is equivalent to minimize $\vec{\alpha}^2$. On the lattice this is done by maximization of the following functional: 
\begin{equation}
\sum\limits_{x,\mu}\left\vert \Tr U^g_\mu(x)\right\vert^2 \to \mathrm{max}
\label{eq:part_ii:maximal_center_gauge}
\end{equation}
with respect to gauge transformations $g$. 
Once a gauge field configuration is brought into the `maximal center gauge', the second step is to perform the center projection\,---\,each link is replaced by its nearest center element \cite{DelDebbio:1998luz}. In case of the SU(2) group the non-Abelian part $\vec{\alpha}$ is set to zero and the Abelian part $\alpha_0$ is set to either $1$ or $-1$.
As a result of the center projection one obtains a lattice in which all links belong to the center of the gauge group, i.e.~all
links are center elements. With all links being equal to (trivial or non-trivial) center elements, the only non-trivial field configurations are center vortices. To see this, consider a line of non-trivial center elements on a two-dimensional lattice (purple lines on the left panel of Fig.~\ref{fig:part_ii:center_projected_vortices}, trivial center elements are not shown). The Wilson loop calculated around
one of the ends of this line (the black contour) yields $(-1)$,
while Wilson loops which do not encircle precisely one of the ends of the string of non-trivial center elements yield the trivial center element $1$. Therefore, 
[cf.~Eq.~(\ref{eq:part_ii:center_vortex_definition})] 
there is a center vortex at the end of such a one-dimensional domain of non-trivial center elements. 

On a three-dimensional lattice one finds surfaces of non-trivial center elements (see the right panel of Fig.~\ref{fig:part_ii:center_projected_vortices}) and the center vortices in this case are located at the boundaries of such surfaces. In this case, they form closed loops on the dual lattice or closed strings of plaquettes being equal to a non-trivial center element on the original lattice. 
Similarly, on a four-dimensional lattice, center vortices form closed surfaces.
In general, on a $D$-dimensional lattice, the center projected vortices are given by the $D-2$-dimensional boundaries of $D-1$-dimensional domains 
of links being equal to a non-trivial center element. Strictly speaking 
these boundaries live 
on the dual lattice and are linked by
plaquettes with non-trivial center value on the original lattice. 
 Because of the Bianchi identity center 
vortices have to be closed. Finally, one should also note that the Wilson loop can be used as a vortex counter. 
In the center projected theory, one finds $W(C)=z^N$, where $N$ is the number of center vortices piercing 
any area encircled by the loop $C$.

It is also interesting to consider 
the physical implication of the center projection. Originally, center vortices are smooth configurations of the gauge 
field. 
These objects have a finite thickness of order $0.8$\,fm as measured on the lattice \cite{Greensite:2011zz}. In the procedure of maximal center gauge fixing and center projection, thick center vortices are replaced by the thin ones on the
$Z(N)$ lattice.\footnote{The continuum version of the thin center projected vortices were referred to as {\em ideal}
center vortices in ref. \cite{Engelhardt:1999xw}.}

On the lattice, one can remove center vortices by hand \cite{deForcrand:1999our}. To remove vortices from the lattice each 
link $U_\mu(x)$ is multiplied by its center-projected image
$Z_\mu(x)$: $U_\mu(x)\rightarrow U_\mu(x)Z^\dagger_\mu(x)$. This adds an (oppositely oriented)
center-projected vortex on top of the physical center vortex and their effects cancel. 

\begin{figure}[t]
\centering
\includegraphics[width=0.5\linewidth]{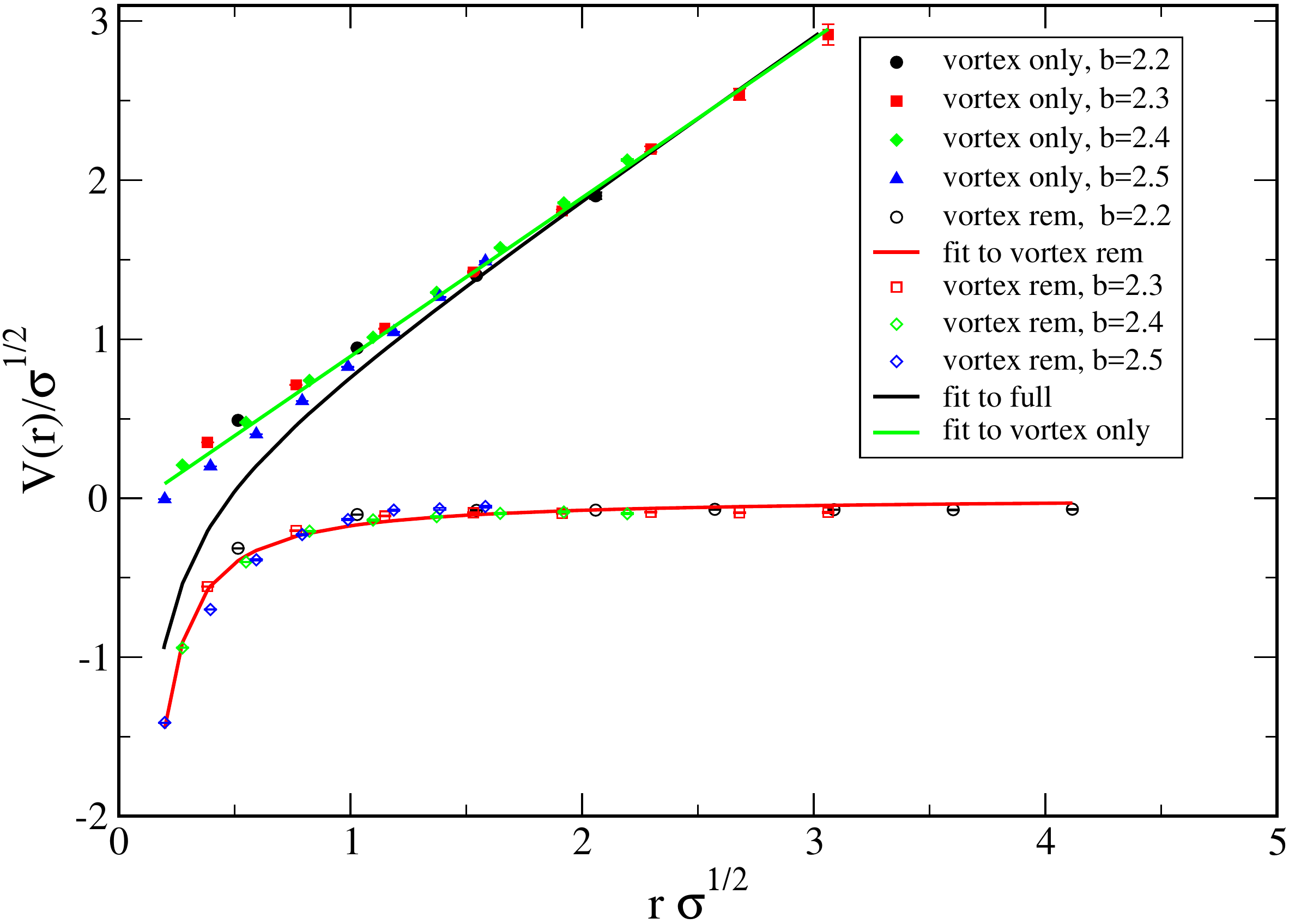}
\caption{Heavy quark potential in SU(2) gauge theory (black line); the potential after removal of center vortices (red line); center vortex contribution (green line).
From \cite{Langfeld:2003ev}.}
\label{fig:part_ii:su_2_potetnial}
\end{figure}

Lattice calculations provide strong evidence that center vortices are the field configurations responsible for confinement. Figure \ref{fig:part_ii:su_2_potetnial} shows the static quark potential for the SU(2) group together with the potentials one obtains after center projection and after center vortex removal, respectively. The full potential grows linearly at long distances and behaves Coulomb-like at 
short distances. 
When center vortices are removed the heavy quark potential becomes flat at large distances and hence loses its confining property. 
In the center projected theory, i.e.~when only the center vortices are kept from the ensemble of gauge fields (and converted into the center projected vortices), one loses the Coulombic part of the potential and finds just the confining (linearly rising) part.
This shows 
that center vortices are indeed responsible for the confining part of the heavy quark potential and thus for confinement.
\subsection{The random vortex model}
\begin{figure}[t]
        \centering
	\includegraphics[width=0.4\linewidth]{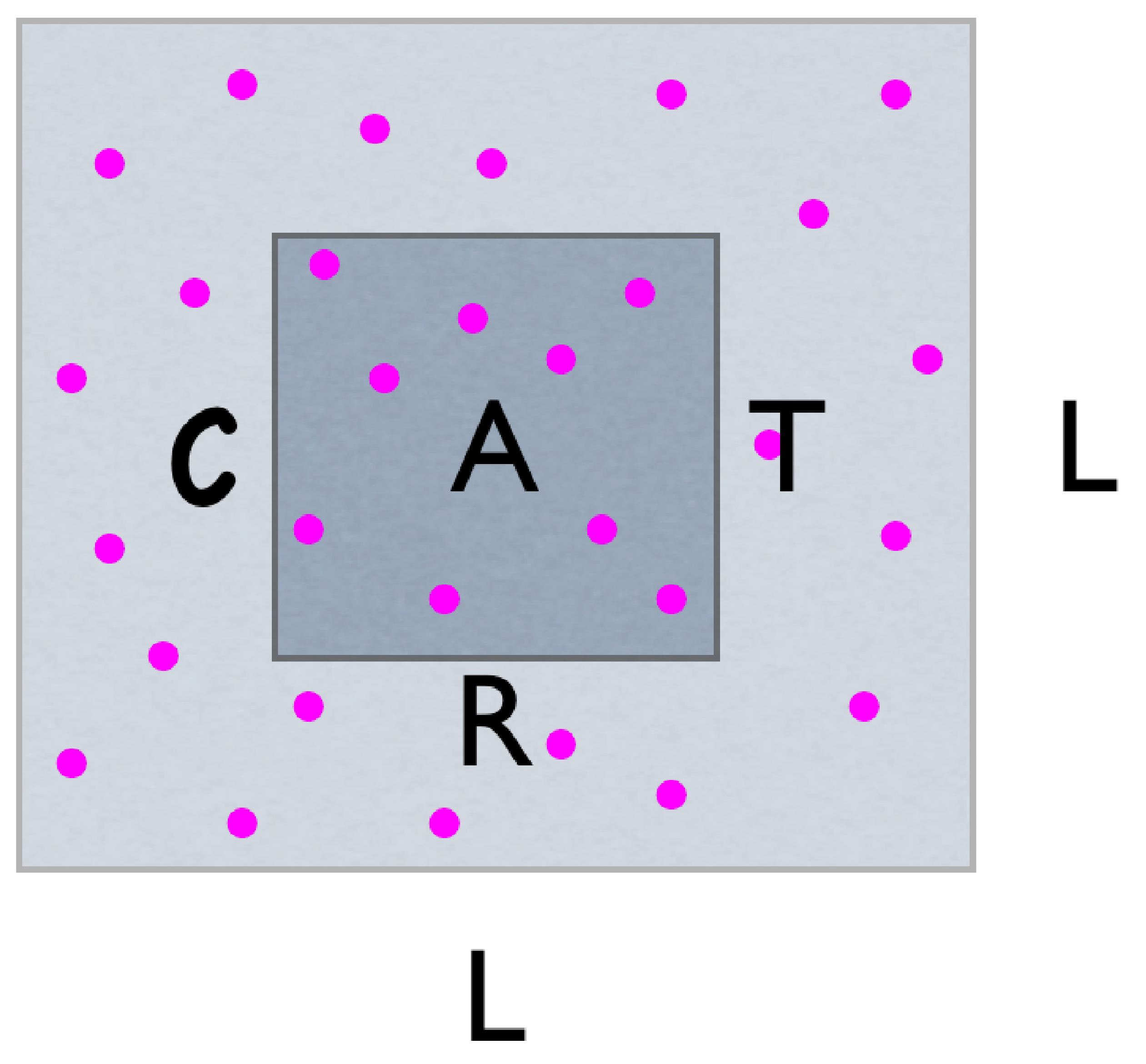}
		\caption{Two-dimensional slice of the lattice universe containing a rectangular Wilson loop $C$ enclosing the area A.
		In the plane center vortices show up as points representing the intersections of the center vortices with the two-dimensional plane.
		}
	\label{fig25}
\end{figure}

The confining properties of center vortices can be exhibited by means of a simple random vortex model \cite{Engelhardt:1998wu}. Consider an ensemble of randomly distributed center vortices, such that their intersection points with a two-dimensional plane in space-time can be found at random and uncorrelated locations. Assume that the space-time is a hypercube of length $L$ and consider its two-dimensional slice of area $L^2$, containing a Wilson loop which circumscribes 
an area $A$, see fig. \ref{fig25}. 
The probability that one vortex pierces this area is $P_1^1=A/L^2$, independently of its location. Accordingly, the probability that the vortex does not pierce that area is $P^1_0=1-P_1^1=1-A/L^2$. When there are $N$ center vortices piercing a slice of the Universe, the probability that $n$ of these pierce the Wilson loop is binomial,
\begin{equation}
P^n_N= \binom{N}{n} \,\left(\frac{A}{L^2}\right)^n \left(1-\frac{A}{L^2}\right)^{N-n},
\label{eq:part_ii:random_vortex_probability}
\end{equation}
because center vortices, as classical configurations of a gauge field in the Yang--Mills functional integral, are distinguishable. As shown above each vortex contributes a factor $-1$ to the Wilson loop. Hence, its average can be evaluated as
\begin{eqnarray}
\langle W\rangle &=& \sum\limits_{n=0}^N(-1)^nP_N^n=\sum\limits_{n=0}^N
\binom{N}{n} \left(-\frac{A}{L^2}\right)^n \left(1-\frac{A}{L^2}\right)^{N-n} \nonumber\\
&=&\left(1-2\frac{A}{L^2}\right)^N=\left(1-2\rho\frac{A}{N}\right)^N \stackrel{N, L \to \infty}{\longrightarrow} \exp(-2\rho A),
\label{eq:part_ii:wilson_loop_in_random_vortex_model}
\end{eqnarray}
where in the last step the size of the universe $L$ and the number of vortices $N$ have been sent to infinity with constant planar density $\rho=N/L^2$. Hence, the random vortex picture provides an area-law for the Wilson loop with the string tension $\sigma_\mathrm{rvm}=2\rho$. 
Lattice calculations of the vortex density yield $\rho\approx 3.4\, \text{fm}^{-2}$, 
which leads to $\sigma_\mathrm{rvm}\approx (521\, \mathrm{MeV})^2$ \cite{Engelhardt:1998wu}. This result overestimates ``experimental''
string tension, which is 
$(440\, \mathrm{MeV})^2$ and which sets the scale in the lattice calculation. 
This is due to the lack of correlations between vortices in the random vortex model. Correlations between the vortices make them less random and thus reduce the string tension.

Lattice calculations show that the density of the center vortices detected by center projection in the maximal center gauge shows the proper scaling \cite{Langfeld:1997jx} with the lattice spacing, 
which proves that these vortices are indeed physical objects, which survive in the continuum  limit.

\begin{figure}[t]
        \centering
	\includegraphics[width=0.4\linewidth]{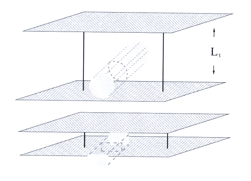}
	\includegraphics[width=0.4\linewidth]{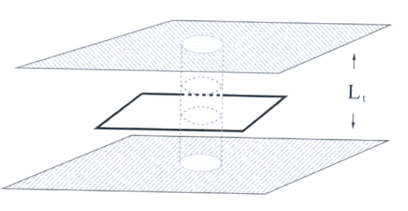}
	\caption{The deconfinement phase transition in the center vortex picture. Left panel: center vortices in the confined  phase; 
	right panel: center vortices in the deconfined phase, which align parallel to the (Euclidean) time axis.}
	\label{FX1}
\end{figure}

The center vortex picture gives not only a natural explanation of 
confinement, that is an area-law falloff of the Wilson loop, but also explains the deconfinement phase transition. In the finite-temperature  quantum 
field theory the time dimension becomes compactified and its length gives the inverse of the temperature, $L_t=1/T$. Center vortices have finite thickness $d\approx 0.8$\,fm, and at sufficiently high temperature space-like  vortices do no longer fit in the lattice universe and they align along the time axis, see Fig.~\ref{FX1}. This happens when  $L_t\approx d$ from which the deconfinement temperature is estimated as $T_c\approx 1.25\, \text{fm}^{-1}\approx 250\, \text{MeV}$ \cite{Engelhardt:1999fd}. 

Due to the alignment of the vortices parallel to the time axis,
in a three-dimensional time-space volume center vortices do no longer percolate.
Therefore the deconfinement phase transition might be seen as a transition from a phase of large vortices percolating through space-time to a phase in which vortices are small and do not percolate.

\begin{figure}[t]
\sidecaption[t]
\includegraphics[width=0.5\linewidth]{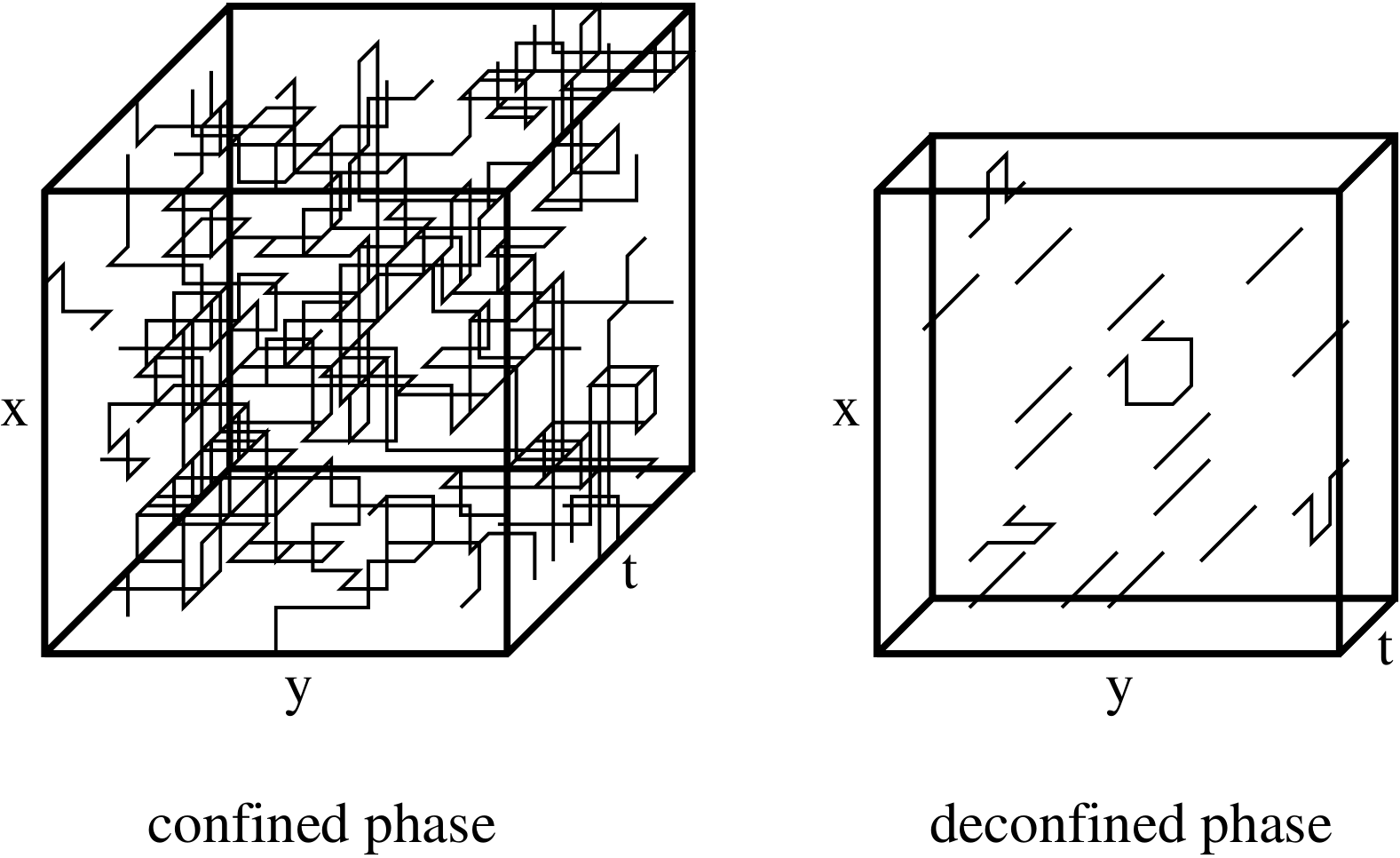}\hfill\null
\caption{Typical vortex configurations in the confining (left panel) and the
deconfined phase (right panel). From \cite{Engelhardt:1999fd}.}
\label{fig:part_ii:vortex_distribution_1}
\end{figure}

\begin{figure}[t]
\centering
\includegraphics[width=0.4\linewidth,trim={0 2cm 0 5cm},clip]{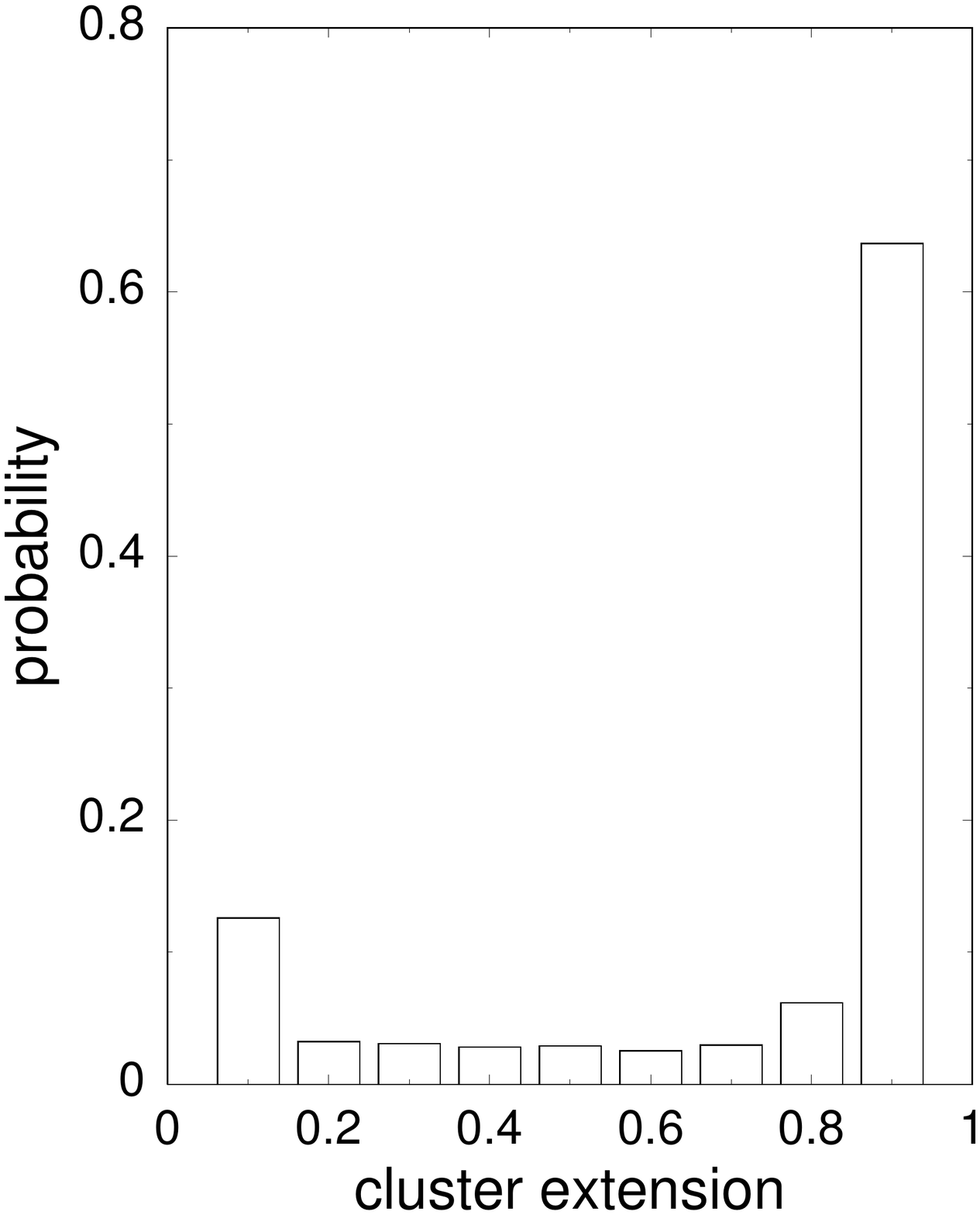}\qquad
\includegraphics[width=0.4\linewidth,trim={0 2cm 0 5cm},clip]{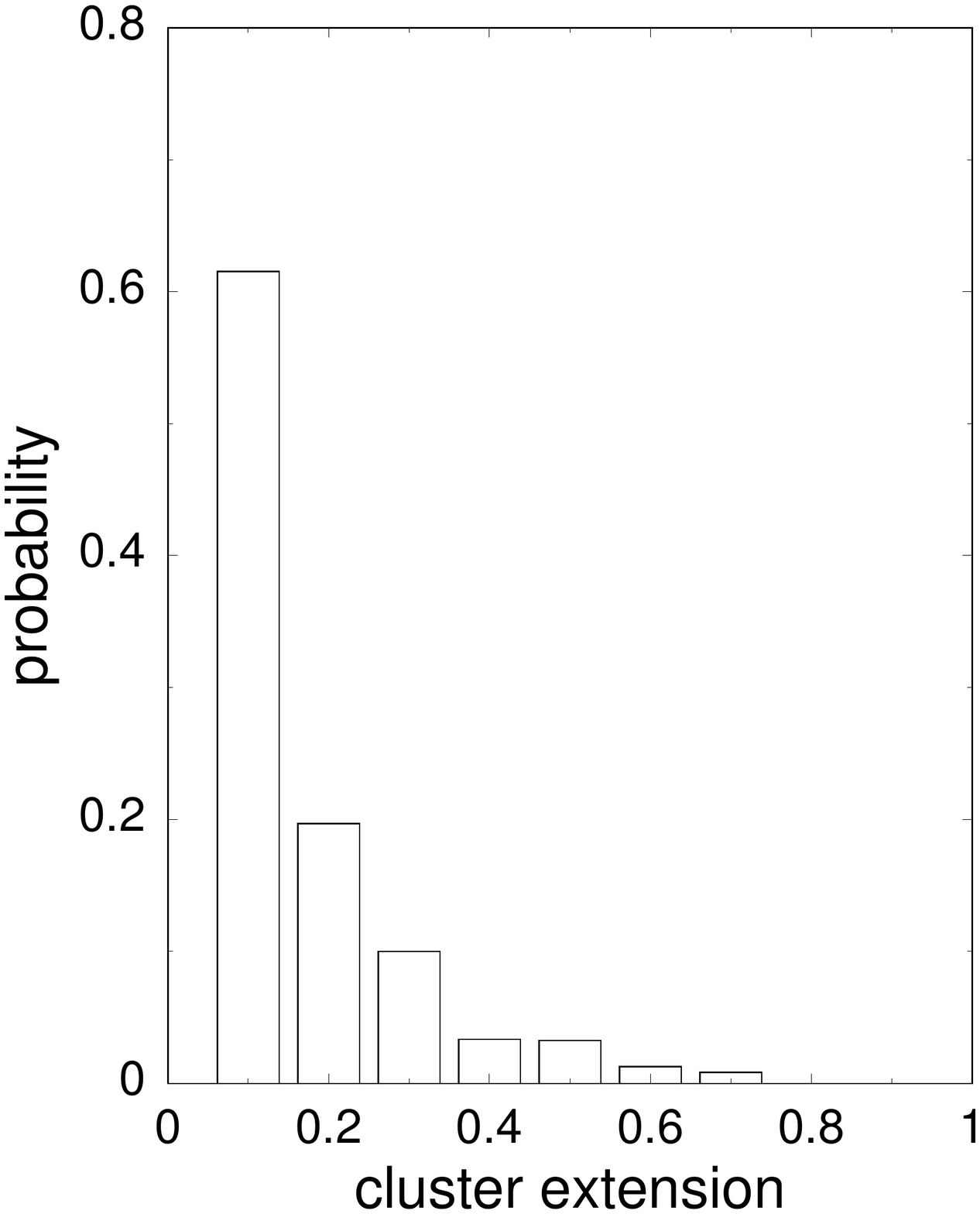}
\caption{Vortex material distributions in the confined phase (left panel) and in the deconfined phase (right panel) 
obtained in SU(2) lattice gauge calculations. Cluster extensions are
normalized to the extension of the lattice. 
From  \cite{Engelhardt:1999fd}.}
\label{fig:part_ii:vortex_distribution_2}
\end{figure}

\begin{figure}[t]
	\centering
	\includegraphics[width=0.7\linewidth]{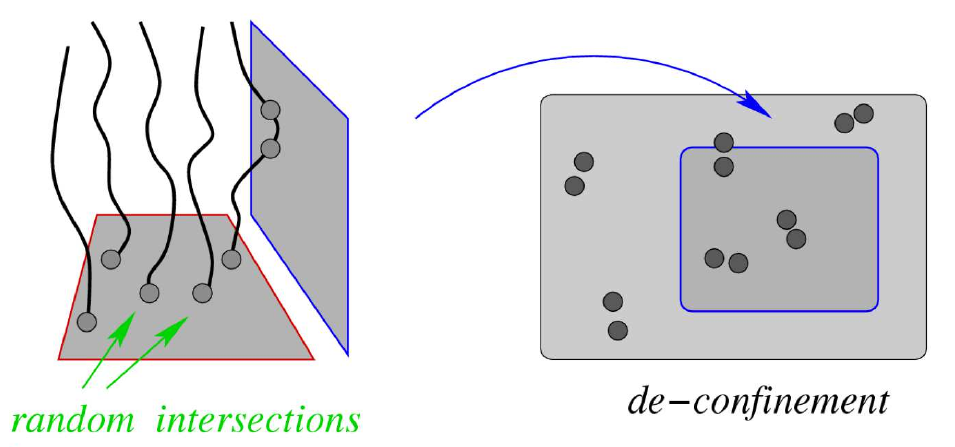}
		\caption{Center vortices in the deconfined phase. Left: alignment of the center vortices 
			parallel to the time-axis, which runs upwards. Right: intersection points of center vortices with a time-space plane, which contains a (temporal) Wilson 
			loop. }
	\label{FX2}
\end{figure}

This picture is supported by lattice results, shown in Figs.~\ref{fig:part_ii:vortex_distribution_1} and \ref{fig:part_ii:vortex_distribution_2}. As one can see, in the confined phase vortices form large clusters, which extend over the size of the temporal dimension, leading to an area law of both spatial and temporal Wilson loops. On the other hand, in the deconfined phase vortices form small clusters, aligned mostly along the time direction. 
Because of fluctuations in the cluster length, there is still a possibility that some vortices cross a {\em temporal}
Wilson loop, see Fig.~\ref{FX2}. In this case the intersection points, however, are correlated\,---\,they occur pairwise, thus giving no 
contribution to the Wilson loop when they occur in the interior of the Wilson loop. The only non-trivial contribution comes from the 
edges of the loop, when one of the intersection points is located outside the loop\,---\,this results in a perimeter law of the 
{\em temporal} Wilson loop. On the other hand intersection points of center vortices with a {\em spatial} plane
 are still uncorrelated, which leads to the area law for the {\em spatial} Wilson loop and thus to a non-vanishing spatial string tension.

The center vortex picture of the QCD vacuum can be extended to higher color
groups SU($N > 2$). For SU($N$), the center is $Z(N)$ and there are $N-1$ non-trivial center elements 
$z_k = \ee^{\ii \frac{2\pi}{N}\,k}$, $k=1,2,\ldots,N-1$.
Accordingly, there are $N-1$ different species of center vortices. Below, we will confine ourselves to the generic case of 
SU(3) \cite{Engelhardt:2003wm, Quandt:2004gy}. The gauge group SU(3) has two non-trivial center elements, $z_1=\exp(2\pi \I/3)$ and $z_2=\exp(4\pi \I/3)$. This means that there are two species of center vortices in SU(3) gauge theory, which, however,
are not independent. Since $z_2=(z_1)^2$ a $z_2$ 
vortex can split into two $z_1$ vortices and conversely, two $z_1$ vortices can merge into one $z_2$ vortex. 
Since also $z_1  = z^2_2$ a $z_1$ vortex can split into two $z_2$ vortices. 
Therefore, in case of three colors center vortices can not only cross, but also branch and merge.

To investigate the role of vortex branching and merging in the deconfinement phase transition, in Refs.~\cite{Engelhardt:2003wm, Quandt:2004gy} a $SU~(3)$ random vortex model 
was constructed and studied numerically. 

\begin{figure}[t]
\includegraphics[width = 0.32\linewidth]{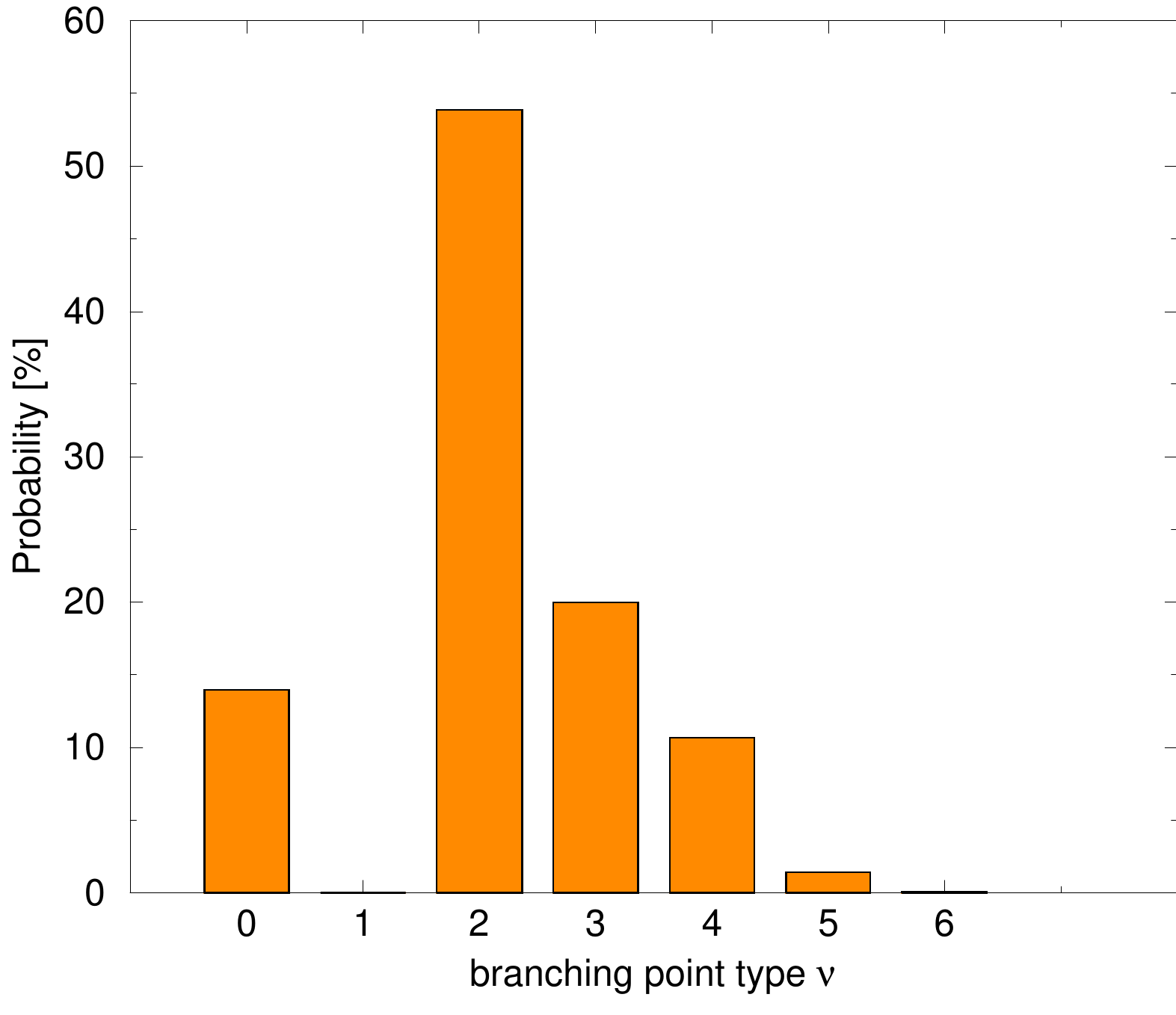}
\includegraphics[width = 0.32\linewidth]{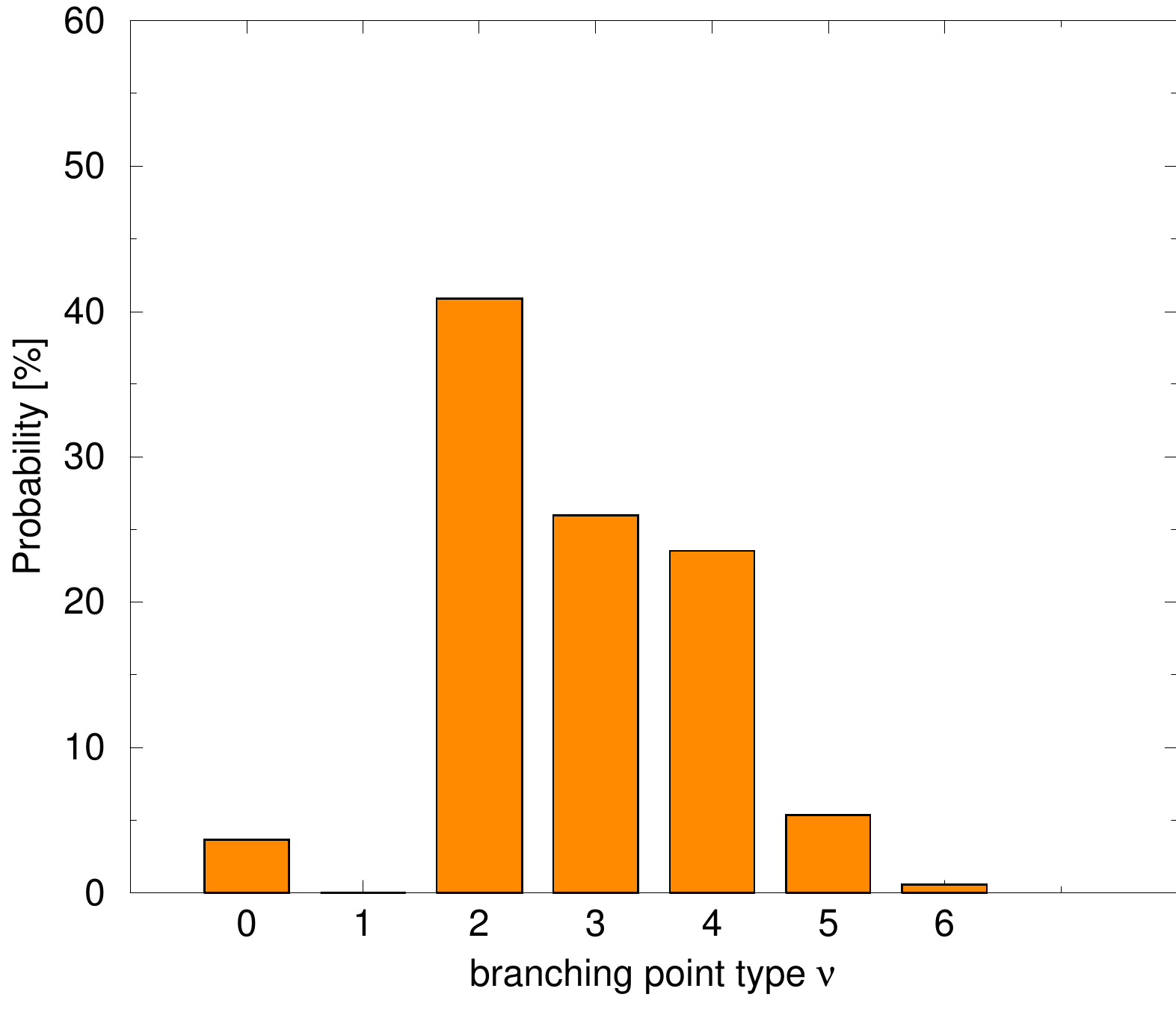}
\includegraphics[width = 0.32\linewidth]{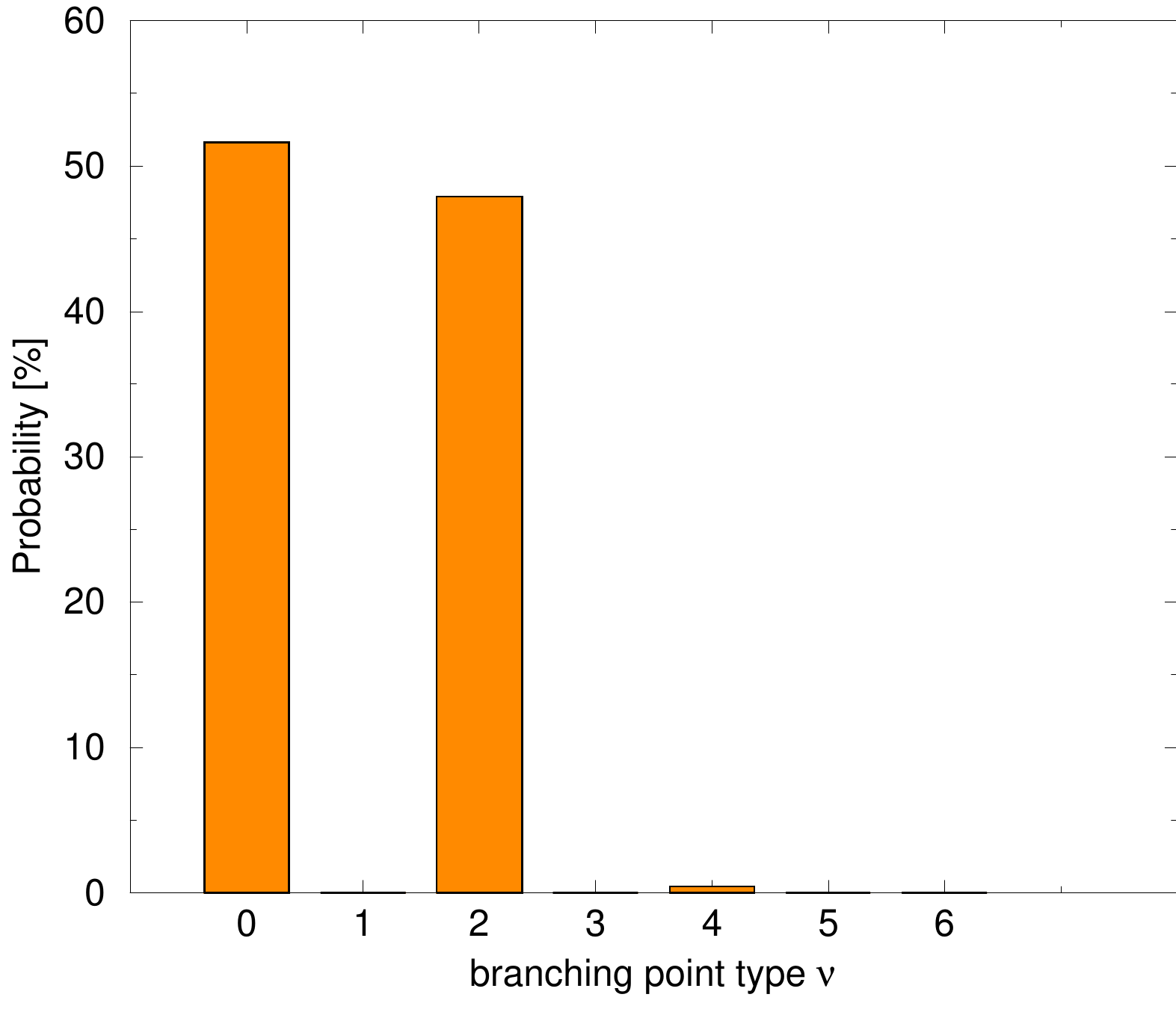}
\caption{Branching point distributions in three-dimensional slices of space-time. The 
left panel shows the results obtained in the confined phase ($T\approx 0$) while the 
middle and right panels give the distributions in the deconfined phase. The middle panel refers to a temporal slice
while the right panel shows the corresponding result for a spatial slice. From  \cite{Engelhardt:2003wm}.}
\label{fig:part_ii:vortex_splitting}
\end{figure}

At a given lattice link a center vortex can be characterized by the number $\nu$ of vortex \mbox{plaquettes}
meeting at this link. 
The following cases can be distinguished:
\begin{itemize}
\item $\nu=0$ -- there is no vortex at the given link
\item $\nu=1$ is impossible  since it would lead to an open vortex 
\item $\nu=2$ -- usual center vortex
\item $\nu=3$ -- vortex splitting
\item $\nu=4$ -- vortex crossing
\item $\nu=5=2+3$ -- vortex splitting
\item $\nu=6=2+4=3+3$ -- vortex crossing and splitting
\end{itemize}
Vortex branching has not been studied in the SU(3) lattice theory yet, but was studied within the effective vortex model of 
Refs.~\cite{Engelhardt:2003wm, Quandt:2004gy}. Distributions of the branching number $\nu$ in three-dimensional slices of 
the four dimension lattice universe obtained in this model are shown in Fig. \ref{fig:part_ii:vortex_splitting}. The left panel shows this distribution in the confined phase -- one can see that links are mainly attached to ordinary vortices ($\nu=2$), but there is also a significant contribution from vortex branching ($\nu=3$). Middle 
and right panels show the distributions in the deconfined phase for a {\em temporal} and a {\em spatial}, respectively, 
slice of the four dimension Euclidean universe. The {\em temporal} slice is the three-dimensional space $\RR^3$ at a fixed time while a 
{\em spatial} slice arises when
 one space  coordinate is kept fixed, i.e.~a spatial slice is 
a three-dimensional space spanned by the time and two spatial axes. In the temporal slice
 (the middle panel), in which the time coordinate is fixed, 
 vortex branching slightly increases compared to the confined phase (left panel). 
 (Note that the spatial string tension also increases.) In the deconfined phase in a
 spatial slice (right panel), in which one space coordinate is fixed, vortex splitting and vortex crossing vanish entirely 
 (and the temporal string tension also vanishes). Therefore 
the vortex branching behaves as an order parameter for the 
SU(3) deconfinement phase transition. The open question is 
whether the first order character of this transition is related to the vortex branching. 
In the case of the SU(2) group vortices can only cross and the transition is second order, while for $N\geq 3$ vortices can also split as well as merge and the transition is first order.

\subsection{Topology of center vortices}\label{sect3.4}
We have shown that the center vortex picture provides a qualitative explanation of the
confinement and deconfinement phase transition.  In the remaining part of this lecture I
will discuss the topological properties of center vortices and explore their role in 
the spontaneous breaking of chiral symmetry in QCD. 

Topological properties of gauge fields are characterized by the so-called topological charge 
(Pontryagin index or second Chern number)
\begin{equation}
\nu=\frac{1}{16\pi^2}\int \D^4x\, \Tr(F_{\mu\nu}\tilde{F}_{\mu\nu}),
\label{eq:part_ii:topological_charge_definition}
\end{equation}
where $\tilde{F}_{\mu\nu}=\frac{1}{2}\varepsilon_{\mu\nu\kappa\lambda}F_{\kappa\lambda}$ is the tensor dual to the field strength tensor $F_{\mu\nu}$. The topological charge can be also expressed in terms of the electric, $E_i^a=F^a_{0i}$,  and the  magnetic, $B^a_i=\tilde{F}_{0i}^a,$ fields,
\begin{equation}
\nu=\frac{1}{4\pi^2}\Tr\int \D^4x\, \vec{E}\cdot\vec{B}.
\label{eq:part_ii:topological_charge_e_b}
\end{equation}
Two components of center-projected vortices can be distinguished\,---\,magnetic and electric ones. Magnetic vortices form closed lines of magnetic flux in the three-dimensional space. These vortex loops
 evolve in time direction for a
  finite time interval and form closed 
  two-dimensional surfaces in four-dimensional space-time. On the other hand, electric vortices form closed surfaces in three-dimensional space which exist only at a single time instant and with the electric field directed along the surface normal. 
  The presence of both types of vortices
   is obviously needed to have a non-vanishing Pontryagin index (\ref{eq:part_ii:topological_charge_e_b}). 
   
\begin{figure}[t]
\centering
\includegraphics[width=0.3\linewidth]{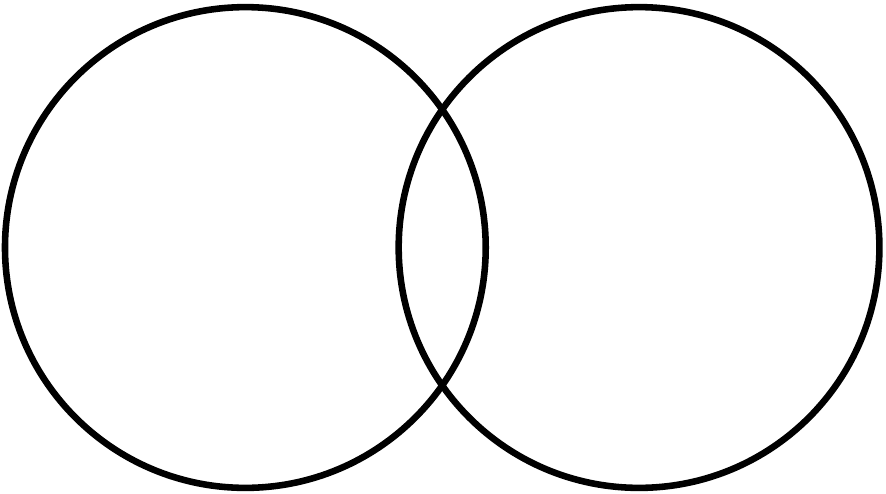}
\caption{Two-dimensional analog of two intersecting closed surfaces in $D = 4$. Closed loops intersect always in an even number of points.}
\label{FX3}
\end{figure}

   The study
   of the topological properties of center vortices is most conveniently done in the continuum where the 
   gauge potential  of a (closed)  center vortex surface $\partial \Sigma$ can be represented in $D$ space-time dimensions as 
   \cite{Engelhardt:1999xw, Reinhardt:2001kf} 
   \begin{equation}
   \label{1115-x1}
   \cA_\mu (x, \Sigma) = 2 \pi \tilde{\omega} \il_\Sigma d^{D - 1} \tilde{\sigma}_\mu \delta^{(D)} (x - \tilde{x} (\sigma)) \, .
   \end{equation}
   Here $\tilde{\omega} = \tilde{\omega}^a h_a$  (with $h_a$ the  (hermitian) generators of the Cartan algebra) is a coweight 
   satisfying 
   \begin{equation}
   \label{1121-22-a}
   e^{i 2 \pi \omega} = z \, ,
   \end{equation}
   $\delta^{(D)} (x)$ is the $D$-dimensional $\delta$-function and $d^{D - 1} \tilde{\sigma}_\mu$ is the $D - 1$ dimensional 
   (dual) surface element in $D$-dimension. 
   Furthermore $\tilde{x}^\mu (\sigma)$ is a parametrization of the $D - 1$-dimensional volume $\Sigma$ enclosed by the 
   vortex surface $\partial \Sigma$. An alternative continuum  representation of the gauge potential of a center vortex surface 
   $\partial \Sigma$ is given by
   \begin{equation}
   \label{1130-x2}
   \cA_\mu (x, \partial \Sigma) = 2 \pi \tilde{\omega} \il_{\partial \Sigma} d^{D - 2} \tilde{\sigma}_{\mu \nu} 
   \partial^x_\nu D (x - \tilde{x} (\sigma)) \, ,
   \end{equation}
   where $d^{D - 2} \tilde{\sigma}_{\mu \nu}$ is the $D - 2$ dimensional (dual) surface element  in $D$-dimension and $D (x)$ 
   denotes the Green function of the $D$-dimensional Laplacian
   \begin{equation}
   \label{1137-x3}
   - \partial_\mu \partial^\mu D (x) = \delta^{(D)} (x)  \, .
   \end{equation}
   Since the coweights $\tilde{\omega} = \tilde{\omega}_a h_a$ live in the Cartan  algebra the non-Abelian part of the 
   field strength of a center vortex vanishes
   \begin{equation}
   \label{1143-33}
   \cF_{\mu \nu} (x, \partial \Sigma) = \partial_\mu \cA_\nu (x) - \partial_\nu \cA_\mu (x) \, .
   \end{equation}
   Both representations (\ref{1115-x1}) and (\ref{1130-x2}) yield, of course, the same field strength
   \begin{equation}
   \label{1148-x3}
   \cF_{\mu \nu} (x, \partial \Sigma) = 2 \pi \tilde{\omega} \il_{\partial \Sigma} d^{D - 2} \tilde{\sigma}_{\mu \nu} 
   \delta^{(D)} (x - \tilde{x} (\sigma) \, .
   \end{equation}
   According to Eq.~$($\ref{eq:part_ii:topological_charge_e_b}$)$, each intersection point\footnote{In $D = 4$ surfaces intersect 
   generically in points. This is analogous to the intersection of lines in $D = 2$, see Fig. \ref{FX3}.} between an electric and a magnetic
   part of a vortex contributes to $\nu$. Using eq. (\ref{1148-x3}) one shows that the topological charge of a center vortex
   can be expressed by the self-intersection number $I (\partial \Sigma, \partial \Sigma)$
of the vortex surface $\partial \Sigma$ \cite{Engelhardt:1999xw}, \cite{Reinhardt:2001kf}
\begin{equation}
\nu=\frac{1}{4}I(\partial\Sigma,\partial\Sigma),
\label{eq:part_ii:topological_charge_self_intersection}
\end{equation}
where the intersection number between two surfaces $S_1$ and $S_2$ in $D = 4$ is defined by 
\begin{equation}
\label{5651}
I (S_1, S_2) = \frac{1}{2} \il_{S_1} \dd\sigma_{\mu \nu} \il_{S_2} \dd\tilde{\sigma}'_{\mu \nu} \, \delta\bigl(\bar{x} (\sigma) - \bar{x} (\sigma')\bigr) .
\end{equation}
Here $\bar{x}^\mu (\sigma)$ 
denotes a parametrization of the vortex surface $S_1$, $\dd \sigma_{\mu \nu}$ is an infinitesimal area element in four dimensions and
$\dd \tilde{\sigma}_{\mu \nu} = \frac{1}{2} \epsilon_{\mu \nu \kappa \lambda} \dd \sigma_{\kappa \lambda}$ is its dual. 

It is not difficult to see that Eq.~(\ref{eq:part_ii:topological_charge_self_intersection}) yields integer topological charge. To see this let us assume that the vortex sheet $\partial \Sigma$ consists of an electric and a magnetic piece
\begin{equation}
\label{6643-aa}
\partial \Sigma = \partial \Sigma_e + \partial \Sigma_m .
\end{equation}
From the definition (\ref{5651}) of the intersection number follows 
\begin{align}
\label{669-ss*}I (\partial \Sigma, \partial \Sigma) &= I (\partial \Sigma_e, \partial \Sigma_m) + I (\partial \Sigma_m, \partial \Sigma_e) \nonumber\\
&= 2 I (\partial \Sigma_e, \partial \Sigma_m)
\end{align}
since $I (\partial \Sigma_e, \partial \Sigma_e) =  0 = 
I (\partial \Sigma_m, \partial \Sigma_m)$. As explained above, electric or magnetic vortex sheets cannot intersect with themselves 
in points in $D = 4$. With (\ref{669-ss*}) we obtain from (\ref{eq:part_ii:topological_charge_self_intersection}) 
\begin{equation}
\label{6785-gg}
\nu = \frac{1}{2}  I (\partial \Sigma_e, \partial \Sigma_m) \, .
\end{equation}
Since closed surfaces intersect in an even number of intersection points (see Fig.~\ref{FX3}) $\nu$ is indeed integer valued.

\begin{figure}[t]
\centering
\includegraphics[width=1.0\linewidth]{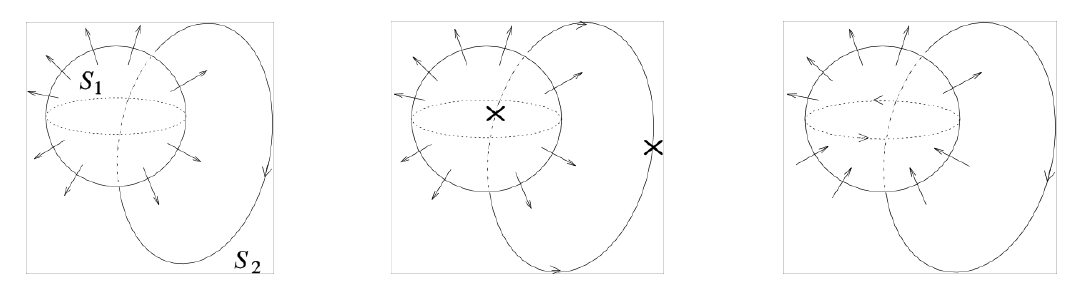}
\caption{Intersecting electric and magnetic vortices. Left panel: both vortices are
oriented; middle panel: oriented electric and non-oriented magnetic vortices; right panel
oriented magnetic and non-oriented electric vortices.}
\label{FX6}
\end{figure}%
Moreover, the intersection number and thus
 the topological charge depends also on the orientation of a vortex surface. Assume that electric and magnetic vortices are both oriented, see Fig.~\ref{FX6}: then for any intersection point between the electric vortex surface and the 
 magnetic vortex line which gives a positive contribution to the topological charge there is one
  intersection point which gives the opposite contribution. In general, one can show that the self-intersection number of {\em oriented}
  surfaces in four-dimensional space is always zero and hence the topological charge of an oriented center vortex vanishes. To be topologically non-trivial, a center vortex surface has to be non-oriented.
  
  The middle and the right panel of Fig.~\ref{FX6} show the intersection of an oriented an a non-oriented center vortex. 
  In the latter
  case the two intersection points add coherently to the intersection number, which is two in this example.
  
\begin{figure}[t]
	\sidecaption[t]
	\centering
	\includegraphics[width=0.3\linewidth]{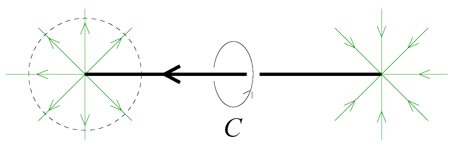}
		\includegraphics[width=0.3\linewidth]{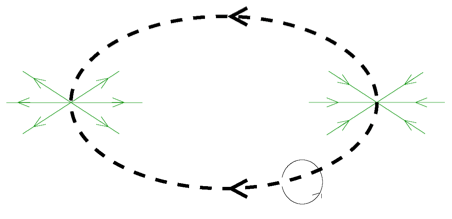}
	\caption{Splitting a Dirac string (left panel) into two center vortices (right  panel)}
	\label{FX4}
\end{figure}

The orientation of center vortices is irrelevant for their confining properties. It is, however, crucial for their topological properties and thus for spontaneous breaking of chiral symmetry.  What makes a center vortex surface non-oriented? The answer is magnetic monopoles. Consider a monopole--anti-monopole pair, connected with the Dirac string, see Fig.~\ref{FX4}. 
The Dirac string itself is not observable. However, it represents a magnetic flux  which is twice the flux of a center vortex. Therefore the Dirac string can be split into two center vortices resulting in  a center vortex loop with two magnetic monopoles on it. 
As can be seen from Fig.~\ref{FX4}, the magnetic monopoles change the orientation of the vortex flux. 

In four dimensions, center vortices are closed surfaces while the trajectories of
 magnetic monopoles form closed loops, placed on center vortex surfaces after center projection. It can be shown rigorously that the topological charge of a center vortex is 
 given by the linking number between the monopole loop and the 
vortex surface \cite{Reinhardt:2001kf}:
\begin{equation}
\nu=\frac{1}{4} \, L(C_{\mathrm{monopole}},\partial\Sigma_{\mathrm{vortex}}).
\label{eq:part_ii:topological_charge_monopole}
\end{equation}
It should come as
no surprise that the magnetic monopoles are here required for a non-zero topological charge $\nu$. This is because in the Abelian gauges $\nu$ can be entirely expressed in terms of the magnetic monopole content of the gauge fields. 
In the Polyakov gauge (see section \ref{section4.5}) the topological charge can be expressed as  \cite{Reinhardt:1997rm}
\begin{equation}
\nu=\sum_i n_im_i \, ,
\end{equation} 
where $m_i$ is the integer valued magnetic charge of a monopole and $n_i$ is an integer related to the Dirac string, and the summation 
runs over all magnetic monopoles contained after Abelian projection in the gauge field configuration considered.
\begin{figure}[t]
\sidecaption[t]
\includegraphics[width=0.6\linewidth]{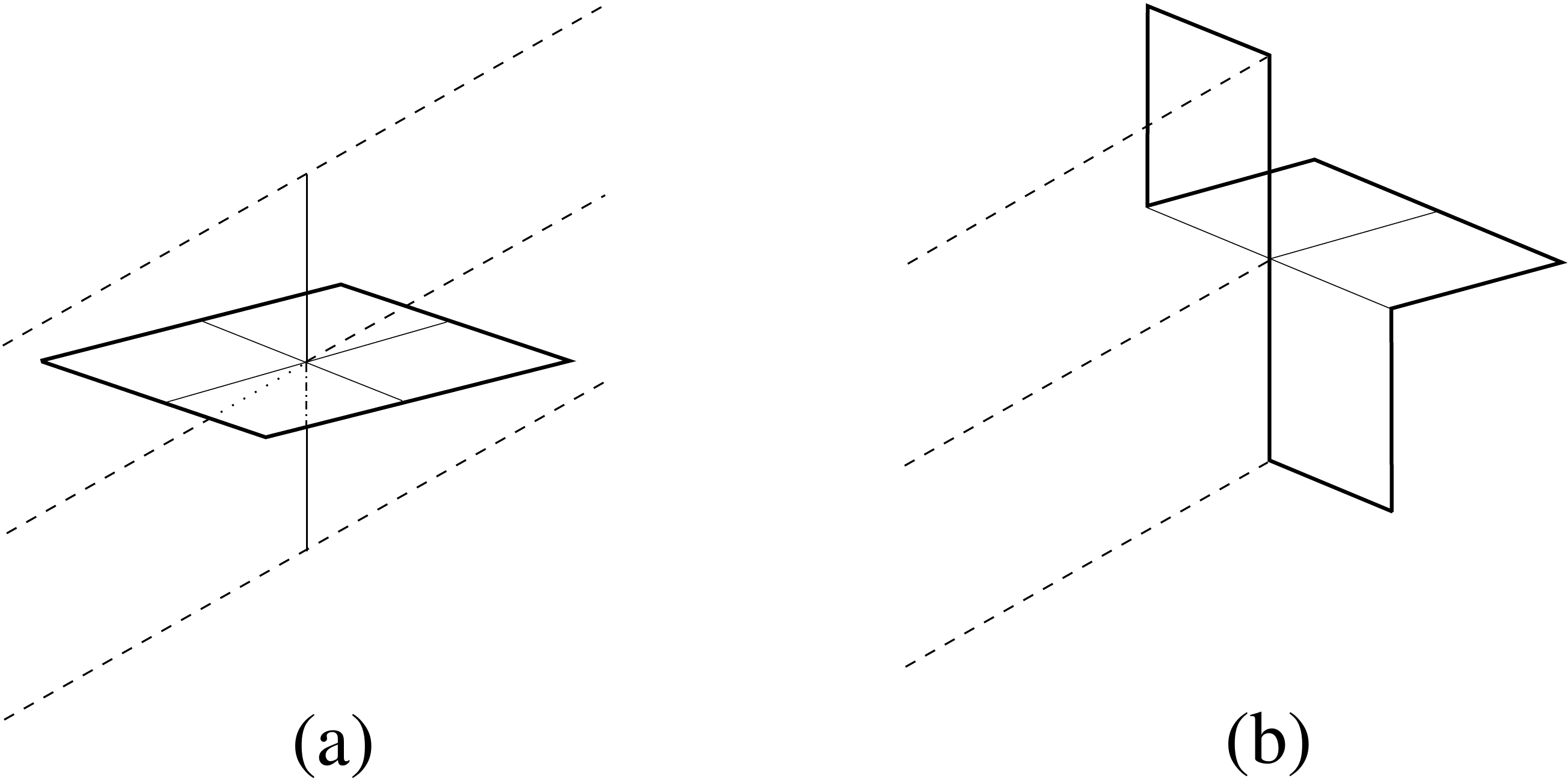}
\caption{Illustration of (a) a transverse intersection point and (b) a writhing or 
twisting point. The dashed lines indicate the fourth dimension (time direction). From  \cite{Reinhardt:2002qm}.}
\label{fig:part_ii:intersection_points}
\end{figure}

There are two types of intersection points \cite{Reinhardt:2002qm}: the transverse intersection point and the writhing point, also called twist. Transverse or generic intersection points are those were in Eq.~(\ref{5651}) 
$x^\mu (\sigma) = x^\mu (\sigma')$ for $\sigma \neq \sigma'$. They contribute $\pm \frac{1}{2}$ to the topological charge 
$\nu$ (\ref{eq:part_ii:topological_charge_self_intersection}). Writhing points contribute to the intersection number
$I$ (\ref{5651}) for $\sigma = \sigma'$. This means that at a writhing point the vortex surface twist in such a way that the tangent
vectors to the 
vortex surface elements emanating from
the writhing point span the full four dimensional space. They contribute less than
$1/2$ to the topological charge $| \nu |$. 
 Lattice realizations of these points are shown in Fig.~\ref{fig:part_ii:intersection_points}, where the left panel corresponds to a
 transverse intersection point and the right panel to a
 writhing point. Dashed lines show the time direction. The topological charge on the lattice can be expressed as
\begin{equation}
\nu \equiv \frac{1}{4} I(\partial\Sigma,\partial\Sigma)= \sum\limits_{\mathrm{site}} \nu_{\mathrm{site}}
\label{eq:part_ii:topological_charge_lattice}
\end{equation}
where \cite{Engelhardt:2000wc}
\begin{equation}
\nu_{\mathrm{site}}=\frac{1}{32}\#_{\mathrm{site}},
\label{eq:part_ii:top_charge_site}
\end{equation}
and $\#_{\mathrm{site}}$ is the number of pairs of plaquettes which share this site and which are completely orthogonal to one another (i.e.~their tangent vectors span the four-dimensional space). In case of transverse intersection points there are four plaquettes which extend in one space dimension and the time dimension and four plaquettes which extend in spatial direction only. Using the prescription $($\ref{eq:part_ii:top_charge_site}$)$ one finds that $\nu=\frac{1}{32}\times 4\times 4=\frac{1}{2}$. For the writhing point
shown in Fig.~\ref{fig:part_ii:intersection_points}b one finds 
two plaquettes extending in one space dimension and in time and two plaquettes extending in spatial directions only. 
This yields $\nu=\frac{1}{32}\times 2\times 2=\frac{1}{8}$.

\begin{figure}[t]
\sidecaption[t]
\includegraphics[width=0.5\linewidth]{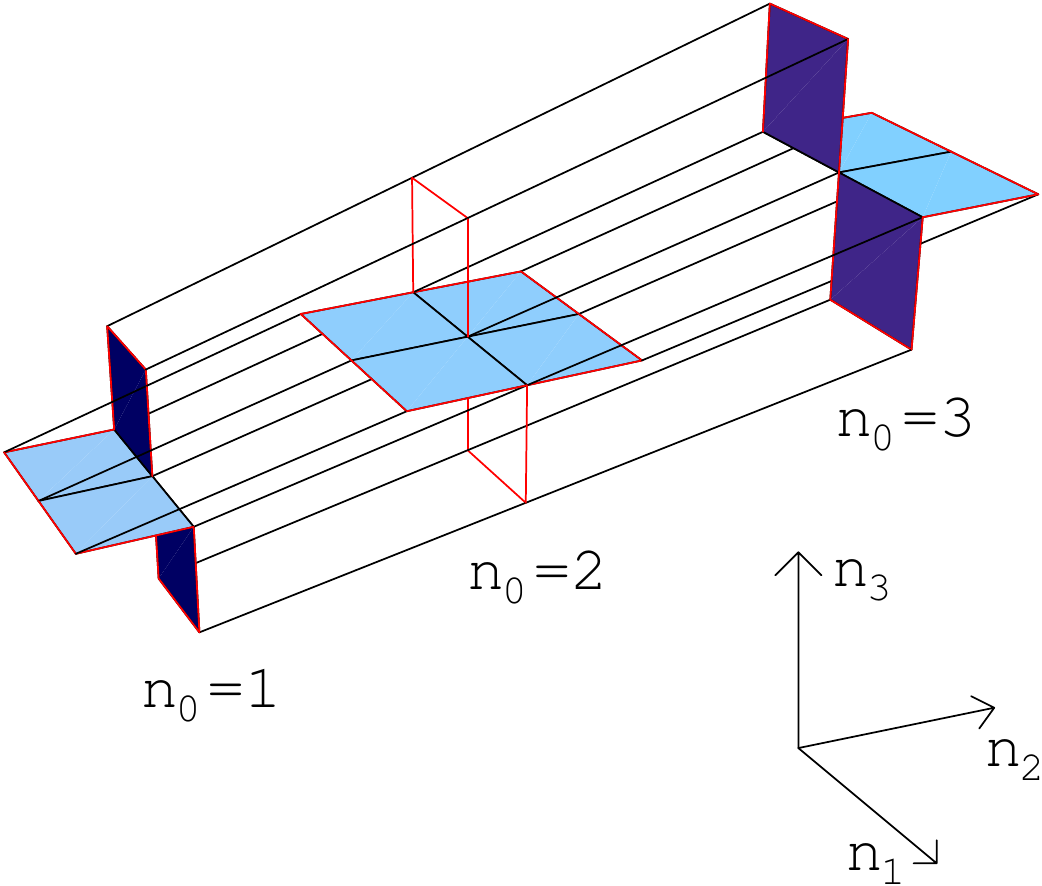}
\caption{Lattice realization of a generic center vortex surface. $n_0$ -- denotes the lattice time. 
 At each lattice time $n_0$, shaded plaquettes are part of the vortex surface. 
 These plaquettes are connected to plaquettes running in the time direction. The two non-shaded plaquettes at 
 $n_0 = 2$ are not part of the vortex but the two sets of links bounding them are. From  \cite{Engelhardt:2000wc}.}
\label{fig:part_ii:generic_vortex}
\end{figure}

\begin{figure}[b]
\centering
\includegraphics[width=0.8\linewidth]{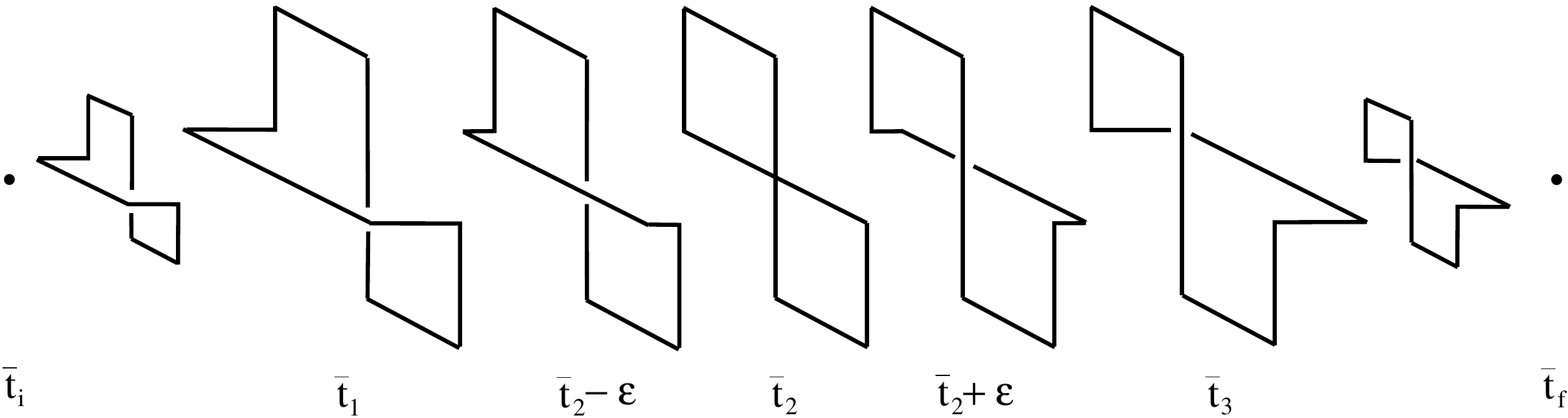}
\caption{The center vortex surface of  Fig.~\ref{fig:part_ii:generic_vortex} shown in $\RR^3$ at different time instants. 
From  \cite{Reinhardt:2001kf}.}
\label{fig:part_ii:vortex_evolution}
\end{figure}

Figure \ref{fig:part_ii:generic_vortex} shows the time evolution (along the $n_0$ axis) of a generic center vortex on the lattice, which has both 
intersection and writhing points, while Figure \ref{fig:part_ii:vortex_evolution} shows a sequence of snapshots of this vortex in $\RR^3$ where this vortex shows up as a (time-dependent) closed loop:  At some time the vortex loop emerges 
from the vacuum, and during its evolution it can intersect with itself. Finally,  the loop disappears. 
On our spatial manifold $\RR^3$, 
the four-dimensional transverse intersection points show up 
as (time-dependent) true crossings of vortex loop segments, while the writhing points become ``fractional crossing'' or turning points of vortex line
segments. 

The time-evolution of a center vortex in $\RR^3$ is sufficient to calculate its topological charge, which can be expressed  as \cite{Reinhardt:2001kf} 

\begin{equation}
\nu=\frac{1}{4}\int \D t \, \partial_t W_r(C(t)),
\label{eq:part_ii:topological_charge_writhing_number}
\end{equation}
where $W_r(C(t))$ is the writhing number, defined as the self-linking number of the closed loop $C$
\begin{equation}
 W_r(C(t))=L(C(t),C(t))
\end{equation}
with $L (C_1, C_2)$ being Gauss' linking number (\ref{proc_765_G21}). 
 For more details see Ref.~\cite{Reinhardt:2001kf}.

\subsection{Chiral symmetry breaking}

Below we show that center vortices also induce spontaneous breaking of chiral symmetry.

Consider the eigenvalue equation for the Dirac operator with massless quarks,
\begin{equation}
\Dl \Psi \equiv \gamma^\mu D_\mu\Psi=\lambda\Psi,
\label{eq:part_ii:dirac_eigenvalues}
\end{equation}
where $D_\mu=\partial_\mu-A_\mu \, , A_\mu = g A^a_\mu T_a$, is the covariant derivative (see 
footnote \footref{foot18} on page \pageref{foot18}). 
When the eigenvector $\Psi$ with eigenvalue $\lambda$ is multiplied by  $\gamma_5$ 
one obtains a new eigenvector with the eigenvalue $-\lambda$. This follows from the fact that 
$\gamma_5$ anticommutes with all Dirac matrices $\gamma^{\mu = 1,2,3, 4}$. 
Therefore the states $\Psi$ and $\gamma_5 \Psi$ are linearly independent and, hence, non-zero eigenvalues 
of the Dirac operator come in pairs $\pm \lambda$. For the zero modes (eigenvectors with $\lambda = 0$) 
it follows from $\{ \gamma^\mu, \gamma_5 \} = 0$ that $\Dl \Psi = 0$ implies also $\Dl \gamma_5 \Psi = 0$. 
Therefore the zero modes can be chosen as left- or right-handed modes, such that 
\begin{equation}
\label{961-39}
\Dl \Psi_{L/R} = 0 \, , \qquad \gamma_5\Psi_{L/R}=\mp \Psi_{L/R}\, , \qquad \Psi_{L/R}=P_{L/R}\Psi_{L/R} \, ,
\end{equation}
 where $P_{L/R}=\frac{1}{2}(1\mp\gamma_5)$ are the left/right projectors. From the {\em Atiyah--Singer index theorem}
 \begin{equation}
  \nu [A]  =n_L-n_R 
 \end{equation}
 follows that the topological charge $\nu$ of a field configuration $A$ is given by the difference between the numbers $n_{L/R}$ of left- and right-handed zero modes. This means that the zero modes of the Dirac operator are related to the topology of gauge fields. 
 In topologically non-trivial field configurations, $\nu\neq 0$, quarks must have zero modes.
 
\begin{figure}[t]
\parbox{.46\linewidth}{\centering\includegraphics[width=\linewidth]{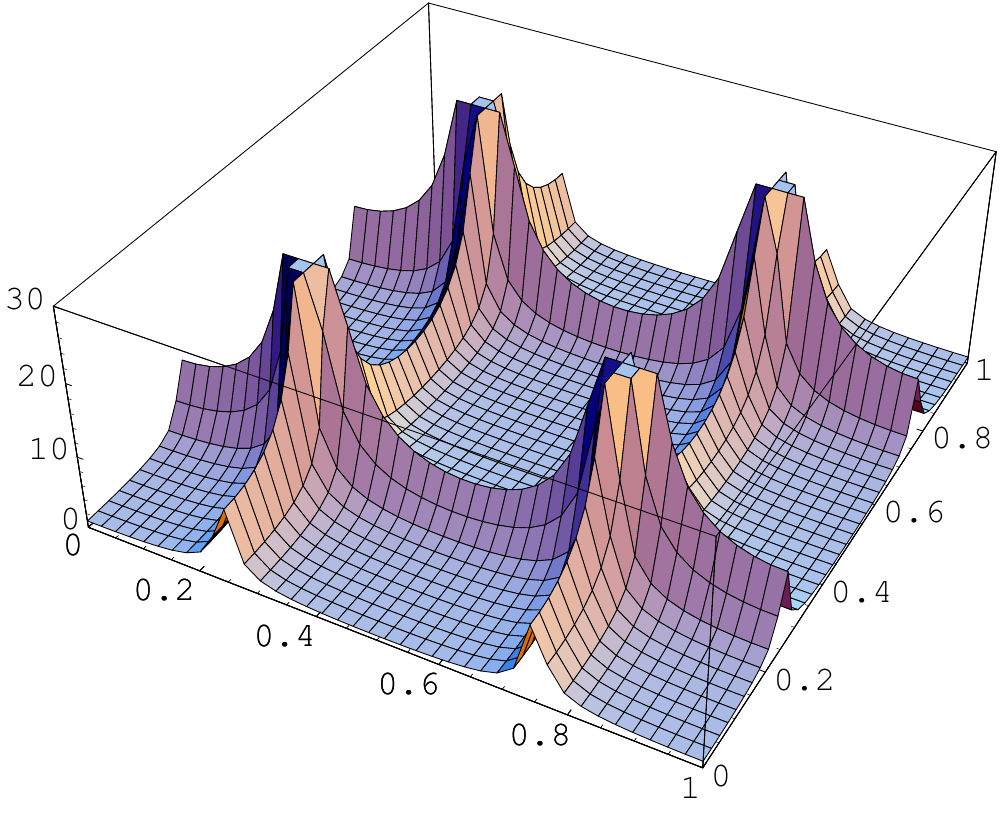}\\(a)}\hfill
\parbox{.46\linewidth}{\centering\includegraphics[width=\linewidth]{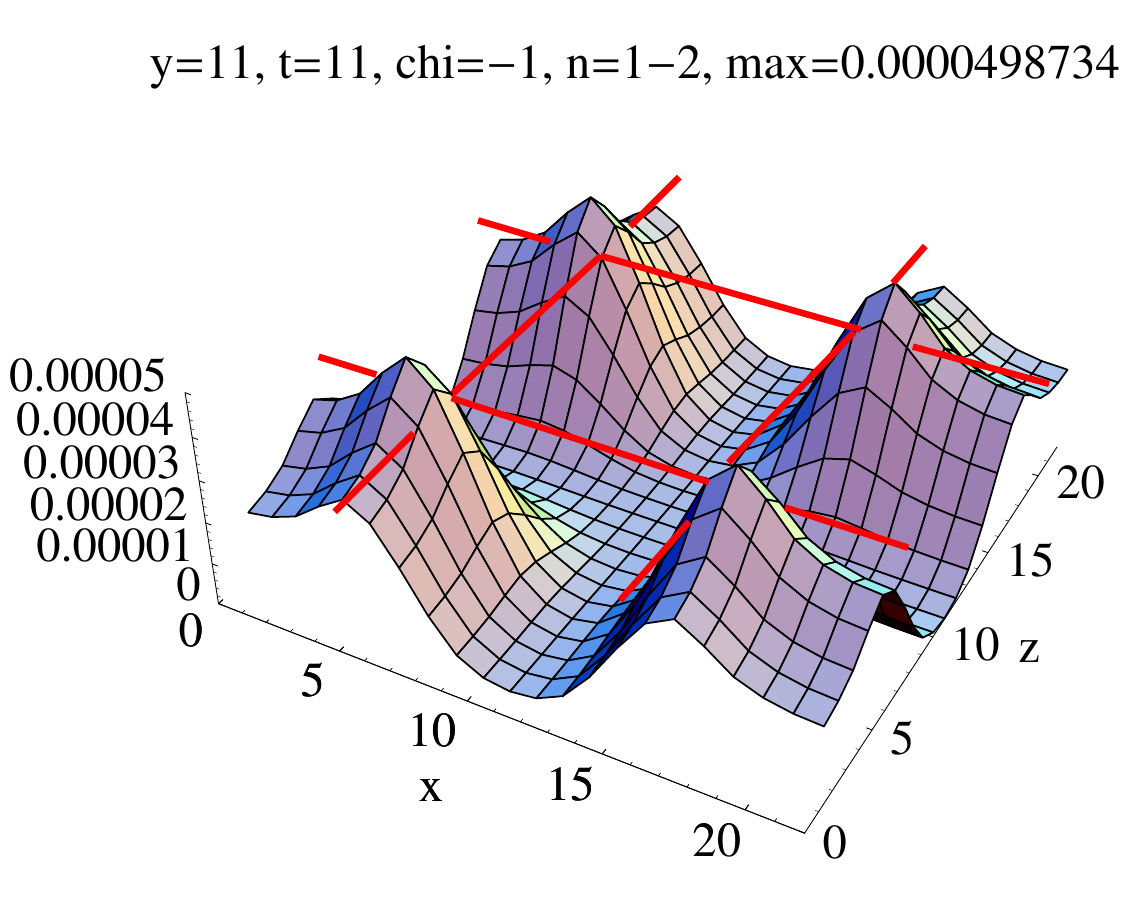}\\(b)}
\caption{The probability density of the zero mode of quarks plotted against the two-dimensional cut of the four-dimensional universe.
(a) Results obtained in the continuum limit with center vortices as a background field, taken from Ref.~\cite{Reinhardt:2002cm}.
(b) Corresponding lattice result, taken from Ref.~\cite{Hollwieser:2011uj}.}
\label{fig:part_ii:zero_mode_density}
\end{figure}

It is interesting to study the quark zero modes emerging in the presence of center vortices.  
As shown above, in order to be topologically non-trivial the vortices must intersect. Figure~\ref{fig:part_ii:zero_mode_density}a shows the probability density of Dirac zero modes in the continuum limit, calculated in the background field of two pairs of intersecting vortex sheets, plotted for a two-dimensional slice of the four-dimensional universe \cite{Reinhardt:2002cm}. In this cut center vortices appear as intersecting lines. One can see that zero modes are concentrated along 
these vortex 
sheets and the concentration is largest at the intersection points. Therefore center vortices act as quark guides in the QCD vacuum. Similar calculations have been done on the lattice (see Ref.~\cite{Hollwieser:2011uj}) and the result is shown in Fig.~\ref{fig:part_ii:zero_mode_density}b. As one can see, it is qualitatively similar to the continuum limit but the probability density 
is smeared out on the lattice, as it should, due to the finite lattice spacing.

\begin{figure}[tbh]
\centering
\begin{tabular}{c@{\qquad\qquad}c@{\qquad\qquad}c}
\includegraphics[height=40ex]{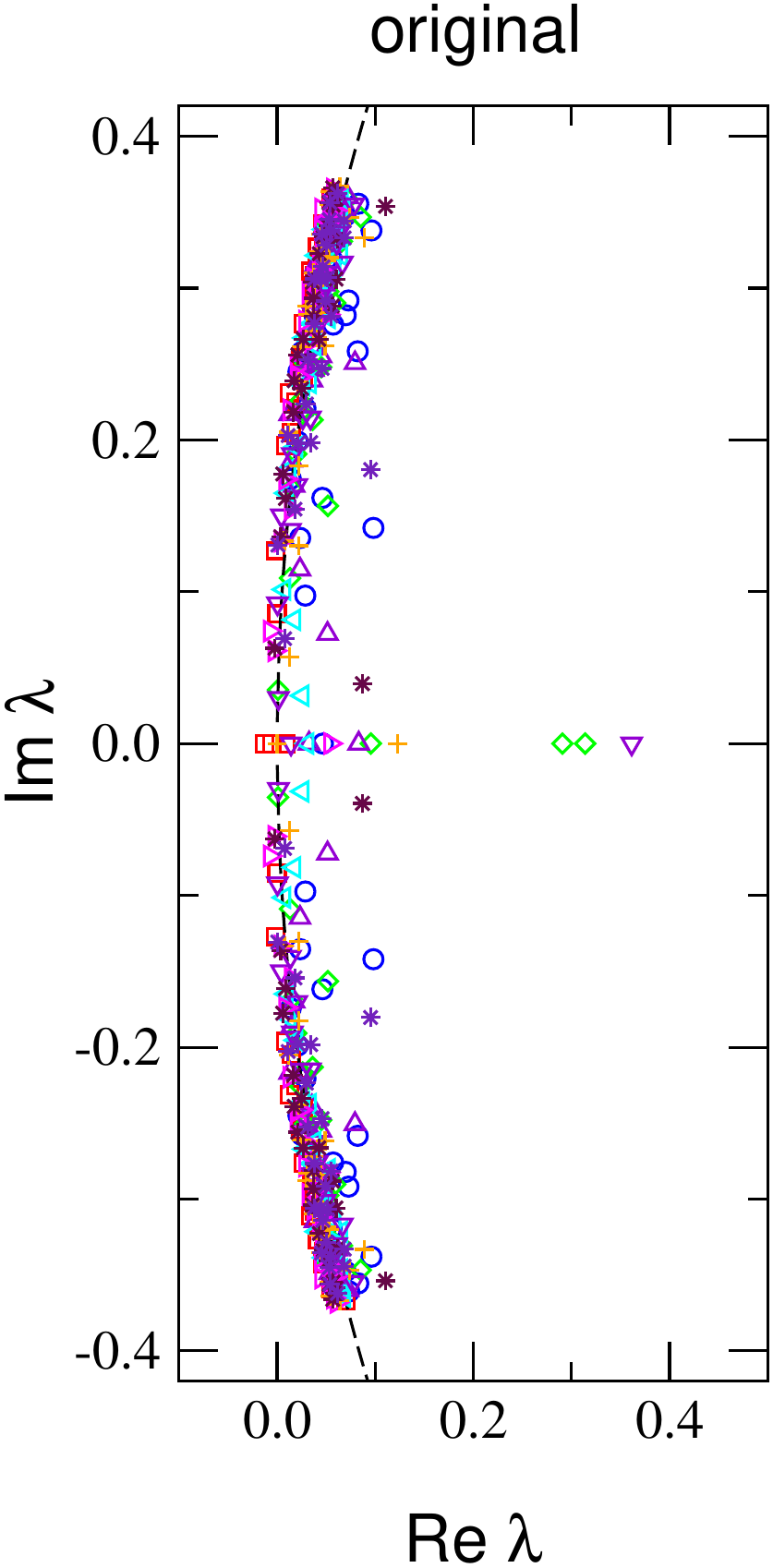} & \includegraphics[height=40ex]{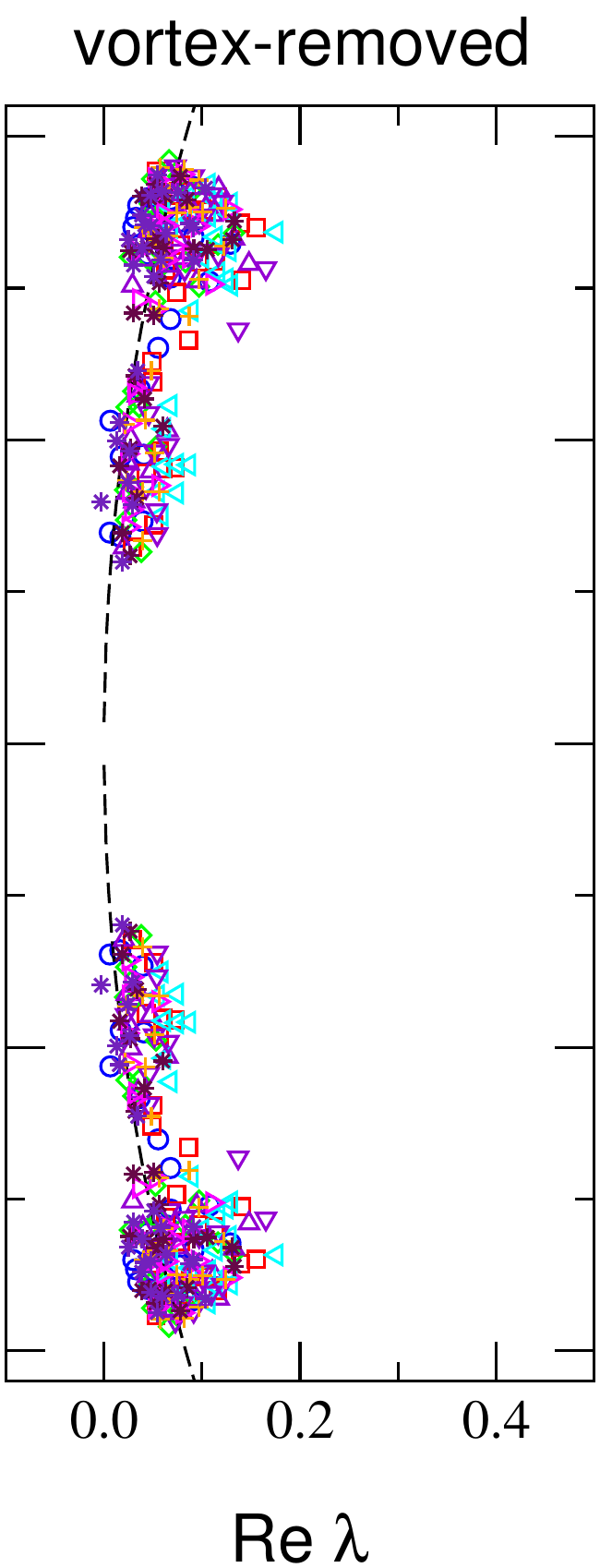} & \includegraphics[height=40ex]{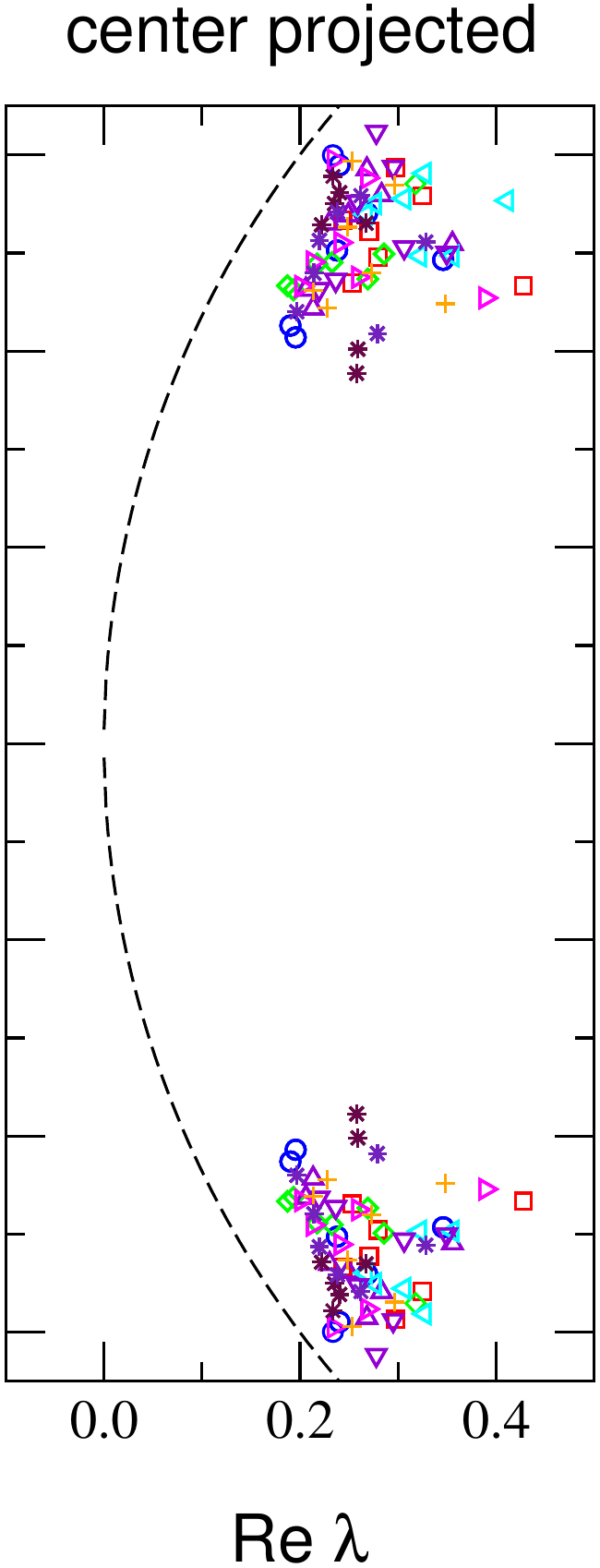} \\
\qquad\quad(a) & (b) & (c)
\end{tabular}
\caption{Low lying eigenvalues of the Dirac operator in the complex plane obtained in SU(2) gauge theory. Each symbol corresponds to different lattice configuration. (a) Original spectrum. (b) Spectrum after removal of center vortices. (c) Spectrum after the center projection. From \cite{Gattnar:2004gx}.}
\label{fig:part_ii:dirac_eigenmodes}
\end{figure}

\begin{figure}[t]
\centering
\includegraphics[width=0.7\linewidth]{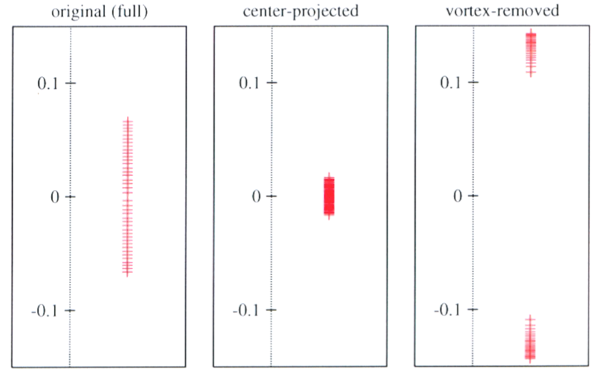}
\caption{The first 20 asqtad Dirac eigenvalue pairs from a $16^4$ lattice at $\beta_{LW} = 3.3$ from Ref.~\cite{Hollwieser:2008tq}.}
\label{FX5}
\end{figure}

Spontaneous breaking of chiral symmetry requires a non-vanishing density of quark (near) zero modes.
This follows from the Banks--Casher relation
\begin{equation}
\langle \bar{q}q\rangle=-\pi\rho(0) \, ,
\label{eq:part_ii:banks_casher}
\end{equation}
which relates the quark condensate $\langle \bar{q}q\rangle$, the order parameter of chiral symmetry breaking, to 
the spectral density $\rho (\lambda) = \dd N_\lambda / \dd \lambda$, where $\dd N_\lambda$ is 
the number of states with eigenvalues in an interval $[\lambda,\lambda+\dd\lambda]$. To have a non-vanishing quark condensate, one needs a non-zero density of near zero Dirac modes. 
The eigenmodes of the Dirac operator have been calculated for the gauge field configurations generated on the lattice. 
Figure~\ref{fig:part_ii:dirac_eigenmodes}a shows the 50 smallest eigenvalues of the Dirac
operator for 10 different configurations for
the gauge group SU(2).
As we can see, the density of near-zero eigenmodes is non-zero and from the Banks--Casher relation (\ref{eq:part_ii:banks_casher}) 
follows that the quark condensate is also non-vanishing. When one removes center vortices from the gauge field configurations considered in Fig.~\ref{fig:part_ii:dirac_eigenmodes}a,
a gap opens up in the Dirac spectrum around zero virtuality $\lambda = 0$ (Fig.~\ref{fig:part_ii:dirac_eigenmodes}b),
and hence the quark condensate vanishes. This suggests that center vortices are responsible for the spontaneous breaking of the chiral symmetry. 
If this is indeed the case 
one would expect that the density of quark models near $\lambda = 0$ increases after center projection.  Surprisingly, the gap is even larger than in the case of vortex removal, see Fig.~\ref{fig:part_ii:dirac_eigenmodes}c. The reason is that the Dirac operator used in these calculations is not sensitive enough to see the (rather singular) center projected vortices. Lattice calculations with a more sophisticated Dirac operator yield the expected result\,---\,after center projection only near-zero modes are left \cite{Hollwieser:2008tq}, see Fig.~\ref{FX5}. This shows that the center vortices are indeed the dominant IR degrees of freedom which are responsible not only for quark confinement but also for the spontaneous chiral symmetry breaking.

\subsection{Center vortex dominance}
\begin{figure}[t]
\parbox{0.5\linewidth}{\vspace*{0pt}\includegraphics[width=\linewidth]{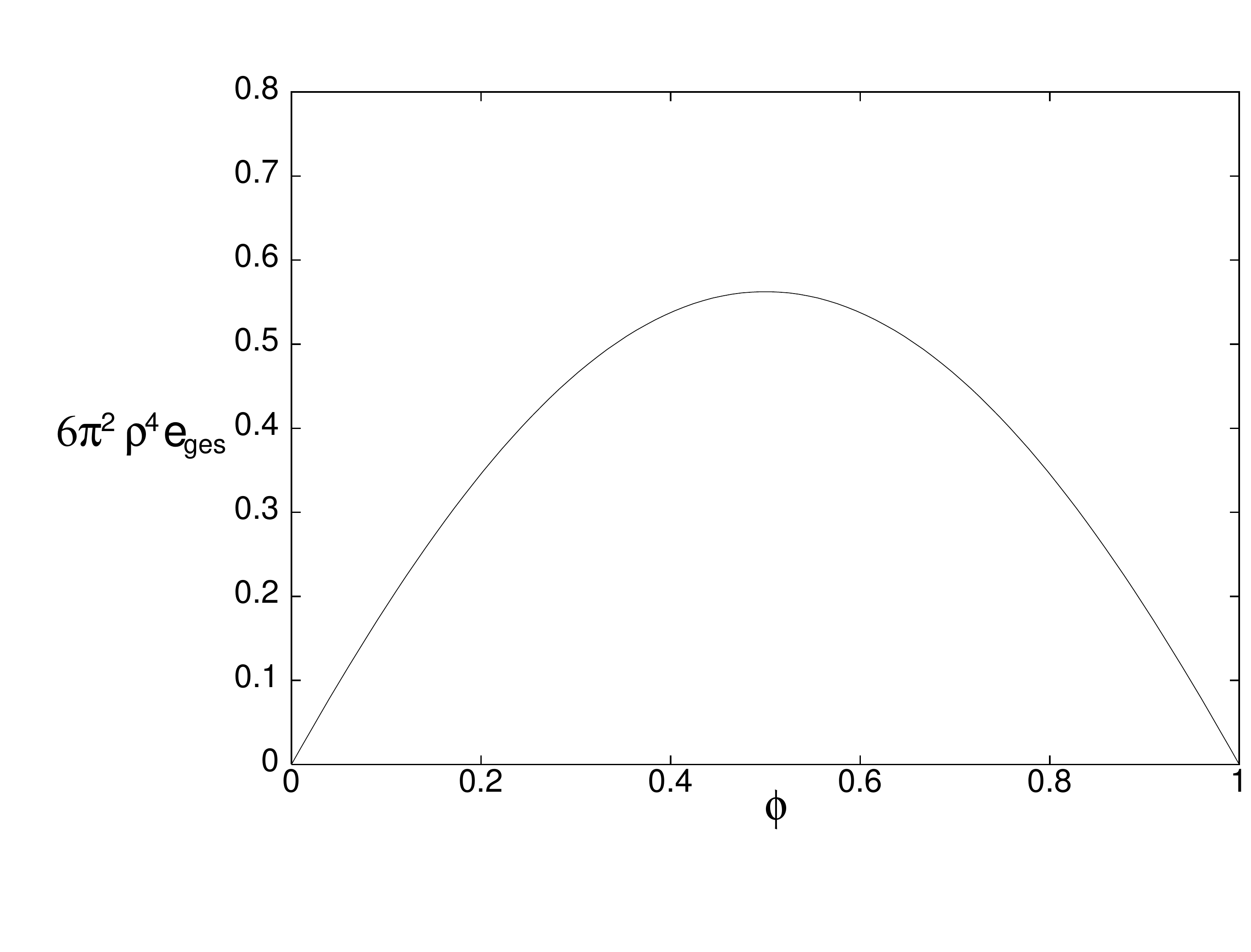}}\hfill
\parbox{0.4\linewidth}{\vspace*{0pt}\includegraphics[width=\linewidth]{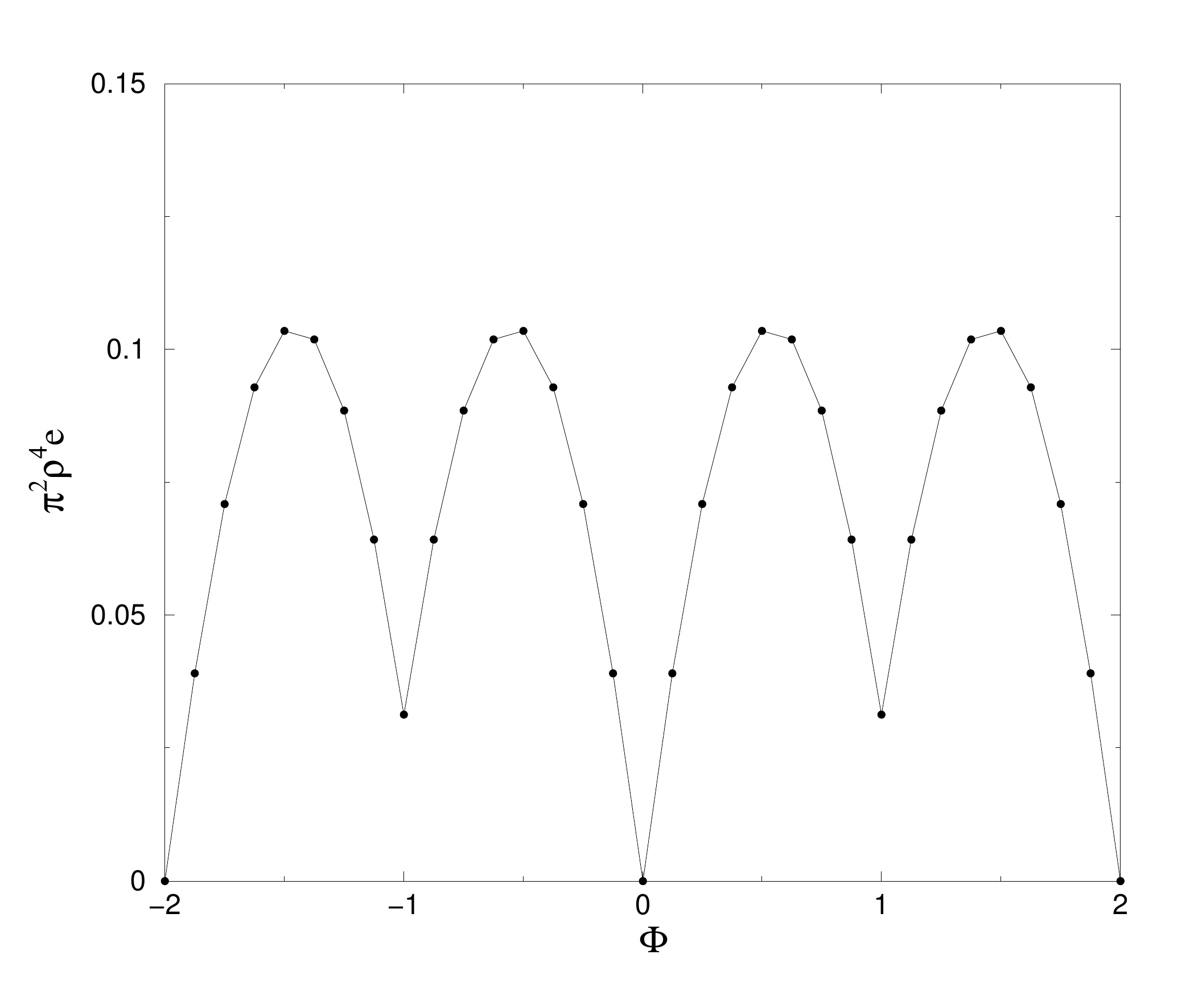}}
\caption{Energy density as a function of the magnetic flux. (left) Pure gauge system. (right) System with both gauge and fermion fields. From \cite{Lange:2003ti}.}
\label{fig:part_ii:vortex_energy_density}
\end{figure}

One may ask: Why are center vortices the dominant IR degrees of freedom? What distinguishes center vortices from other gauge field 
configurations, for instance from vortices with a flux different from that of center vortices? The latter question was investigated in Ref.~\cite{Lange:2003ti}, where the energy density of a straight magnetic vortex was calculated at one-loop level as function of its flux $\phi$ for pure Yang--Mills theory and for QCD. The result is shown in Fig.~\ref{fig:part_ii:vortex_energy_density}. The flux is normalized such that $\phi=1$ corresponds to the the flux of the center vortex, while $\phi = 0$ refers to the perturbative vacuum. In the pure Yang--Mills  case one finds that the energy density of the center vortex is the same as the one of the vacuum, 
while all other fluxes have higher energy. In the presence of 
quarks the energy density corresponding to the center vortex flux exceeds the energy density of the perturbative vacuum, but the center vortex flux still represents a local minimum. This provides a qualitative 
explanation of the center vortex dominance in the IR sector of QCD as compare to other flux tube configurations.

\subsection{Conclusions}
In this lecture the center vortex picture of the QCD vacuum has been presented. Center vortices seem to be dominant infra-red configurations 
of the gauge field and provide appealing pictures of 
confinement (center vortices percolating through 
space-time lead to an area law for the Wilson loop) and the deconfinement phase 
transition (as the temperature increases, center vortices align in the temporal direction which leads to a 
vanishing temporal string tension).

Topological properties of center vortices and their relevance for
the spontaneous breaking of 
chiral symmetry have also been discussed. The topological charge of an oriented vortex sheet vanishes and to be topologically non-trivial, a vortex sheet has to be non-oriented. Non-orientability of vortex 
sheets is caused by magnetic monopoles, which change the direction of the vortex flux. 
The topological charge is concentrated on vortex intersection points, which give fractional contributions to the 
topological charge. Nevertheless the total topological charge is integer valued. 
Finally, the quark zero modes are concentrated on center vortices and, in particular, at their
  intersection points. Furthermore, the low-lying Dirac modes disappear when center vortices are removed from the Yang--Mills ensemble 
  while only near zero modes survive after center projection. This
   shows that center vortices are responsible for chiral symmetry breaking.

   
\section{Hamiltonian approach to QCD in Coulomb gauge}

\subsection{Introduction}
In the previous lectures we have discussed two scenarios of quark confinement: magnetic monopole condensation (dual Mei\ss ner effect), 
which utilizes the maximal Abelian gauge, and center vortex condensation, which was established by lattice calculations in
the maximal center gauge. In this lecture the Gribov--Zwanziger mechanism of confinement is presented. This picture of 
confinement is formulated in Coulomb gauge and naturally emerges within a Hamiltonian approach as I
 will show in this lecture. I will also exhibit various connections between the different
 pictures of 
 confinement. Before I expand the Hamiltonian approach to QCD in Coulomb gauge let me give some arguments why this approach is advantageous in non-perturbative studies.
 
Nowadays the most popular approach to quantum field theory is the path integral formulation, 
which is definitely advantageous in perturbation theory, which can be formulated in terms of Feynman diagrams. 
Furthermore, the functional integral formulation is also the basis for the numerical lattice calculation, which 
are fully non-perturbative. On the other hand in ordinary (non-relativistic) quantum mechanics it is usually much simpler 
to solve the Schr\"odinger equation than calculating the corresponding functional integral. 
The hydrogen atom is a good example. 
One can easily find the exact solution of the Schr\"odinger equation for this problem, while 
it is an extremely difficult task to find the exact
solution within the path integral formalism. 
Therefore, for non-perturbative studies in continuum quantum field 
theory 
we expect the Hamiltonian approach to be more efficient than approaches based on the functional integral formulation 
like for e.g. Dyson--Schwinger equations or functional renormalization group flow equations. 
The Hamiltonian approach to QCD has been worked out mainly in collaboration with C.~Feuchter, W.~Schleifenbaum, D.~Cam\-pa\-gnari and P.~Vastag \cite{Feuchter:2004mk,Reinhardt:2004mm,Schleifenbaum:2006bq,Campagnari:2010wc,Vastag:2015qjd}.
For didactic reasons I will develop the Hamiltonian approach first for pure Yang--Mills theory. 

\subsection{Canonical quantization of Yang--Mills theory}
The starting point for the canonical quantization of Yang--Mills theory is the classical action,
\begin{equation}
S=\frac{1}{4g^2}\Tr\int \D^4x F_{\mu\nu}F_{\mu\nu},
\label{eq:part_iii:yang_mills_action}
\end{equation}
where 
\begin{equation}
F_{\mu\nu}=\partial_\mu A_\nu-\partial_\nu A_\mu+[A_\mu,A_\nu]
\label{eq:part_iii:field_strength_tensor}
\end{equation}
is the field strength tensor. 
In the canonical quantization the components of the gauge field $A_\mu^a(x)$ itself serve as coordinates. The canonical momenta conjugate to the spatial 
components of the gauge field are given by the color electric field
\begin{equation}
\Pi_i^a(\vx)=\frac{\delta S}{\delta \dot{A}_i^a(\vx)}=E_i^a(\vx) \, ,
\label{eq:part_iii:conjugate_momentum_defininition}
\end{equation}
while the momenta conjugate 
to the temporal components of the gauge field vanish, $\Pi_0^a=0$. This causes 
a problem in the canonical quantization. To avoid
 this problem one can choose Weyl gauge, $A_0^a=0$, in which the classical Yang--Mills Hamiltonian becomes
\begin{equation}
H=\frac{1}{2}\int \D^3x \bigl[\vec{\Pi}^a(\vx) \cdot \vec{\Pi}^a(\vx)+\vec{B}^a(\vx) \cdot \vec{B}^a(\vx) \bigr].
\label{eq:part_iii:hamiltonian_in_weyl_gauge}
\end{equation}
In order to quantize the system, on has to replace the canonical conjugate
coordinates and momenta by operators, satisfying the following commutation relations:
\begin{eqnarray}
\left[ A^a_k (\vx) , \Pi^b_l (\vy) \right]&=&\I \delta^{ab} \delta_{kl} \delta(\vx - \vy) \\
\left[ A^a_k (\vx) , A^b_l (\vy) \right]&=&\left[ \Pi^a_k (\vx) , \Pi^b_l (\vy) \right] = 0 .
\label{eq:part_iii:commutation_relations}
\end{eqnarray}
In the ``coordinate representation'' the gauge fields are classical functions while the canonical momentum operator is given by
\begin{equation}
\Pi^a_k (\vx) = \frac{\delta}{\I \delta A^a_k (\vx)}.
\label{eq:part_iii:canonical_momentum}
\end{equation}
In Weyl gauge Gauss' law is lost as an equation of motion and has to be 
imposed as a constraint on the wave functional $\Psi [A]$
\begin{equation}
\hat{D}^{ab}_k (\vx) \Pi^b_k (\vx) \psi [A] = \rho_m^a (\vx) \psi [A],
\label{eq:part_iii:gauss_law}
\end{equation}
where 
\begin{equation}
\hat{D}^{ab}_k(\vx) = \delta^{ab} \partial^x_k + g f^{acb} A^c_k (\vx)
\label{eq:part_iii:covariant_derivative}
\end{equation}
is the covariant derivative in the adjoint representation of the gauge group and $\rho_m$ is the color charge density of the matter fields. 
Moreover, the operator on the left-hand side, $\hat{\vD} \cdot \vec{\Pi}$, is the generator of space-dependent but time-independent gauge transformations. When matter fields are not present, $\rho_m=0$, the right-hand side of Eq.~$($\ref{eq:part_iii:gauss_law}$)$ vanishes and the wave functional must be invariant under time-independent gauge transformations $U(\vx)$, $\psi[A]=\psi[A^U]$.

After quantization the 
Hamiltonian (\ref{eq:part_iii:hamiltonian_in_weyl_gauge}) becomes an operator, acting in the Hilbert space of gauge invariant wave functionals with the scalar product 
\begin{equation}
\langle \Phi\vert \ldots\vert \Psi\rangle =\int \mathcal{D}\vA \Phi^*[\vA]\ldots\Psi[\vA].
\label{eq:part_iii:scalar_product}
\end{equation}
Here $\int \mathcal{D}\vA$ is the functional integral over the time-independent spatial components of the gauge field. The main objective in the Hamiltonian approach is to solve the Schr\"odinger equation
\begin{equation}
H\Psi[A]=E\Psi[A]
\label{eq:part_iii:schroedinger_equation}
\end{equation}
for the wave functional $\Psi [A]$.
Usually one is interested in the wave functional of the vacuum. In $1 + 1$ dimensions the Yang-Mills  Schr\"odinger equation (\ref{eq:part_iii:schroedinger_equation}) can be solved exactly \cite{Reinhardt:2008ij}.  Approximate solutions for the gauge invariant vacuum state have been found in 2+1 dimensions \cite{Haagensen:1995py,Karabali:1998yq} and for a limiting case
 in 3+1 dimensions \cite{Greensite:1979yn}. In general, the construction of the gauge invariant wave functional is extremely difficult and 
 it is much more efficient to choose a specific gauge. 
 A convenient choice for this problem is the Coulomb gauge, $\vec{\partial}\cdot\vA=0$. 
 Implementing the Coulomb gauge by means of the Faddeev--Popov method the scalar product $($\ref{eq:part_iii:scalar_product}$)$ becomes
\begin{equation}
\langle \Phi\vert \ldots \vert\Psi\rangle =\int \mathcal{D}\vA^{\perp}J(\vA^\perp) \Phi^*[\vA^\perp]\ldots\Psi[\vA^\perp],
\label{eq:part_iii:scalar_product_in_coulomb_gauge}
\end{equation}
where the functional integration extends over the transverse part, $\vA^\perp$, of the gauge field only and  
\begin{equation}
J(\vA^\perp)=\Det(-\hat{\vD}\cdot\vec{\partial})
\label{eq:part_iii:faddeev_popov_det}
\end{equation}
is the Faddeev--Popov determinant in Coulomb gauge. 
The Coulomb gauge fixing may be seen as a transition from the Cartesian $\vA$ 
to ``curvilinear'' $\vA^\perp$ coordinates with the Faddeev--Popov determinant corresponding to the Jacobian of this transformation.

Coulomb gauge fixing eliminates the longitudinal components of the (spatial) 
gauge field. The momentum operator, however, still contains transverse and longitudinal components, 
$\vec{\Pi}=\vec{\Pi}^\perp+\vec{\Pi}^\parallel$, where 
the transverse part is still given by eq. (\ref{eq:part_iii:canonical_momentum}), 
$\Pi^{\perp a}_k(\vx)=\delta/\ii\delta A^{\perp a}_k(\vx)$. 
The longitudinal components of the momentum operator can be determined by resolving Gauss' law (\ref{eq:part_iii:gauss_law}), which leads to
\begin{equation}
\vec{\Pi}^\parallel\psi[\vA]=-\vec{\partial}(-\hat{\vD}\vec{\partial})^{-1} \rho \, \psi[\vA],
\label{eq:part_iii:longitudinal_momentum}
\end{equation}
where 
\begin{equation}
\rho^a(\vx) =-f^{abc}A^b_k(\vx)\Pi_k^c(\vx)+\rho_m^a(\vx)
\label{eq:part_iii:color_charge_density}
\end{equation}
is the total color charge density, composed of the color charge density of matter, $\rho_m$, and the color charge density of the gauge field, $\rho^a_g(\vx)=-f^{abc}A^b_k(\vx)\Pi_k^c(\vx)$. The latter exists only in non-Abelian gauge theories. 
It should be emphasized that ~Eq.~\eqref{eq:part_iii:longitudinal_momentum} is not a true operator identity, i.e.
the operator expression  implied by Eq.~(\ref{eq:part_iii:longitudinal_momentum}) for $\vec{\Pi}^{||}$ 
(\ref{eq:part_iii:longitudinal_momentum}) 
is valid only when $\vec{\Pi}^\parallel$ acts on the wave functional. Using $\vec{\Pi} = \vec{\Pi}^\perp + \vec{\Pi}^{||}$ and the relations
$($\ref{eq:part_iii:longitudinal_momentum}$)$ and $($\ref{eq:part_iii:color_charge_density}$)$ one derives from Eq.~(\ref{eq:part_iii:hamiltonian_in_weyl_gauge}) the Yang--Mills Hamiltonian in the Coulomb gauge \cite{Christ:1980ku}
\begin{equation}
H=\frac{1}{2}\int \D^3x \left(J^{-1}\vec{\Pi}^a(\vx)\cdot J\vec{\Pi}^a(\vx)+\vec{B}^a(\vx)\cdot \vec{B}^a(\vx)\right)+H_C\equiv H_{YM}+H_C
\label{eq:part_iii:YM_hamiltonian}
\end{equation}
where
\begin{equation}
H_C=\frac{1}{2}\int \D^3x\int \D^3y J^{-1}\rho^a(\vx)J\left[(-\hat{\vD}\cdot\vec{\partial})^{-1}(-\vec{\partial}^2)(-\hat{\vD}\cdot\vec{\partial})^{-1}\right]^{ab}(\vx,\vy)\rho^b(\vy)
\label{eq:part_iii:coulomb_term}
\end{equation}
is the Coulomb term which arises from the kinetic energy of the longitudinal components of the momentum operator. In the case of QED 
this term reduces to the ordinary Coulomb interaction
 between the electric charge distribution $\rho_m$.
 
The Yang--Mills Hamiltonian in the Coulomb gauge, Eq.~$($\ref{eq:part_iii:YM_hamiltonian}$)$, is more complicated than the original gauge invariant one, Eq.~$($\ref{eq:part_iii:hamiltonian_in_weyl_gauge}$)$. The kinetic energy of the transverse degrees of freedom (the first term of Eq.~$($\ref{eq:part_iii:YM_hamiltonian}$)$) contains the Faddeev--Popov determinant. Moreover, the Coulomb term is a highly non-local object. The Faddeev--Popov determinant is also present in the scalar product $($\ref{eq:part_iii:scalar_product_in_coulomb_gauge}$)$. It is, however, still more convenient to work with the complicated gauge fixed Hamiltonian (\ref{eq:part_iii:YM_hamiltonian}) than with gauge invariant wave functionals. It should also be stressed that 
by implementing 
 Gauss' law in the gauge fixed Hamiltonian gauge invariance has been fully accounted for.
 
\subsection{Variational solution for the Yang--Mills vacuum wave functional}
To solve
 the Yang--Mills Schr\"odinger equation for the gauge-fixed Hamiltonian (\ref{eq:part_iii:YM_hamiltonian}) one can exploit  the variational principle. This was 
 first done by D. Schutte who assumed a Gaussian \emph{ansatz} for the vacuum wave functional \cite{Schutte:1985sd},
\begin{equation}
\Psi[\vA]=\exp \left[-\frac{1}{2}\int \D^3x\int \D^3y\,  A^a_k(\vx) \, \omega(\vx,\vy) \, A^a_k(\vy)\right],
\label{eq:part_iii:gaussian_ansatz}
\end{equation}
and derived a set of coupled 
integral equations for the gluon propagator, the ghost propagator and the Coulomb potential. This set of equations was rederived 
and solved numerically in ref. \cite{Szczepaniak:2001rg}. 
An improved variational approach to the Yang--Mills Schr\"odinger equation has been developed in refs.  \cite{Feuchter:2004mk,Reinhardt:2004mm}. 
This approach differs from previous works 
in: i) the form of the vacuum wave functional, ii) the treatment of the Faddeev--Popov determinant (which turns out to be crucial for the confining properties of the theory) and iii) the renormalization. 
The trial \emph{ansatz} used in ref. \cite{Feuchter:2004mk} for the vacuum wave functional is
\begin{equation}
\Psi[A]=\frac{1}{\sqrt{J(A)}}\exp \left[-\frac{1}{2}\int \D^3x\int \D^3y \, A^a_k(\vx)\, \omega(\vx,\vy)\, A^a_k(\vy)\right],
\label{eq:part_iii:trial_ansatz}
\end{equation}
where $\omega(x,y)$ is the variational kernel which is determined by 
 minimizing the energy $\langle\Psi\vert H\vert\Psi\rangle$. The advantage of this \emph{ansatz} is that the static gluon propagator is essentially
 given by the inverse of the variational kernel:
\begin{equation}
\langle A^a_k(\vx)A^b_l(\vy)\rangle=\delta^{ab}t_{kl}(\vx)\frac{1}{2} \omega^{-1} (\vx, \vy),
\label{eq:part_iii:gluon_propagator}
\end{equation}
\begin{figure}[t]
\includegraphics[width=0.49\linewidth]{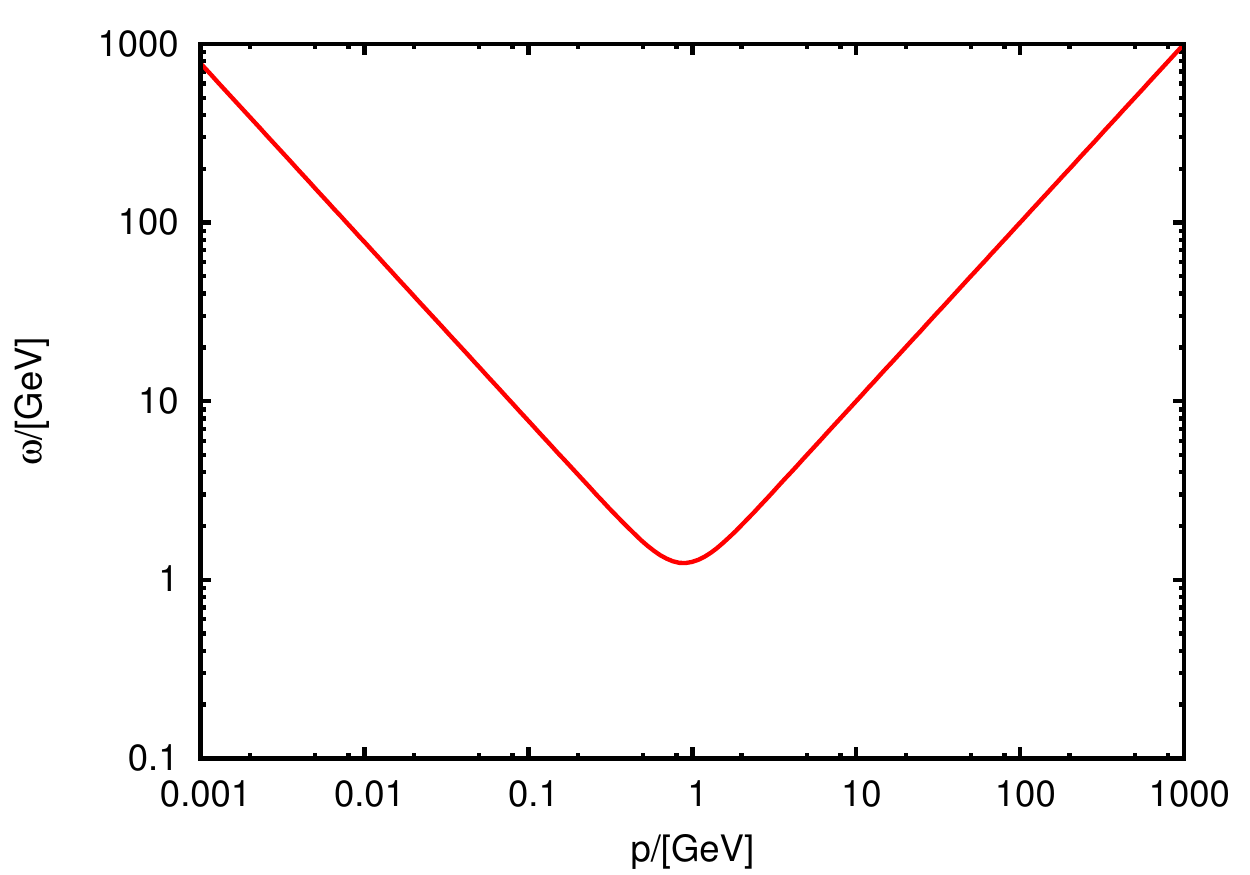}
\includegraphics[width=0.49\linewidth]{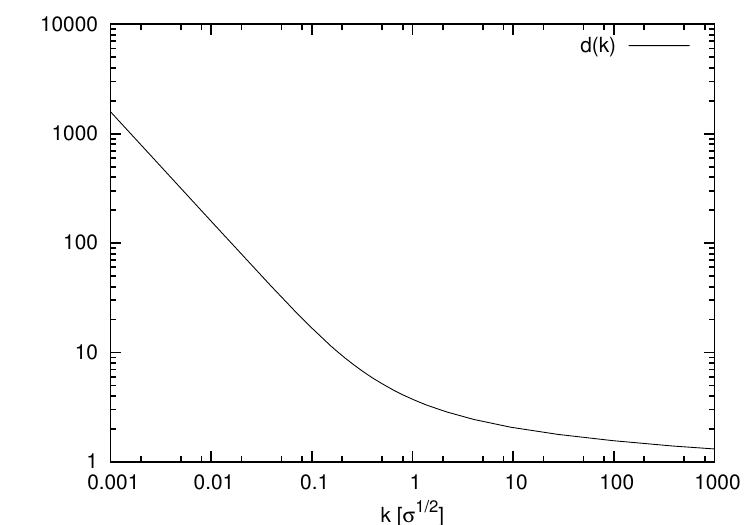}
\caption{The gluon energy $\omega (p)$ (left panel) and the ghost form factor $d$ (right panel) obtained in
	 the variational approach for SU(2) Yang--Mills theory. From  \cite{Epple:2006hv}.}
\label{fig:part_iii:variational_results}
\end{figure}%
where $t_{kl}(\vx)=\delta_{kl}-\partial^x_k\partial^x_l/\partial^2_x$ is the transverse projector. From the form of this propagator follows 
that its Fourier transform $\omega (\vp)$ represents the single-particle gluon energy. Minimization of the energy density with respect to $\omega(\vx,\vy)$ leads to the result shown in the left panel of Fig.~\ref{fig:part_iii:variational_results}. For large momenta the gluon energy behaves like
the photon energy, $\omega(p)\sim \vert \vp\vert$, while at small momenta it diverges, $\omega(p)\sim 1/\vert \vp\vert$. This means that there are 
no free gluons in the infrared, which is a manifestation of gluon confinement.

\begin{figure}[t]
\centering
\includegraphics[width=0.2\linewidth]{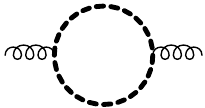}
\qquad \hspace*{2cm}
\includegraphics[width=0.25\linewidth]{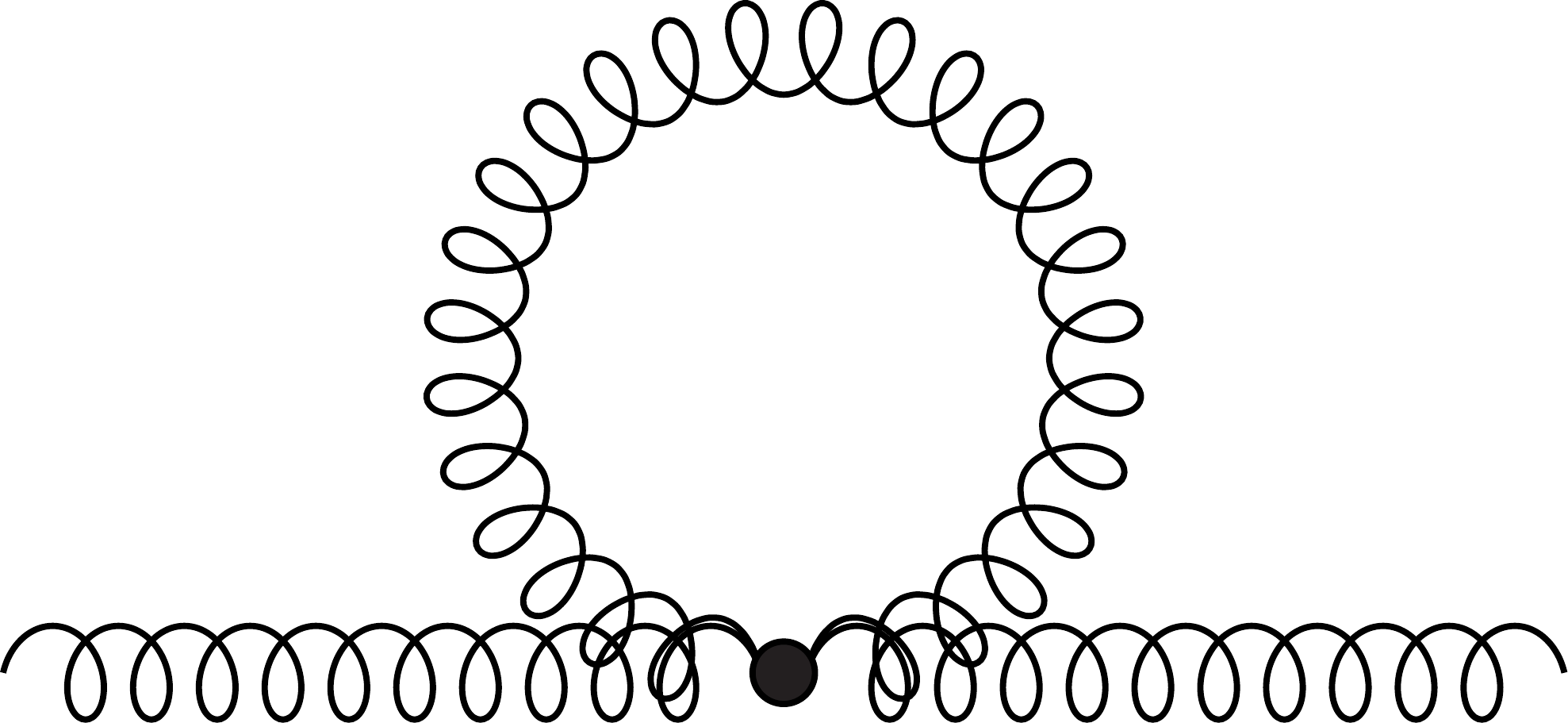}
\caption{Diagrammatic representation of the ghost loop (\ref{1634-GX1}) (left panel) and the tadpole $I_{tad}$ (right panel). A wavy and a dashed, respectively,  line represents the gluon (\ref{eq:part_iii:gluon_propagator}) and ghost (\ref{eq:part_iii:ghost_propagator}), respectively, propagator.}
\label{fig:part_iii:diagrams}
\end{figure}

\begin{figure}[t]
	\centering
	\includegraphics[width=0.5\linewidth]{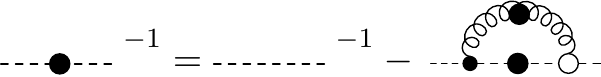}
	\caption{Dyson--Schwinger equation for the ghost propagator.}
	\label{fig:part_iii:diagrams_33}
\end{figure}

The general form of the gluon gap equation obtained by minimizing the energy density is
\begin{equation}
\omega^2(p)=\vp^2+\chi^2(p)+I_\mathrm{tad}+I_\mathrm{C}(p),
\label{eq:part_iii:gap_equation_general}
\end{equation}
where $I_\mathrm{tad}$ is the tadpole, see Fig.~\ref{fig:part_iii:diagrams}, and 
\begin{equation}
\chi^{ab}_{kl} (\vx, \vy) = - \frac{1}{2} \langle \Psi \left| \frac{\delta^2 \ln J [A]}{\delta A^a_k (\vx) \delta A^b_l (\vy)} \right| \Psi \rangle = \delta^{ab} t_{kl} (\vx - \vy) \chi (\vx - \vy) \label{1634-GX1}
\end{equation}
 is the ghost loop, ($t_{ij} (x) = \delta_{ij} - \partial_i \partial_j / \partial^2$ is the transversal projector.)
Furthermore,
 $I_\mathrm{C}(p)$ follows from the Coulomb term. Both $I_{tad}$ and $I_C$ contain UV-divergencies, which are removed 
 by adding  two counter terms to the gauge fixed Yang-Mills Hamiltonian (\ref{eq:part_iii:YM_hamiltonian}), 
 see refs. \cite{Epple:2007ut}, \cite{Reinhardt:2007wh} for details. The gluon gap
 equation (\ref{eq:part_iii:gap_equation_general}) has the form of a relativistic dispersion relation. To calculate the ghost loop $\chi$ one needs the ghost propagator 
 $G = \langle \Psi | ( - \hat{D} 
 \partial)^{- 1} | \Psi \rangle$. Once the vacuum wave functional $| \Psi \rangle$ is known the ghost propagator
 can, in principle, be evaluated. 
 However, this cannot be done in closed form even for the variational \emph{ansatz} (\ref{eq:part_iii:trial_ansatz}). 
  To handle this problem one can expand the inverse of the Faddeev--Popov operator $-\hat{\vD}\cdot \vec{\partial}$ in  
powers of the gauge field $\vA^\perp$ 
 and then resum it after certain approximations. This, however, does not lead to a closed form of the ghost propagator, but to the Dyson--Schwinger equation for this propagator (shown in Fig.~\ref{fig:part_iii:diagrams_33}). In principle, it is neither clear nor trivial that the gap equation (\ref{eq:part_iii:gap_equation_general}) coupled to the ghost Dyson--Schwinger equations does have a 
 solution.\footnote{In principle, there is only one variational equation since we have only one variational kernel, $\omega$. 
 The ghost Dyson--Schwinger equation only comes into the game since we 
 are unable to calculate the ghost propagator with the trial wave functional (\ref{eq:part_iii:trial_ansatz}) in closed form.}\\
It is convenient to represent the ghost propagator as
\begin{equation}
G \equiv \langle \psi | (- \hat{\vD} \bar{\partial})^{- 1} | \psi \rangle =\frac{d(\varDelta)}{-\varDelta},
\label{eq:part_iii:ghost_propagator}
\end{equation}
where $d(\varDelta)$ is called the ghost form factor, containing all the deviations of the gluon propagator from the 
photon propagator. For the latter the form factor is just unity.

Numerical calculations of the coupled gluon gap equation and ghost DSE show
 that in the pure glue sector the tadpole contribution $I_\mathrm{tad}$ and the Coulomb term $I_\mathrm{C}(p)$ can be neglected. 
 The latter, however, has to be included when quarks are present, for it is responsible for chiral symmetry breaking. 
 Hence, in the Yang--Mills sector the gap equation (\ref{eq:part_iii:gap_equation_general}) can be reduced to
\begin{equation}
\omega^2(p)=\vp^2+\chi^2(p),
\label{eq:part_iii:gap_equation_pure_glue}
\end{equation}
with the ghost loop $\chi$ given in the left panel of Fig.~\ref{fig:part_iii:diagrams} in terms of the ghost propagator, which in turn has to be found by solving the ghost Dyson--Schwinger equation, shown in Fig.~\ref{fig:part_iii:diagrams_33}.

The infrared analysis of these two equations has been carried out in Refs.~\cite{Schleifenbaum:2006bq}, where power laws for the gluon energy, $\omega=A/p^\alpha$, and the ghost form factor, $d(p)=B/p^\beta$, have been assumed. Furthermore, the ghost form factor is assumed to fulfill the so-called horizon condition, $d^{-1}(0)=0$, which is the crucial part of the Gribov--Zwanziger confinement scenario (see the dission following eq. (\ref{eq:part_iii:coulomb_potantial_transform}). Assuming also that the ghost-gluon vertex is bare, one finds the following sum rule for the IR exponents:
\begin{equation}
\alpha=2\beta+2-d,
\label{eq:part_iii:sum_rule}
\end{equation}
\begin{figure}[t]
\sidecaption[t]
\includegraphics[width=0.5\linewidth]{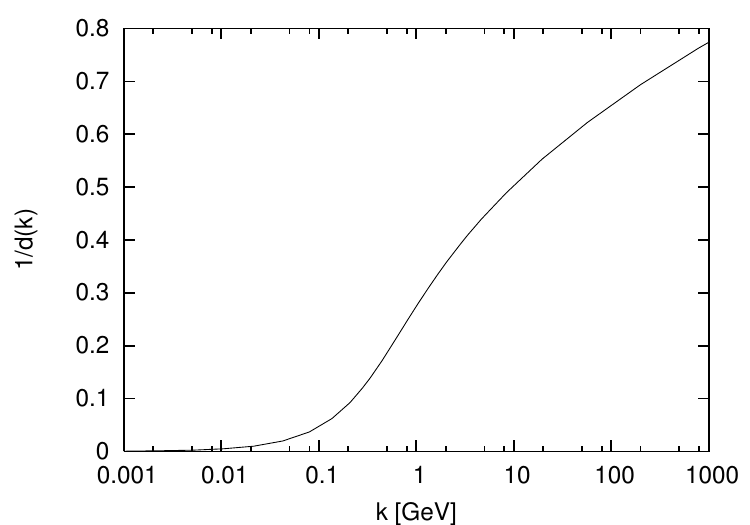}
\caption{Dielectric function of the Yang--Mills vacuum obtained in the variational approach. From  \cite{Reinhardt:2008ek}.}
\label{fig:part_iii:dielectric_function}       
\end{figure}%
\begin{figure}[b]
\parbox{.45\linewidth}{\centering\includegraphics[width=\linewidth]{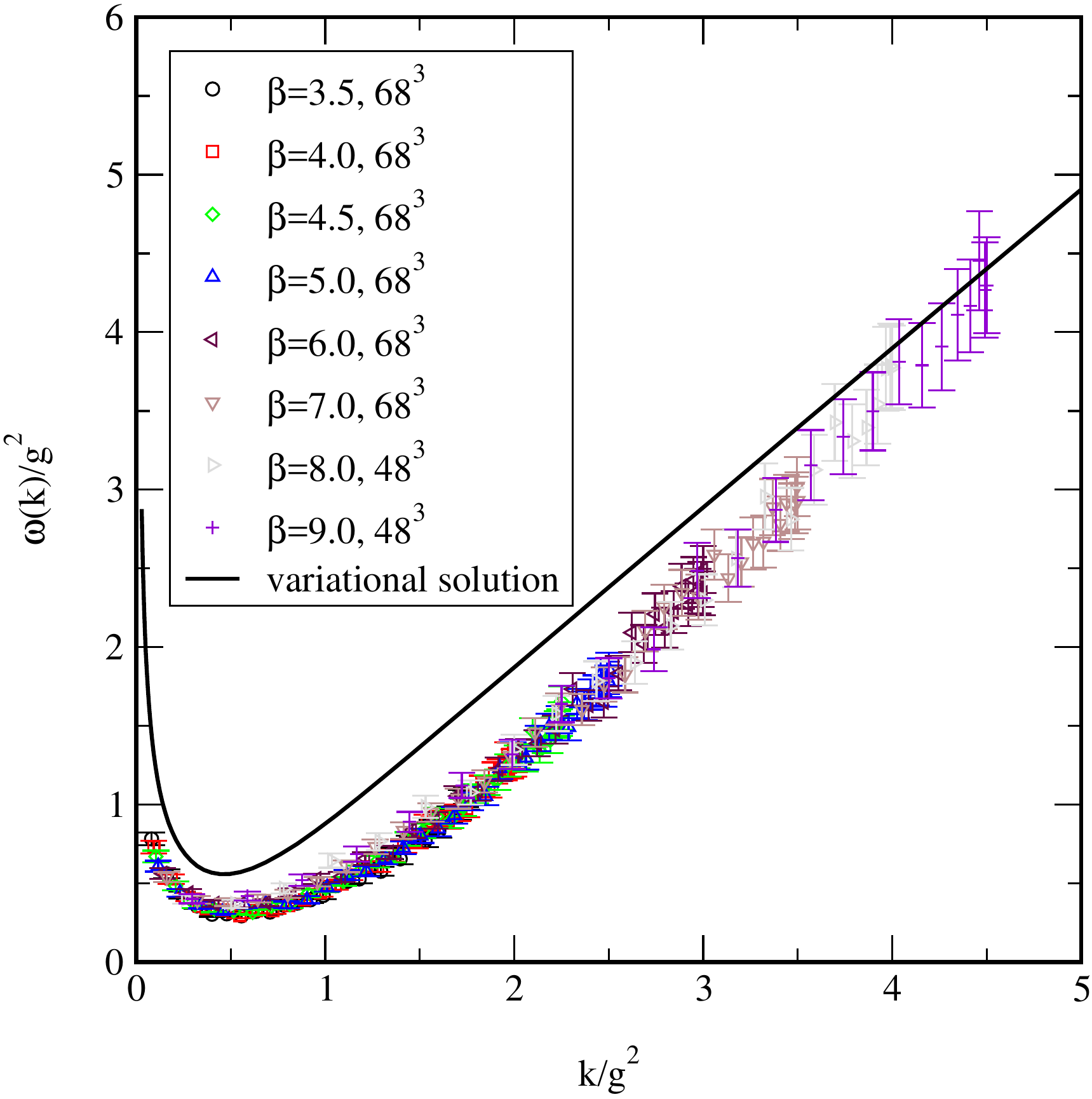}\\(a)}\hfill
\parbox{.45\linewidth}{\centering\includegraphics[width=\linewidth]{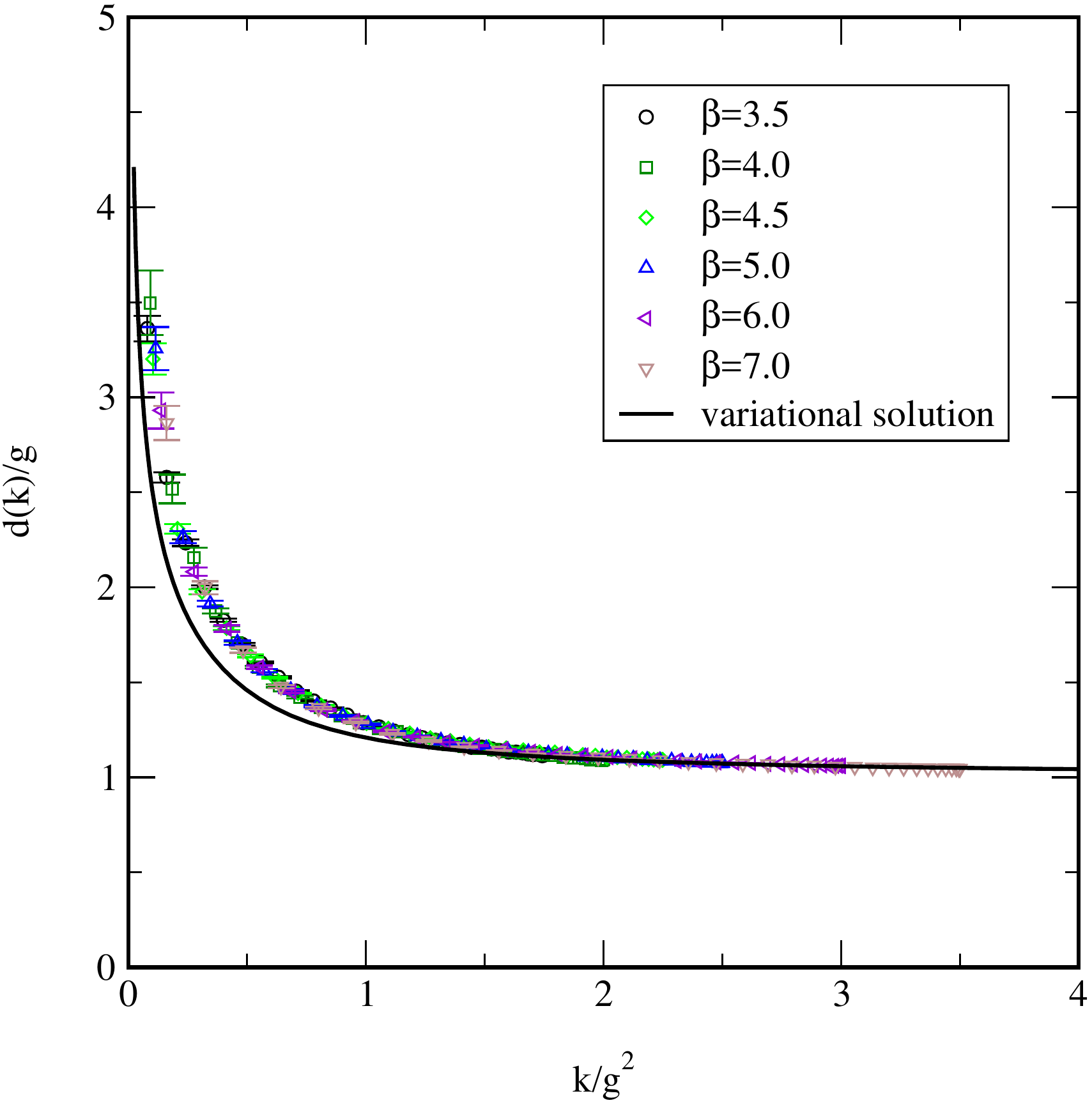}\\(b)}
\caption{Comparison between the lattice data \cite{MoyaertsPhD} and the variational solution (black line) in $2+1$ dimensions for the gluon energy (a) and the ghost form factor (b). From \cite{Feuchter:2007mq}.}
\label{fig:part_iii:comparison}
\end{figure}
\begin{figure}[t]
\includegraphics[width=0.49\linewidth]{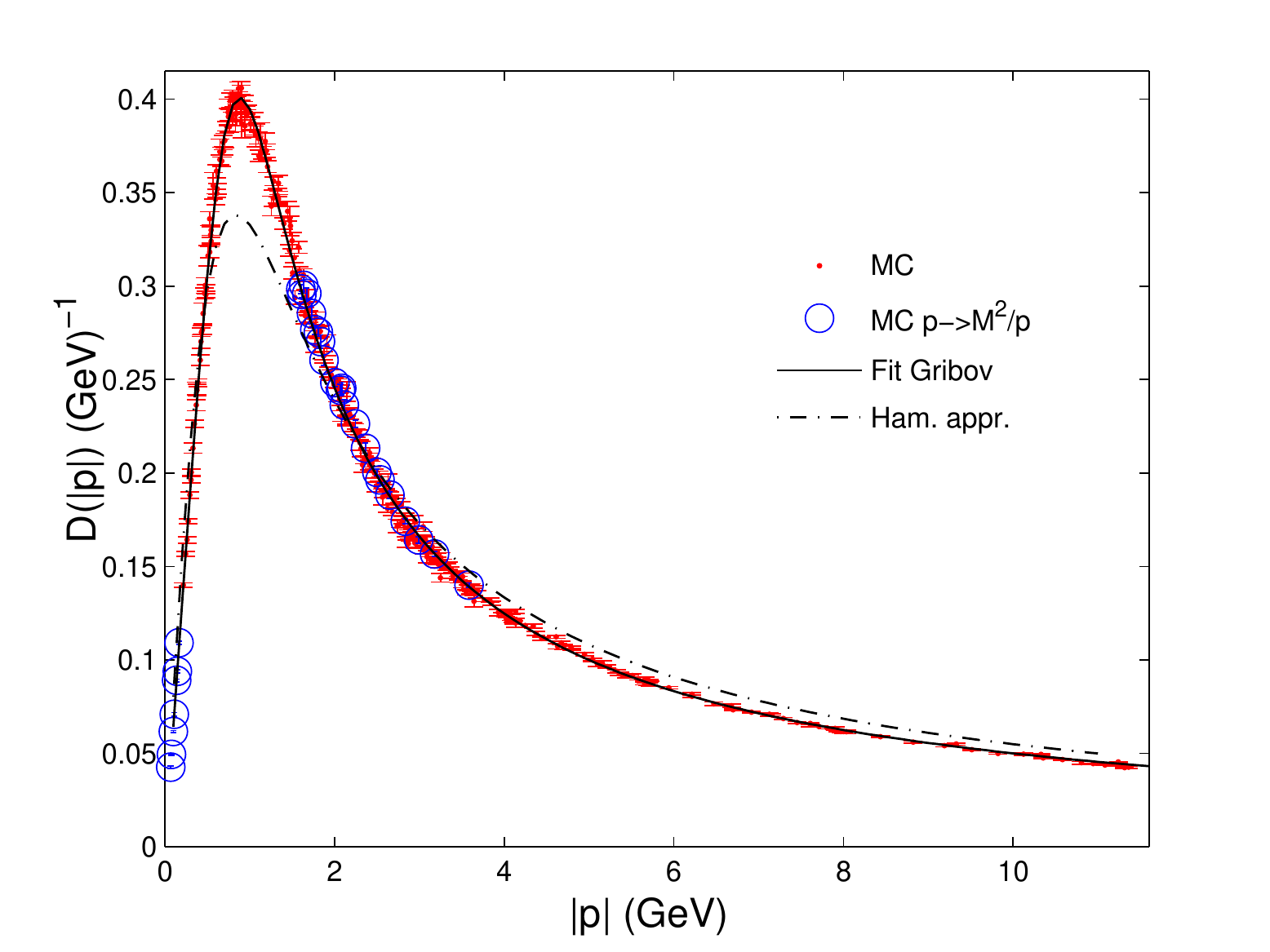}
\includegraphics[width=0.49\linewidth]{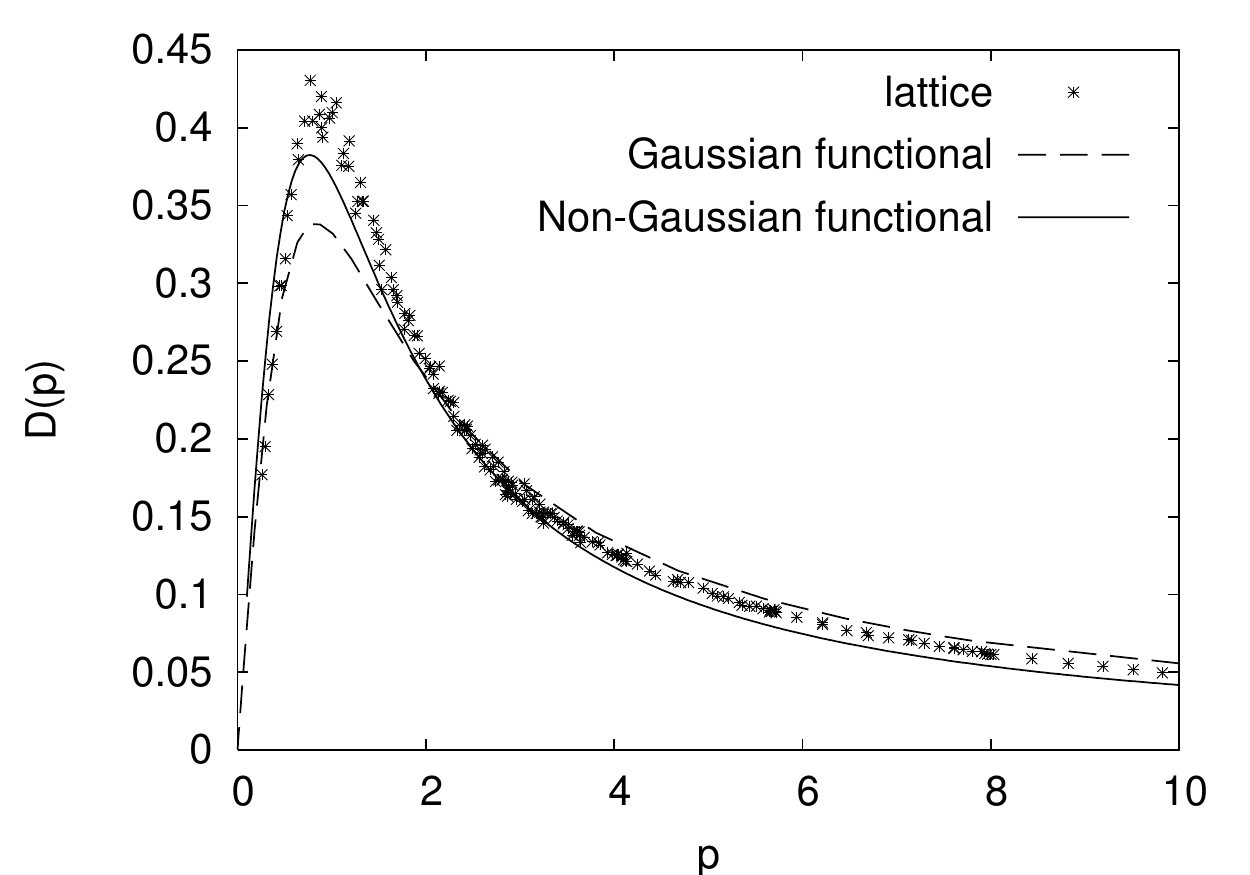}
\caption{Left panel -- The static gluon propagator in $D=3+1$ dimensions calculated on the lattice (red points) and obtained from the Hamiltonian approach (dot-dashed line) with the Gaussian type wave functional (\ref{eq:part_iii:trial_ansatz}). The solid line shows the
	 fit to Gribov's formula (\ref{eq:part_iii:gribov_fromula}). From  \cite{Burgio:2008jr}. Right panel -- the same lattice data compared to the gluon propagator obtained with 
	 Gaussian (dashed line) and non-Gaussian (solid line) wave functionals. From  \cite{Campagnari:2010wc}.}
\label{fig:part_iii:comparison_4d}
\end{figure}%
where $d$ is the number of spatial dimensions. For $d=3$ one finds from the variational equations two solutions for the IR exponents: either $\alpha=\beta=1$ or 
$\alpha=0.6$ and $\beta=0.8$ of which only the first one is physical. 
To find the whole momentum dependence of the gluon energy and the ghost form factor one has to solve the variational equations numerically. The corresponding results are shown in Fig.~\ref{fig:part_iii:variational_results}.

In the Coulomb gauge the ghost form factor has a physical meaning \cite{Reinhardt:2008ek}. Its inverse gives the dielectric function of the Yang--Mills vacuum, $\varepsilon(p)=d^{-1}(p)$. This function calculated in the variational approach of Ref.~\cite{Feuchter:2004mk} is shown in Fig.~\ref{fig:part_iii:dielectric_function}. The horizon condition $d^{- 1} (0) = 0$ implies that the dielectric function vanishes in the infrared regime. This means that there 
are no free color charges since
for $\varepsilon=0$ the electrical displacement $\vD = \varepsilon \vec{E}$ vanishes and from Gauss' law, $\vec{\partial} \cdot \vD = \rho_{free}$, follows that the density of free (color) charges has to vanish. A medium with vanishing dielectric constant is a perfect dielectric or a dual superconductor. In this way the Hamiltonian approach to Yang--Mills theory in the Coulomb gauge establishes the connection to the dual superconductor picture of confinement, discussed in the first lecture.

\subsection{Comparison with the lattice}

To check the quality of the variational approach let us confront the 
obtained propagators with the lattice data. We will confine ourselves to the gauge 
group SU(2). Figure~\ref{fig:part_iii:comparison} shows a comparison between lattice data and the variational solution for the gluon energy (a) and the ghost form factor (b) in $2+1$ dimensions. The variational solution correctly reproduces both infrared and ultraviolet behaviors, 
but somewhat 
deviates from the lattice data in the intermediate momentum regime. The left panel of Fig.~\ref{fig:part_iii:comparison_4d} shows the static gluon propagator $D(p)=(2\omega(p))^{-1}$ in $3+1$ dimensions, obtained on the lattice (points) and using the variational approach (dot-dashed line). The solid line shows a fit to Gribov's formula \cite{Gribov:1977wm} for the gluon energy,
\begin{equation}
\omega(p)=\sqrt{p^2+\frac{M^2}{p^2}},
\label{eq:part_iii:gribov_fromula}
\end{equation}
with $M=0.88$ GeV. Similarly to the $2+1$ dimensional case, the variational approach reproduces correctly high and low momentum regimes and deviates from the lattice results at intermediate momenta. The deviations from the lattice data in the mid-momentum regime can be largely removed
by using a non-Gaussian \emph{ansatz} for the vacuum wave functional \cite{Campagnari:2010wc} (see also Ref.~\cite{Campagnari:2015zsa}),
which reproduces the lattice data much better, see fig. \ref{fig:part_iii:comparison_4d}

In the original lattice calculations of the ghost form factor in $D = 3 + 1$ \cite{Burgio:2012bk}
an infrared exponent of $\beta\approx 0.5$ was found 
while the IR exponent of the gluon energy was determined as
  $\alpha\approx 1$ \cite{Burgio:2008jr}. 
  This result is surprising since it violates the sum rule (\ref{eq:part_iii:sum_rule}). It turns out that the lattice results for the ghost form factor depend on the way the Coulomb gauge is implemented. An alternative method of the gauge fixing leads to
  results compatible with
   $\beta\approx 1$ \cite{Burgio:2016nad}. The gluon propagator, on the other hand, ems to beems to be insensitive to the choice of the gauge fixing method.
   
\begin{figure}[t]
\sidecaption[t]
\includegraphics[width=0.5\linewidth]{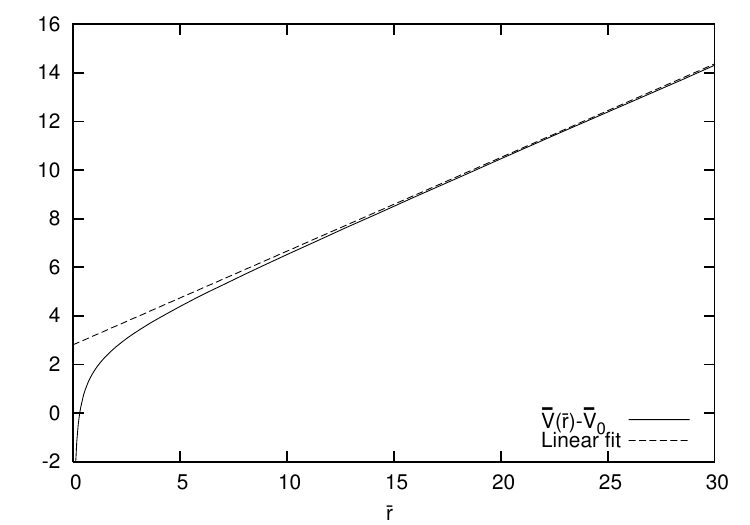}
\caption{The static color charge potential obtained from Eq.~\eqref{eq:part_iii:coulomb_potential}. From  \cite{Epple:2006hv}.}
\label{fig:part_iii:static_charge_potential}
\end{figure}%

\subsection{The non-Abelian Coulomb potential}

So far we have not discussed the Coulomb term, Eq.~ \eqref{eq:part_iii:coulomb_term}, of the gauge fixed Yang--Mills Hamiltonian, Eq.~(\ref{eq:part_iii:YM_hamiltonian}). 
In the presence of matter fields, this term contains a part which is quadratic in the color charge $\rho_m$ -- it represents a two body interaction induced by the Yang--Mills vacuum. The vacuum expectation value of this part, 
\begin{equation}
V_C (\vert\vx- \vy\vert) = g^2 \Big\langle \langle \vx | ( - \hat{\vD} \cdot \vec{\partial})^{- 1} (- \varDelta) (- \hat{\vD} 
\cdot \vec{\partial})^{- 1} | \vy \rangle \Big\rangle,
\label{eq:part_iii:coulomb_potential}
\end{equation}
is the static color charge potential. To simplify its evaluation we use 
the following factorization
\begin{equation}
V_C (\vert\vx- \vy\vert)  \approx \int \D^3 w \int \D^3 z \, G (\vx, \vec{w})\, \langle \vec{w} | - \varDelta | \vec{z} \rangle 
\,G (\vec{z}, \vy) \, ,
\label{eq:part_iii:coulomb_potantial_factorized}
\end{equation}
where $G(\vx,\vy)$ is the ghost propagator, given by Eq.~$($\ref{eq:part_iii:ghost_propagator}$)$. 
Using for $G$ the result of the variational approach one finds the potential
 is shown in Fig.~\ref{fig:part_iii:static_charge_potential}. For small distances it behaves like the ordinary  (QED) 
 Coulomb potential, $V_C(r)\sim 1/r$, 
 while for large distances it rises linearly, $V_C(r)\sim \sigma_Cr$, where $\sigma_C$ is the so-called Coulomb string tension, 
 which can be shown rigorously to represent an upper bound for the Wilson
 string tension, $\sigma_W<\sigma_C$ \cite{Zwanziger:2002sh}.
 On the lattice one finds that $\sigma_C$ is about $2 \ldots 3$ times the Wilson 
 string tension $\sigma_W$. 
 Fourier transforming the potential (\ref{eq:part_iii:coulomb_potantial_factorized})
\begin{equation}
V_C (\vert\vx- \vy\vert) = \int \frac{\D^3p}{(2\pi)^3} \, \E^{\I\vp\cdot(\vx- \vy)} V_C(p),
\label{eq:part_iii:coulomb_potantial_transform}
\end{equation}
\begin{figure}[b]
\includegraphics[width=0.49\linewidth]{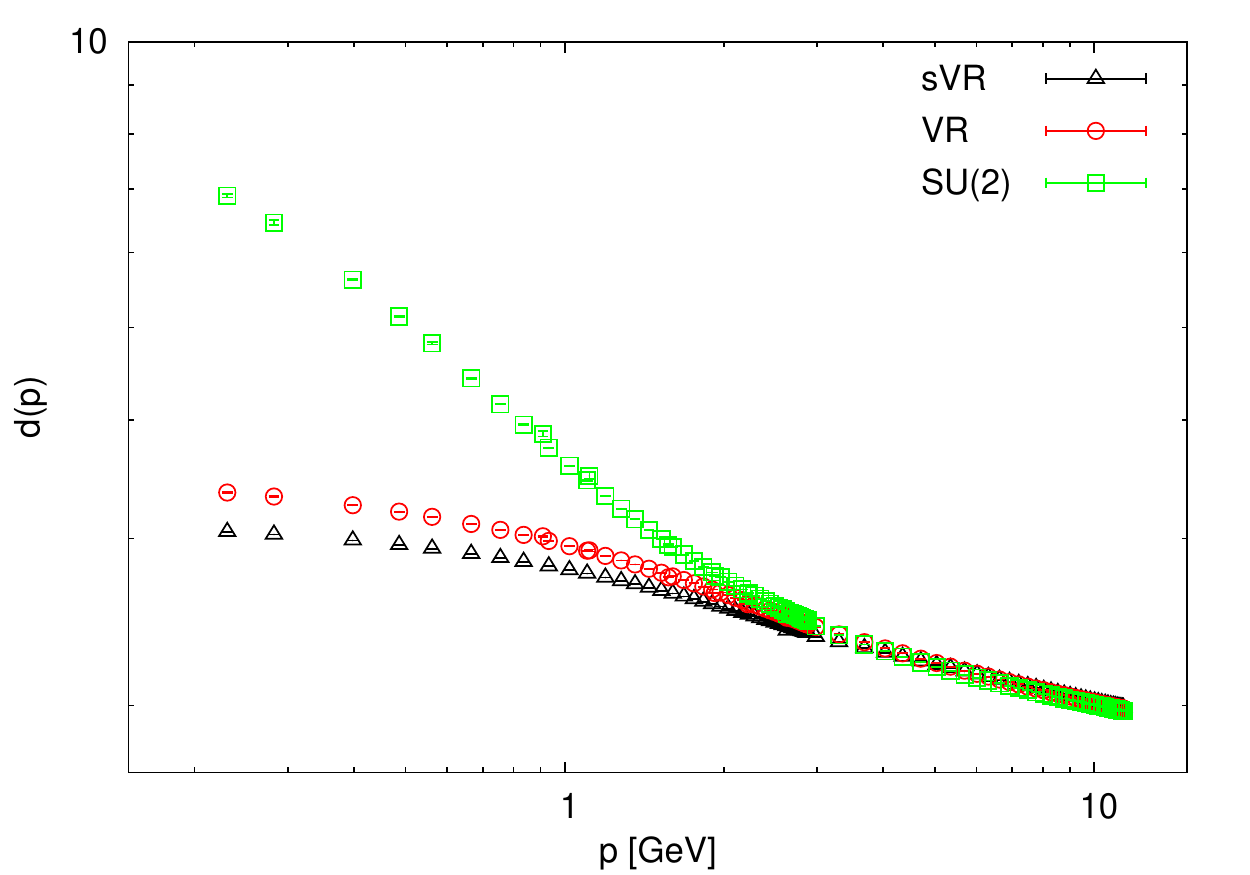}
\includegraphics[width=0.49\linewidth]{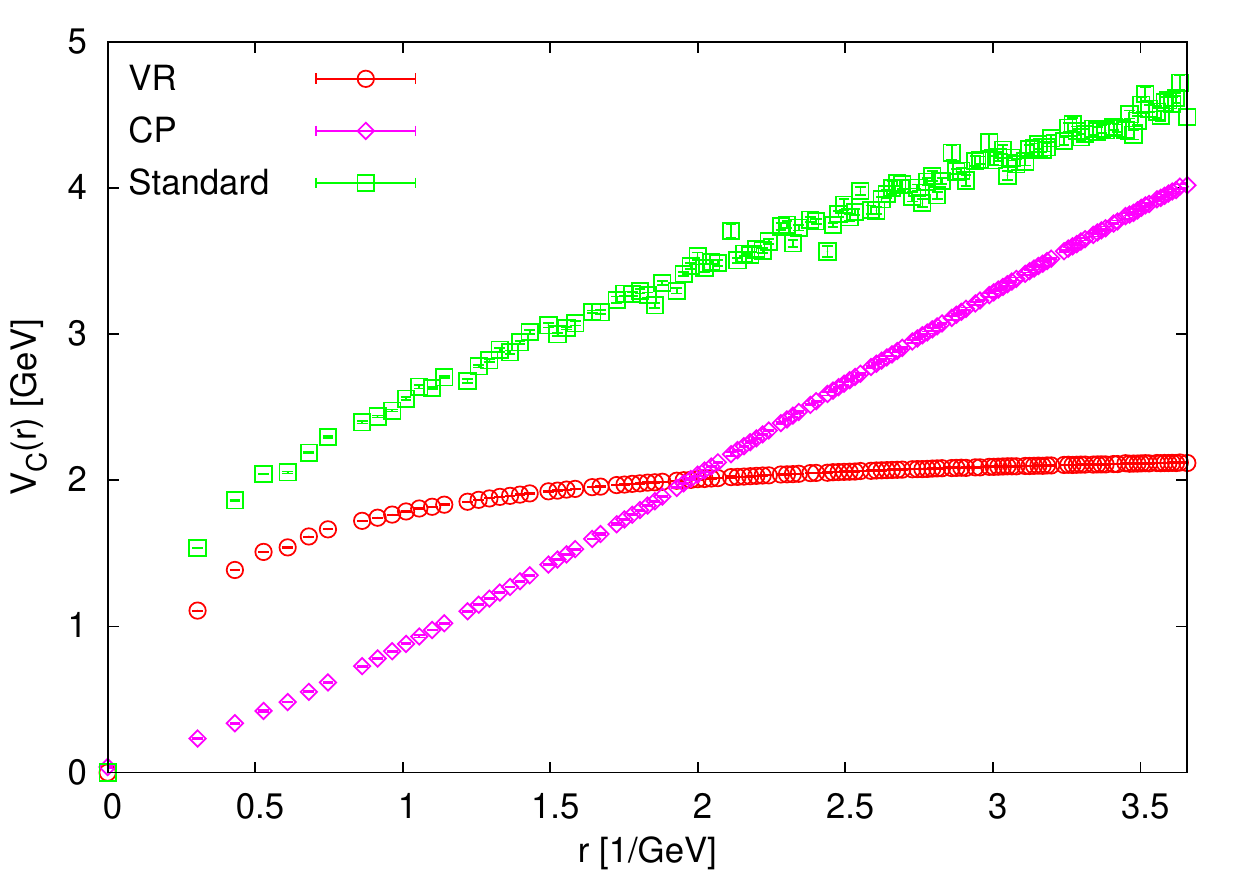}
\caption{ Left panel: the renormalized ghost form factor obtained on the lattice. Green points -- the full result. Red and black points -- the form factor after removal of center vortices (prefix s stands for spatial). Right panel: lattice calculations of the Coulomb potential. Green points -- the full result. Red points -- the potential after removal of center vortices. Purple points -- the potential obtained after the center projection. Both figures from  \cite{Burgio:2015hsa}.}
\label{fig:part_iii:cv_contribution}
\end{figure}%
one finds $V_C(p)= (d(p))^2/p^2$. A linearly rising potential $V_C(p)\sim 1/p^4$ 
requires $d(p) \sim 1/p$, and hence an infrared exponent $\beta=1$, which is obtained for one of our two variational solutions. Such 
a ghost form factor obviously satisfies the horizon condition $d^{-1}(0)=0$.

\begin{figure}[t]
\includegraphics[angle=270,width=0.49\linewidth,clip]{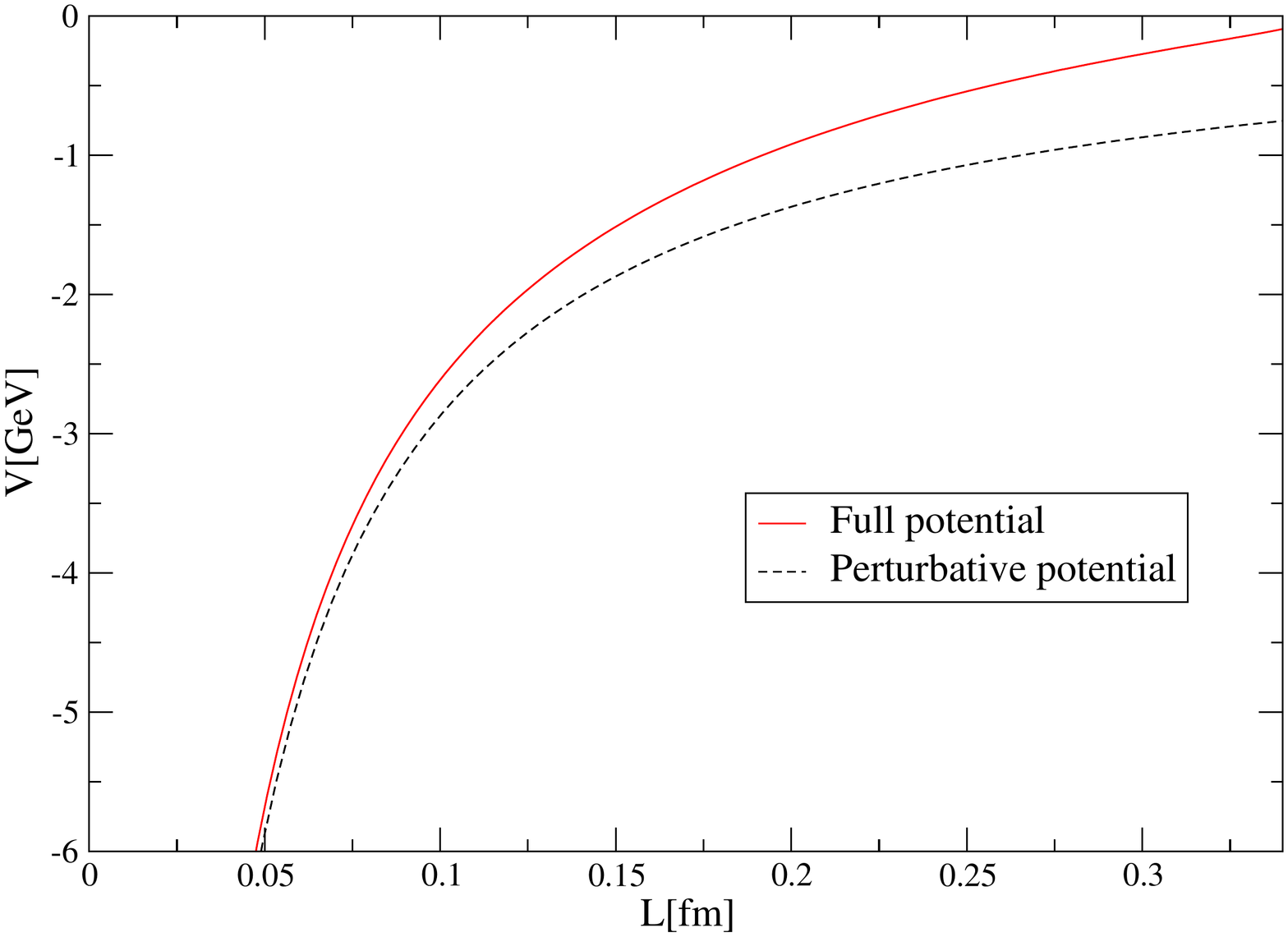}
\includegraphics[angle=270,width=0.49\linewidth,clip]{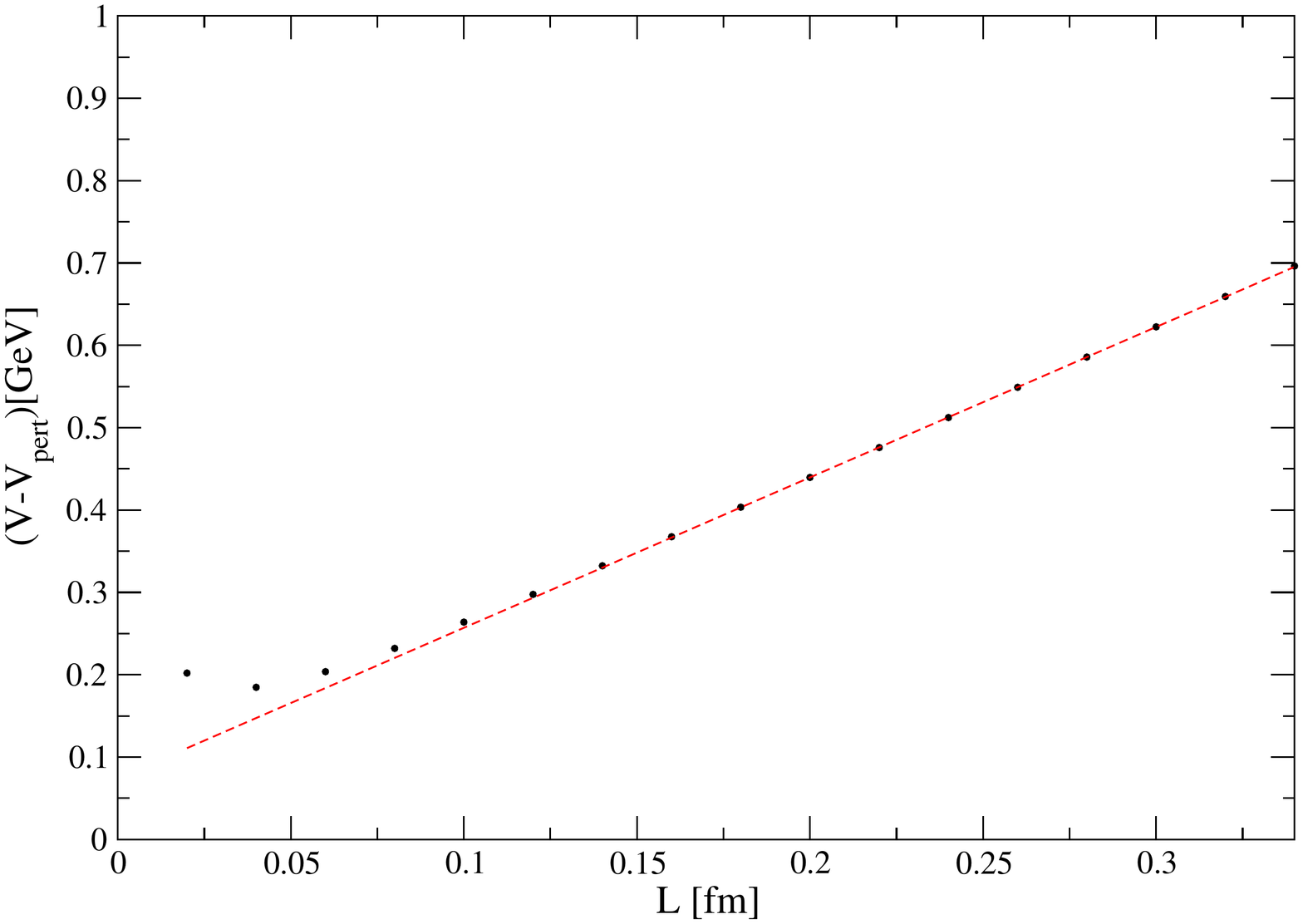}
\caption{(left) The Coulomb potential obtained from the Wilson loop using the Dyson--Schwinger equation. (right) The potential after subtraction of the perturbative part. From  \cite{Pak:2009em}.}
\label{fig:part_iii:potential_from_dse}       
\end{figure}%

As we have seen, the horizon condition is crucial for confinement. An interesting question is therefore: what are the field configurations which trigger the horizon condition? An infrared singular ghost form factor 
arises from field configurations which are on, or near, the Gribov horizon. 
Center vortices and magnetic monopoles can be shown to lie exactly on the 
 Gribov horizon \cite{GreOleZwa05}. Using the lattice methods  presented in the second lecture 
 one can calculate the 
 contributions of center vortices to the ghost form factor  \cite{Burgio:2015hsa}. 
 Figure \ref{fig:part_iii:cv_contribution} shows the ghost form factor obtained in the full lattice gauge theory together with the one obtained when the center vortices are removed from the gauge field ensemble.
  After removal of center vortices the ghost form
  factor is no longer divergent at $\vp\rightarrow 0$ and the 
  confining properties of the theory are lost. A similar result holds for the Coulomb potential (the right panel of Fig.~\ref{fig:part_iii:cv_contribution}) -- after 
removal of center vortices the Coulomb string tension vanishes and the potential is no longer confining. 
It was also shown in ref. \cite{Burgio:2015hsa} that the Coulomb string tension $\sigma_C$ is not 
related to the temporal $\sigma_{W_t}$ but to the spatial Wilsonian string tension $\sigma_{W_s}$.
This explains also the finite temperature behaviour of the Coulomb string tension,
which does not disappear but slightly increases above the deconfinement phase transition, see ref. \cite{Burgio:2015hsa}.

\subsection{Spatial Wilson and 't Hooft loops}

Since $\sigma_C \geq \sigma_{W_t}$
the confining Coulomb potential is a necessary but not sufficient condition for confinement. 
To really show that the variational
 approach yields confinement, one has to calculate the vacuum expectation value of the Wilson loop, the true
  order parameter of confinement. As discussed in the first lecture 
  in the confined phase this quantity exhibits an area-law falloff, while in the deconfined phase it falls off with the perimeter. The 
  calculation of the Wilson loop is, unfortunately, 
  difficult in a continuum theory due to the path ordering in this operator. In Ref.~\cite{Erickson:1999qv} a Dyson--Schwinger equation for the Wilson loop in supersymmetric Yang--Mills theory has been derived. Although 
  this equation is strictly valid only in supersymmetric Yang--Mills theory
  this equation can be used for an approximate evaluation of the Wilson loop
  in the non-supersymmetric theory as well
   \cite{Pak:2009em}. The static potential extracted from the obtained Wilson loop is shown in Fig.~\ref{fig:part_iii:potential_from_dse}. Unfortunately the method used to 
   solve the DSE for the Wilson loop works only up to
    intermediate distances, where the potential is still strongly affected by the Coulomb-like behavior. After subtraction of the Coulombic part
     one obtains the 
linearly rising potential, as shown in the right panel of Fig.~\ref{fig:part_iii:potential_from_dse}

An alternative order parameter (or rather a disorder parameter) of confinement, which is easier to calculate in a continuum theory is the (spatial) 't Hooft loop \cite{tHooft:1977nqb}. This object is the expectation value of
 the operator $\hat{V}(C)$ defined by the following commutation relation
\begin{equation}
\hat{V} (C_1) W (C_2) = z^{L (C_1, C_2)} W (C_2) \hat{V} (C_1),
\label{eq:part_iii:thooft_loop}
\end{equation}
where $C_1$ and $C_2$ are closed curves in $\RR^3$, $W (C_2)$ is the Wilson loop, $z$ is a non-trivial center element of the gauge group and $L (C_1, C_2)$ 
is the Gauss linking number (\ref{proc_765_G21}). As discussed in the first lecture
the 't Hooft loop $- \ln \langle \hat{V} (C) \rangle$ measures the electric flux through a surface enclosed by $C$. It
 exhibits an area law falloff in the deconfined phase and a perimeter law in 
 the confined phase, see Eq.~(\ref{34716}). 
In Ref.~\cite{Reinhardt:2002mb} the following continuum representation of the 
't Hooft loop operator was derived
\begin{equation}
\hat{V} (C) = \exp \biggl[\I \int_{\mathbb{R}^3} \mathcal{A} (C) \Pi \biggr],
\label{eq:part_iii:thooft_loop_cont}
\end{equation}
where $\Pi=\delta/(i\delta A)$ is the momentum operator of the gauge field and $\mathcal{A} (C)$ is the gauge potential of a thin center vortex located at the loop $C$. 
A realization of such a gauge potential is given by
\begin{equation}
\label{1358-57}
{\cal{A}} (C) (\vx) = 2 \pi \mu \il_{\vec{\Sigma}  (C)} \dd \vec{\Sigma}(\vx') \, \delta (\vx - \vx') ,
\end{equation}
where $\Sigma  (C)$ is a surface with boundary $C$ and $\mu$ is a co-weight, which satisfies
\begin{equation}
\label{1364-f57-a}
\ee^{i 2 \pi \mu} = z .
\end{equation}
With the gauge potential (\ref{1358-57}) the 't Hooft loop operator 
(\ref{eq:part_iii:thooft_loop_cont}) becomes
\begin{equation}
\label{1370-58-X2}
\hat{V} (C) = \exp \biggl[\ii2 \pi \mu \int_{\Sigma (C)} \dd \vec{\Sigma}(\vx') \cdot \vec{\Pi} (\vx') \biggr] .
\end{equation}
Since $\vec{\Pi} (\vx)$ is the operator of the electric field, see Eq.~(\ref{eq:part_iii:yang_mills_action}), this representation (\ref{1370-58-X2})
shows that $\hat{V} (C)$ measures indeed the electric flux through $C$. 

When the 't Hooft loop operator acts on the wave functional, it displaces 
 its argument by the center vortex field $\mathcal{A}(C)$:
\begin{equation}
\hat{V} (C) \Psi [A] = \Psi [A + \mathcal{A} (C)] .
\label{eq:part_iii:thooft_loop_action}
\end{equation}
This is obvious if one notices that $\Pi = \delta /\ii\delta A$ is the momentum operator
and recalls a similar relation from the usual quantum mechanics,
$\exp(\ii a\hat{p})\psi(x)=\psi(x+a)$. The 't Hooft loop  has been calculated within the
variational approach in Ref.~\cite{Reinhardt:2007wh} at zero temperature and a perimeter law was found.

\subsection{Hamiltonian approach to QCD in Coulomb gauge}\label{sect4.4}
The Hamiltonian approach to pure Yang--Mills theory presented above can be extended to full QCD. The QCD Hamiltonian in Coulomb gauge is
\begin{equation}
H_\mathrm{QCD}=H_\mathrm{YM}+H_\mathrm{C}+H_{q},
\label{eq:part_iii:qcd_hamiltonian}
\end{equation}
where $H_\mathrm{YM}$ and $H_\mathrm{C}$ are the Yang--Mills Hamiltonian, Eq.~$($\ref{eq:part_iii:YM_hamiltonian}$)$, and the Coulomb term, Eq.~$($\ref{eq:part_iii:coulomb_term}$)$, respectively. The latter
contains now also the color charge density of the quarks
\begin{equation}
\rho_m^a(\vx)=\psi^\dag(\vx)t^a\psi(\vx),
\label{eq:part_iii:quark_charge_density}
\end{equation}
where $\psi(\vx)$ is the quark field and $t^a$ is the $a$th generator of the gauge group in the fundamental representation. The last term of Eq.~$($\ref{eq:part_iii:qcd_hamiltonian}$)$ is the Dirac Hamiltonian of quarks coupled to the spatial gauge field 
\begin{equation}
H_q=\int \D^3x \, \psi^\dag(\vx)\left[\vec{\alpha}(\vp+g\vA)+\beta m_0\right]\psi(\vx),
\label{eq:part_iii:dirac_hamiltonian}
\end{equation}
where $\vec{\alpha}$ and $\beta$ are Dirac matrices.

For the vacuum wave functional of full QCD the following variational \emph{ansatz} 
was used \cite{Vastag:2015qjd} 
\begin{equation}
\label{1409-57}
| \Phi (A) \rangle_\mathrm{QCD} \sim \Psi [A] | \phi [A] \rangle_q  ,
\end{equation}
where $\Psi [A]$ is the wave functional (\ref{eq:part_iii:coulomb_term}) of the Yang--Mills sector and
\begin{equation}
\lvert \phi [A] \rangle_q = \exp \left[ \int \psi^\dagger_+ (s \beta + v \vec{\alpha} \cdot \vA + w 
\beta \vec{\alpha} \cdot \vA ) \psi_- \right] \lvert 0 \rangle_q \, 
\label{eq:part_iii:quark_wave_functional}
\end{equation}
is the wave functional of the quarks coupled to the gauge field. 
Here $\psi^\dagger_+$ and $\psi^\dagger_-$ are the positive and negative energy components of the quark field, respectively, and $s$, $v$, $w$ are variational kernels.
When $v$ and $w$ are set to zero, the wave functional $($\ref{eq:part_iii:quark_wave_functional}$)$ takes 
a form reminiscent of the Bardeen--Cooper--Schrieffer (BCS) wave function
of superconductivity. A wave functional of this type was used in Refs.~\cite{Finger:1981gm, Adler:1984ri,Alkofer:1988tc}.

When one varies 
the vacuum expectation value $\langle \Phi | H_{QCD} | \Phi \rangle$  in the state (\ref{1409-57}), 
$($\ref{eq:part_iii:quark_wave_functional}$)$ with respect to the variational kernels, one finds 
four coupled equations for $s, v$ and $w$, and the gluon energy $\omega$. The equations for $v$ and $w$ can be explicitly solved
 in terms of the scalar kernel $s$ and the gluon energy $\omega$,
\begin{equation}
w(p,q)=f_w[s,\omega]\qquad v(p,q)=f_v[s,\omega],
\label{eq:part_iii:v_w_functions}
\end{equation}
while for
the scalar kernel $s$ and the gluon energy $\omega$ one finds non-linear integral equations referred to as gap equations. 
 The gluon gap equation (\ref{eq:part_iii:gap_equation_general}) of the pure Yang-Mills sector is then modified by 
 additional quark loop terms, see ref. \cite{Campagnari:2016wlt}, while the equation for the scalar kernel $s$ 
has the form
\begin{equation}
s(p)=f_s[s,v,w;p].
\label{eq:part_iii:gap_equation_s}
\end{equation}
\begin{figure}[t]
\includegraphics[width=0.45\linewidth,clip]{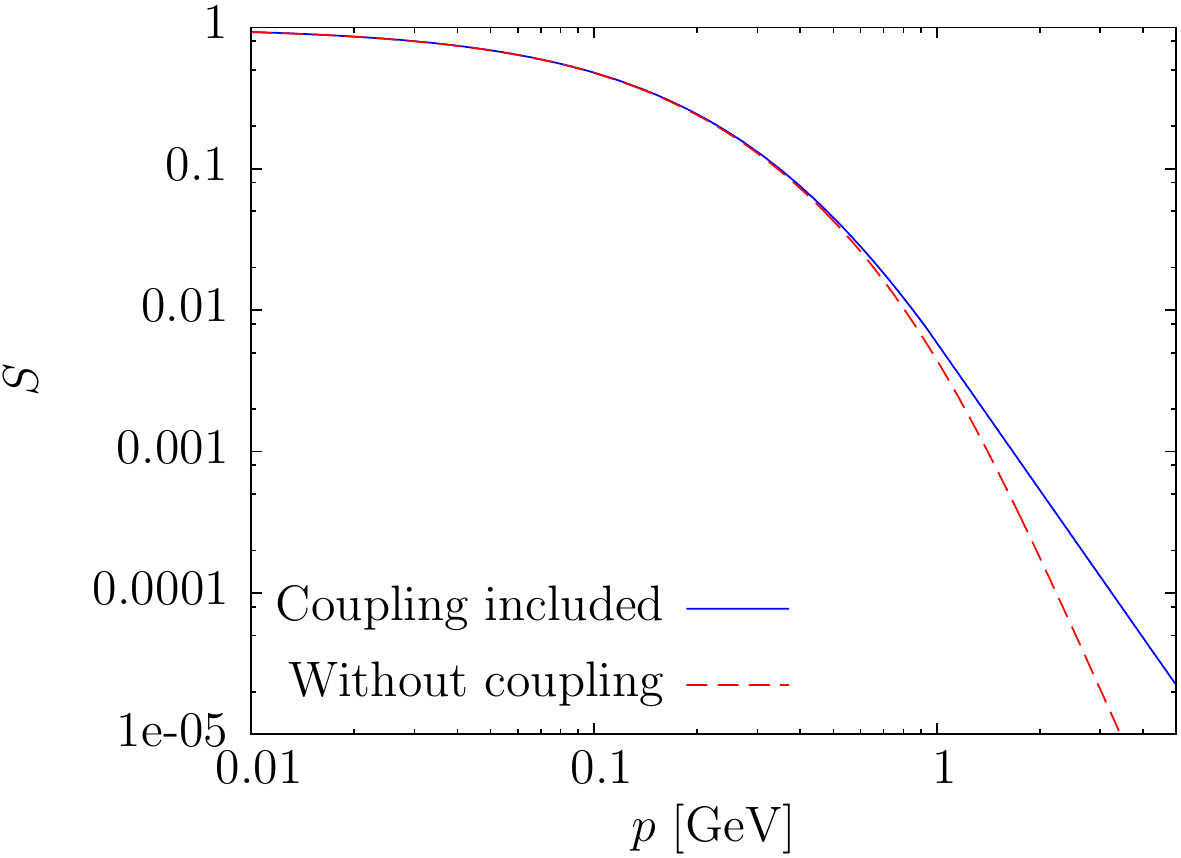}\hfill
\includegraphics[width=0.45\linewidth,clip]{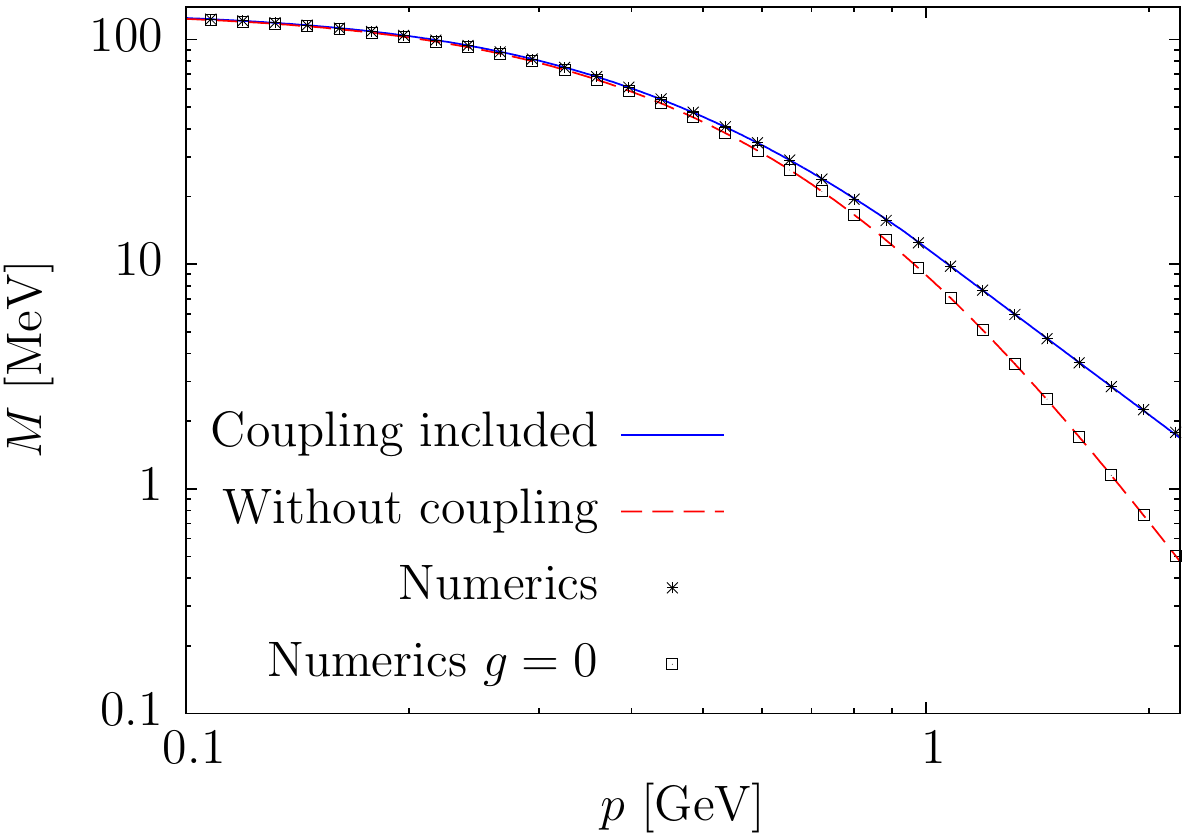}
\caption{Scalar kernel (left panel) and the effective quark mass (right panel), obtained in the variational approach. Red line -- calculation without quark-gluon coupling. Blue line -- calculation with included coupling. From \cite{Campagnari:2016wlt}.}
\label{fig:part_iii:hamiltonian_qcd_plots}       
\end{figure}%
An important advantage of the quark wave functional $($\ref{eq:part_iii:quark_wave_functional}$)$ is that, as opposed to previous \emph{ansatz}es, all the UV divergences in the gap equation $($\ref{eq:part_iii:gap_equation_s}$)$ strictly cancel.
The gap equation (\ref{eq:part_iii:gap_equation_s}) was numerically solved
in Ref.~\cite{Campagnari:2016wlt}  using the Gribov formula 
(\ref{eq:part_iii:gap_equation_pure_glue}) as input for the gluon energy. 
The obtained scalar form factor $s (\rho)$ and the effective quark mass 
\begin{equation}
M(p)=\frac{2ps(p)}{1-s^2(p)},
\label{eq:part_iii:effective_mass}
\end{equation}
are shown in Fig.~\ref{fig:part_iii:hamiltonian_qcd_plots}. 
The red curve corresponds to a calculation without the
 quark-gluon coupling, i.e. with a BCS type wave functional, and the blue line shows the result obtained with this
  coupling included. 
 The inclusion of the coupling of the quarks to the transverse gauge field 
 influences the UV behavior but does not significantly change the IR 
 behavior. 
 This is not surprising, because the IR behavior is dominated by the Coulomb term. There is, however, an increase of $S (p)$ and $M (p)$ in the UV which increases the quark condensate from $(- 185\,\mathrm{MeV})^3$ towards the phenomenological  value of $(- 235\,\mathrm{MeV})^3$. Let us also stress that 
 without the Coulomb term there is no spontaneous breaking of chiral symmetry as the numerical calculations show \cite{Campagnari:2016wlt}. 
 
\subsection{Alternative Hamiltonian approach to finite temperature QFT}\label{section4.5}
In Refs.~\cite{Reinhardt:2011hq, Heffner:2012sx} the variational approach to Yang--Mills theory in Coulomb gauge was extended to finite temperatures by making a quasi-particle \emph{ansatz} 
for the density operator $\exp (- L H)$
of the grand canonical ensemble where the quasi-particle energy was determined by minimizing the free energy. The resulting variational equations could be solved analogously to the ones at zero temperature. There is, however, a more efficient way to treat Yang--Mills theory at finite temperature within the Hamiltonian approach.
The motivation comes from the Polyakov loop (\ref{334-13b})
\beq
P[A_4](\vx) = \frac{1}{d_r} \tr \, \mathcal{P} \exp\left[\ii g \int_0^{L} \dd x^4 \, A_4 (x^4, \vx) \right] , \label{Gl: Polyakovloop}
\eeq
where $A_4 = A_4^a t^a$ is the temporal gauge field in the fundamental representation, $\mathcal{P}$ is the path ordering prescription 
and $L = 1/T$ is the length of the compactified Euclidean time axis,
which represents the inverse temperature. 
Furthermore $d_r$  is the dimension of the representation, which serves as normalization factor.
The Polyakov loop cannot be calculated straightforwardly in the Hamiltonian approach due to the unrestricted time interval and the
use of the Weyl gauge $A_0 = - i A_4 = 0$. Both problems are overcome in the more efficient Hamiltonian approach to finite-temperature 
quantum field theory developed in Ref.~\cite{Reinhardt:2016xci}. This novel approach does not require an \emph{ansatz} for the density operator
of the grand canonical ensemble and allows the evaluation of the Polyakov loop. In this approach, one exploits the $O(4)$ invariance to interchange the Euclidean time axis with one spatial axis. The temporal (anti-)periodic boundary conditions to the fields become then spatial boundary conditions, while the 
new (Euclidean) time axis has infinite extent as is required within the Hamiltonian approach (see below). The upshot is that the partition function at finite temperature $L^{- 1}$ is entirely 
given by the ground state calculated on the spatial manifold $\R^2 \times S^1(L)$, where $S^1(L)$ is a circle with length $L$. The whole thermodynamics of the theory is then encoded in the vacuum calculated on the partially compactified spatial manifold $\R^2 \times S^1(L)$. 
This approach was used to study Yang--Mills theory at finite temperature \cite{Heffner:2015zna},
to calculate the Polyakov loop within the Hamiltonian approach \cite{Reinhardt:2013iia} and to evaluate
the so-called dual quark condensate \cite{Reinhardt:2016pfe}. Let us briefly sketch its main features.

\subsubsection{Finite temperature from the compactification of a spatial dimension}

Consider finite-temperature quantum field theory in the standard functional integral approach. Here the finite temperature is 
introduced by going to Euclidean space and compactifying the Euclidean time dimension by imposing periodic and antiperiodic boundary 
conditions for Bose $A$ and Fermi $\psi$ fields, respectively,
\begin{subequations} \label{656-32}
\begin{align}
A(x^4 = L/2) &= A(x^4 = -L/2) \, , \\
\psi(x^4 = L/2) &= -\psi(x^4 = - L/2) \, .
\end{align}
\end{subequations}
The length of the compactified dimension $L$ represents then the inverse temperature $T^{-1} = L$. One can now exploit the $O(4)$ invariance of the Euclidean Lagrangian 
to rotate the Euclidean time axis $x^4$ into a spatial
axis and, correspondingly, one spatial axis into the Euclidean time axis. Of course, thereby all vector quantities transform in the 
same way. For example, we can choose the transformation:
\begin{alignat}{5}
\label{G43}
& x^4 &\to x^3 \, , &\quad \quad A^4 &\to A^3 \, , &\quad \quad \gamma^4 &\to \gamma^3 \, , \nonumber\\
& x^3 &\to x^2 \, , &\quad \quad A^3 &\to A^2 \, , &\quad \quad \gamma^3 &\to \gamma^2 \, , \nonumber\\
& x^2 &\to x^1 \, , &\quad \quad A^2 &\to A^1 \, , &\quad \quad \gamma^2 &\to \gamma^1 \, , \nonumber\\
& x^1 &\to x^4 \, , &\quad \quad A^1 &\to A^4 \, , &\quad \quad \gamma^1 &\to \gamma^4 \, .
\end{alignat}
After this rotation we are left with the spatial  periodic and antiperiodic boundary conditions
\begin{subequations} \label{G44}
\begin{align}
A(x^3 = L/2) &= A (x^3 = - L/2) \, , \nonumber\\
\psi(x^3 = L/2) &= - \psi (x^3 = - L/2) \, .
\end{align}
\end{subequations}
As a consequence of the $O(4)$ rotation our spatial manifold is now $\R^2 \times S^1(L)$ instead of $\R^3$ while the temporal manifold is $\R$ independent of the temperature, i.e.~the temperature is now encoded in one spatial dimension while time has infinite extension. We can now apply the usual canonical Hamiltonian approach to this rotated space-time manifold. As the new time axis has infinite extension 
$\ell \to \infty$, the partition function is now given by
\beq
Z(L) = \lim\limits_{\ell \to \infty} \mathrm{tr} \exp(- \ell H(L)) \, , \label{G45}
\eeq
where $H(L)$ is the usual Hamiltonian obtained after canonical quantization, however, now defined on the spatial manifold $\R^2 \times S^1(L)$. Taking the trace in the basis of the exact eigenstates of the Hamiltonian $H(L)$, we obtain for the partition function (\ref{G45})
\beq
Z(L) = \lim\limits_{\ell \to \infty} \sli_n \exp (- \ell E_n (L)) = \lim\limits_{\ell \to \infty} \exp (- \ell E_0 (L)) \, . \label{G46}
\eeq
The full partition function is now obtained from the ground state energy $E_0 (L)$ 
calculated on the spatial manifold $\R^2 \times S^1(L)$. Introducing the energy density $e(L)$ on $\R^2 \times S^1(L)$ by separating the volume $L \ell^2$  of the spatial manifold from the energy we have
\beq
E_0(L) = L \ell^2 e(L) \, . \label{G47}
\eeq
For the physical pressure
\beq
P = \frac{1}{L} \frac{\partial \ln Z}{\partial V} \, , \qquad V = \ell^3 \label{949-949}
\eeq
one finds from (\ref{G46})
\beq
P = - \frac{\partial (V e (L))}{\partial V} =  - e(L)  - V \frac{\partial e (L)}{\partial V} \, , \label{G48}
\eeq
while the physical energy density 
\begin{equation}
 \label{pro_1801}
 \varepsilon = \frac{\langle H \rangle}{V} = - \frac{1}{V} \frac{\partial \ln Z}{\partial L} + \frac{\mu}{V} 
 \frac{1}{L} \frac{\partial \ln Z}{\partial \mu}
\end{equation}
is obtained as 
\beq
\varepsilon = \frac{\partial (L e (L))}{\partial L} - \mu \frac{\partial e (L)}{\partial \mu} \, . \label{G49}
\eeq
To distinguish this quantity from the (negative) Casimir pressure $e(L)$ [Eq.~(\ref{G48})], which also appears as an energy density in our formalism after the transformation [Eq.~(\ref{G43})], we will denote $e(L)$ as 
\textit{pseudo-energy density}. 
Note also that the last term in eq. (\ref{G48}) vanishes for generic non-interacting systems such that $P = - e (L)$.

Finally, after the $O(4)$ rotation, Eq.~(\ref{G43}), the finite chemical potential $\mu$ enters the single-particle Dirac Hamiltonian $h$ in the form
\beq
h(\mu) = h(\mu=0) + \ii \mu \alpha^3 \, , \label{G50}
\eeq
where $\alpha^3$ is the third Dirac matrix and $h(\mu=0)$ is the usual single particle Dirac Hamiltonian.

\subsubsection{Free Bose and Fermi gases}\label{4.5.1}

\begin{figure}
\parbox{.48\linewidth}{\centering\includegraphics[width=\linewidth]{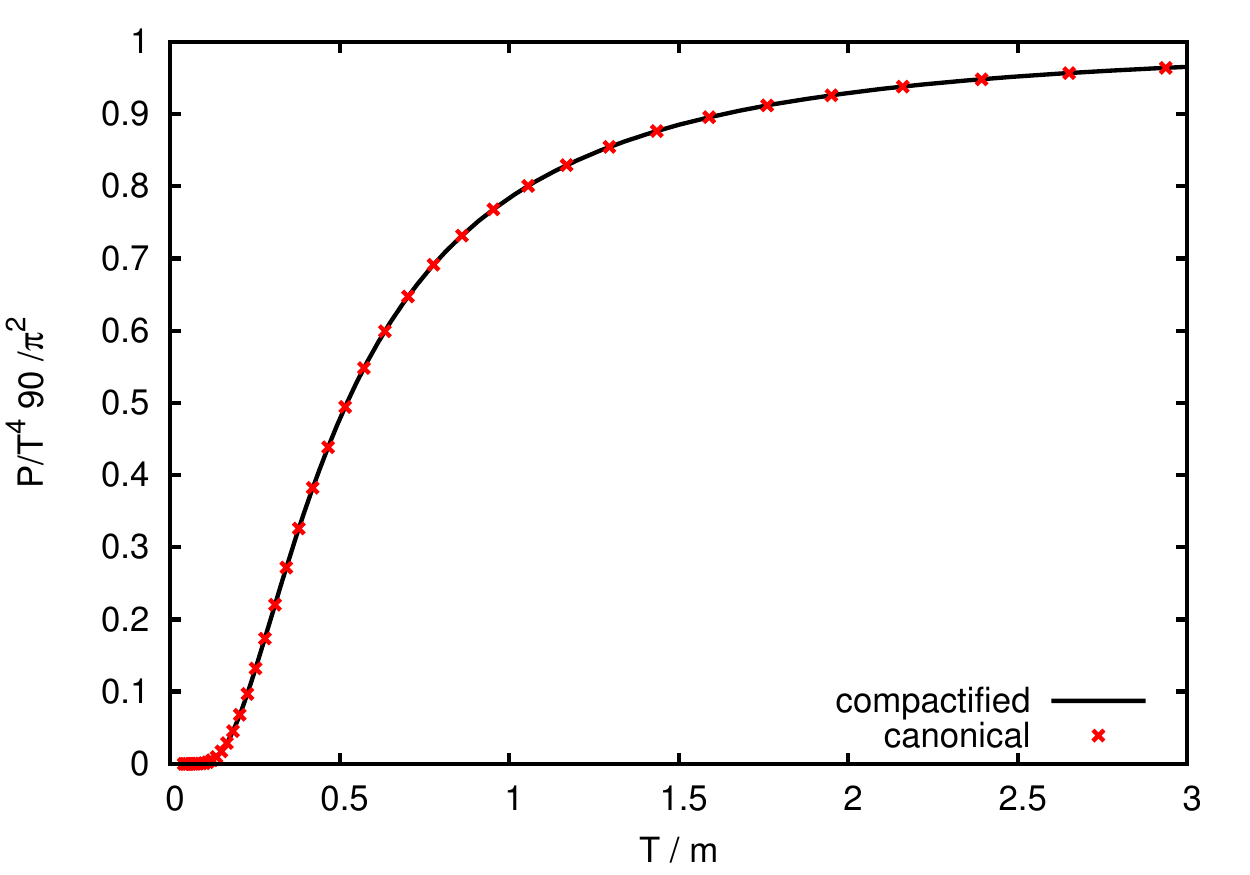}\\(a)}
\hfill
\parbox{.48\linewidth}{\centering\includegraphics[width=\linewidth]{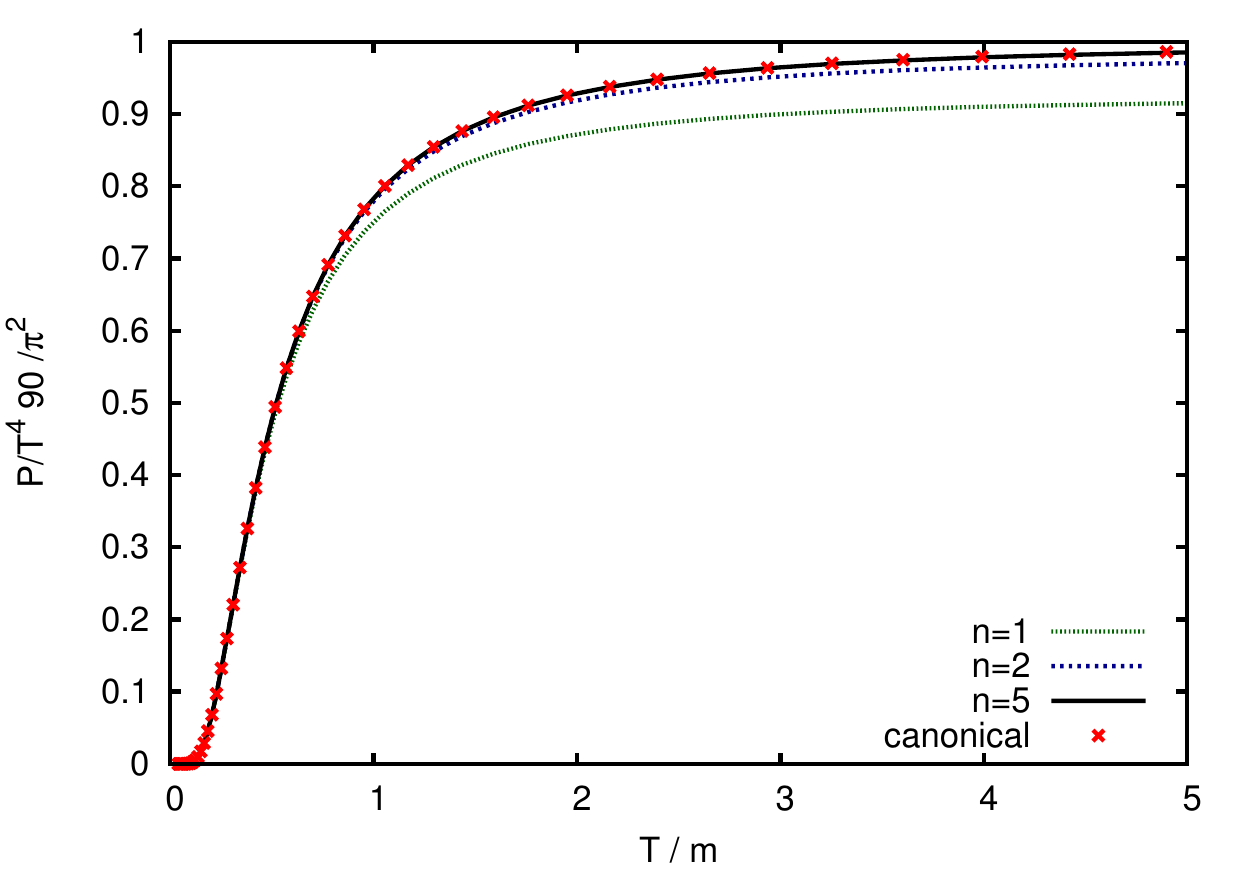}\\(b)}
\caption{The pressure of a free massive Bose gas (a) calculated from Eq.~(\ref{G55}) (full curve) and from the grand canonical ensemble (\ref{G51}) (crosses). (b) The pressure when the summation index in Eq.~(\ref{G55}) is restricted to $|n| = 1, 2$ and $5$.}
\label{fig-10}
\end{figure}

To illustrate the above approach let us first consider a relativistic Bose gas with dispersion relation $\omega(p) = \sqrt{\vp^2 + m^2}$, where we assume for simplicity a vanishing chemical potential. The 
thermodynamical pressure obtained from the grand canonical ensemble for such a system is given by $(L = T^{- 1})$
\beq
P = \frac{2}{3} \int \frac{\dd^3 p}{(2 \pi)^3} \frac{p^2}{\omega (p)} n (p) \, , \quad \quad n (p) = \frac{1}{\exp(L \omega (p)) - 1} \, , \label{G51}
\eeq
where $n(p)$ are the finite temperature Bose occupation numbers. On the other hand,
for the ideal Bose gas with dispersion relation $\omega (p) = \sqrt{\vp^2 + m^2}$ one finds 
the pseudo-energy density on the spatial manifold $\R^2 \times S^1(L)$ \cite{Reinhardt:2016xci}
\beq
e(L) = \frac{1}{2} \int \frac{\dd^2 p_\perp}{(2 \pi)^2} \frac{1}{L} \sli^\infty_{n = - \infty} \sqrt{\vp^2_\perp + p_n^2 + m^2} \, , 
\quad \qquad p_n = \frac{2 n \pi}{L} \, , \label{G52}
\eeq
where $p_n$ are the bosonic Matsubara frequencies. This quantity does not look at all like the negative of the pressure (\ref{G51}), 
as it should by Eq.~(\ref{G48}). In fact, as it stands $e (L)$ (\ref{G52}) 
is ill defined: the integral and the sum are both divergent. To make it mathematically well defined, we first use the proper-time regularization of the square root,
\beq
\sqrt{A} = \frac{1}{\Gamma \lk - \frac{1}{2} \rkx} \lim\limits_{\Lambda \to \infty}\left[\il^\infty_{1/\Lambda^2} \dd \tau \,\tau^{-\frac{1}{2}}\, \exp (- \tau A) - 2 \Lambda + \mathcal{O}(\Lambda^{-1})\right] \, . \label{G53}
\eeq
The divergent constant appears because the limit $\Lambda \to \infty$ of the incomplete $\Gamma$-function $\Gamma \left( - \frac{1}{2}, \Lambda \right)$ is not smooth; it drops out when taking the difference to the zero-temperature case after Eq.~(\ref{G55}) below.  To extract the zero temperature part of eq. (\ref{G52}) for the Matsubara sum we use the Poisson resummation formula,
\beq
\frac{1}{2 \pi} \sli^\infty_{k = -\infty} \exp(\ii k x) = \sli^\infty_{n = - \infty} \delta (x - 2 \pi n) , \label{G54}
\eeq
by means of which one derives the relation
\begin{equation}
\frac{1}{L} \sli^\infty_{n = - \infty} f (p_n) = \frac{1}{2 \pi} \il^\infty_{- \infty} d z f (z) \sli^\infty_{k = - \infty} e^{ikzL} \, .
\label{2080-GX2}
\end{equation}
Inserting this relation into eq. (\ref{G52}) we obtain
\begin{equation}
e (L) = \frac{1}{2} \int \dbar^3 p \sqrt{\vp^2  + m^2} \sli^\infty_{k = - \infty} e^{ikL p_3} \,  , 
\label{2084GX3}
\end{equation}
where the integration variable  $z$ in (\ref{2080-GX2}) was renamed $p_3$ and interpreted as third component of the 3-momentum $\vp = \vp_\perp  + p_3 \ve_3$. In eq. (\ref{2084GX3}) $\int \dbar^3 p = \int d^3 p / (2 \pi)^3$ is the usual integration measure in flat momentum space $\RR^3$. Obviously, the $k = 0$ term is just the zero temperature part of the vacuum energy density, which is an (infinite) temperature independent (and thus irrelevant) constant, which has to be omitted from the thermodynamical quantities. Then with the replacement (\ref{G53}) (taking the limit $\Lambda \to \infty$ thereby skipping the divergent  piece $\Lambda$) we find from (\ref{2084GX3}) 
\begin{equation}
e (L) = \frac{1}{\Gamma (- \frac{1}{2})} \il^\infty_0 d \tau \tau^{- 1} \int \dbar^3 p e^{- \tau (\vp^2 + m^2)} \sli^\infty_{k = 1} \cos (k L p_3) \, . \label{2089-GX4}
\end{equation}
After performing the momentum integrals 
the proper-time integral can also be carried out, yielding for the pseudo energy density (\ref{2089-GX4}) 
\beq
e(L) = - \frac{1}{ \pi^2} \sli^\infty_{n = 1} \lk \frac{m}{n L} \rkx^2 K_2(n L m) \, , \label{G55}
\eeq
where $K_\nu(z)$ is the modified Bessel function and we have used $\Gamma (- \frac{1}{2}) = -  2 \sqrt{\pi}$. The individual terms $(n \neq 0)$ are all finite and also their sum converges. This sum, however, cannot be carried out analytically for massive bosons (the same
applies to the integral in the grand canonical expression (\ref{G51}) for the pressure). In the zero-mass limit 
the expression Eq.~(\ref{G55}) can be worked out analytically. Using the asymptotic form of the Bessel function
\begin{equation}
 \label{proc_1850}
 K_\nu (z) = \frac{1}{2} \Gamma (\nu) \left( \frac{1}{2} z \right)^{- \nu} \, , \qquad z \to 0 \, , \qquad \nu > 0
\end{equation}
we find from (\ref{G55}) 
\begin{equation}
 \label{proC_1855}
 e (L) = - \frac{1}{\pi^2} T^4 \zeta (4) \, , \qquad T = L^{- 1} \, ,
\end{equation}
where
\beq
\label{proc_1860}
\zeta (x) = \sli^\infty_{n = 1} \frac{1}{n^x}
\eeq
is the Riemann $\zeta$-function. With $\zeta (4) = \pi^4 / 90$ we find for the pressure $P  = - e (L)$ of massless bosons
\beq
P = \frac{\pi^2}{90} T^4 , \label{G56}
\eeq
which is Stefan--Boltzmann law, the correct result also obtained from the grand canonical ensemble. For massive bosons the evaluation of the sum in Eq.~(\ref{G55}) as well as the evaluation of the integral in Eq.~(\ref{G51}) have to be done numerically. The result is shown in Fig.~\ref{fig-10}a. As expected the pressure calculated from the compactified spatial dimension reproduces the result of  the usual grand canonical ensemble. Figure~\ref{fig-10}b shows the various contributions to the pressure. It is seen that only a few terms in the sum of Eq.~(\ref{G55}) are necessary to reproduce the result of the grand canonical ensemble to good accuracy.

In the case of the relativistic \emph{Fermi} gas with dispersion relation $\omega (p) = \sqrt{\vp^2 + m^2}$ the energy density on $\R^2 \times S^1(L)$  is given by
\beq
e(L) = - 2 \int \frac{\dd^2 p_\perp}{(2 \pi)^2} \frac{1}{L} \sli^\infty_{n = - \infty} \sqrt{\vp^2_\perp + (p_n + \ii \mu)^2 + m^2} \, , \quad \quad p_n = \frac{2 n + 1}{L} \pi \, , \label{G57}
\eeq
where we have now included a non-vanishing chemical potential $\mu$. To make this expression mathematically well-defined one has to resort again to  the proper-time regularization and Poisson resummation technique sketched above. The result is
\beq
e(L) = \frac{2}{\pi^2} \sli^\infty_{n = 0} \cos \left[ n L \lk \frac{\pi}{L} - \ii \mu \rkx \right] \lk \frac{m}{n L} \rkx^2 K_{- 2} (n L m) \, . \label{G58}
\eeq
Again, the term with $n = 0$ represents the zero temperature vacuum energy density, which is divergent and has to be removed. As before, this expression can only be calculated in closed form for massless particles. 
Furthermore for the remaining sum to converge, an analytic continuation $\ii \mu L  \to \bar{\mu} \in \R$ 
is required to carry out the sum
\beq
\sli^\infty_{n = 1} (-1)^n \frac{\cos (n \bar{\mu})}{n^4} = \frac{1}{48} \left[ - \frac{7}{15} \pi^2 + 2 \pi^2 \bar{\mu}^2 - \bar{\mu}^4 \right] \, . \label{G59}
\eeq
Continuing back to real chemical potentials one finds for the pressure $P = - e (L)$
\beq
P = \frac{1}{12 \pi^2} \left[ \frac{7}{15} \pi^4 T^4 + 2 \pi^2 T^2 \mu^2 + \mu^4 \right] \, , \label{G60}
\eeq
which is the correct result obtained also from the usual grand canonical ensemble.

In Ref.~\cite{Heffner:2015zna}, the above approach was used to study Yang--Mills theory at finite temperature. For this purpose,
it was only necessary
to repeat the variational Hamiltonian approach on the spatial manifold $\R^2 \times S^1 (L)$.  Due to the one compactified spatial dimension the three-dimensional integral equations of the zero-temperature case are replaced by a set of two-dimensional integral equations distinguished by different Matsubara frequencies. Below, I will use this approach to calculate the effective potential of the Polyakov loop, the order parameter of confinement.

\subsubsection{The Polyakov loop}

Consider SU($N$) gauge theory at finite temperature, where the temperature is introduced by the usual periodic boundary condition in the temporal direction (\ref{656-32}). Gauge
transformations preserving this boundary conditions need to be periodic only up to an element $z$ of the center $Z(N)$ 
of the gauge group,
\beq
U(x^4 = L) = z U(x^4 = 0) \, , \quad \quad z \in Z (N) \, . \label{G61}
\eeq
Since there are $N$ center elements, this theory has a residual global $Z(N)$ symmetry, which remains after local 
gauge fixing. However, there are quantities which are sensitive to such a $Z(N)$ symmetry transformation. The most prominent  example is the Polyakov loop (\ref{Gl: Polyakovloop}). A gauge transformation of the form (\ref{G61}) multiplies the Polyakov loop by the center element $z$, i.e.
\beq
P[A^U_4] = z P [A_4] \, . \label{G63}
\eeq
As discussed in sect. \ref{sect133}
the expectation value of the Polyakov loop
\beq
\langle P[A_4](\vx) \rangle \sim \exp \lk - L (F_q (\vx) - F_0) \rkx \label{G64}
\eeq
is related to the free energy $F (\vx) - F_0$ of a static color point charge located at $\vx$ 
where $F_0$ is the (divergent) free energy of the vacuum 
\cite{Svetitsky:1985ye}. In a confining theory this quantity has to be infinite since there are no free color charges, while in a deconfined phase it is finite. Accordingly we find for the expectation value of the Polyakov loop
\beq
\langle P [A_4] (\vx) \rangle \begin{cases}
                               = 0 &\quad \text{confined phase,} \\
                               \neq 0 &\quad \text{deconfined phase.}
                              \end{cases} \label{G65}
\eeq
From Eq.~(\ref{G63}) follows that a state which is invariant with respect to the global center transformation, has a 
vanishing expectation value of the Polyakov loop. Hence, 
in the deconfined phase the $Z(N)$ center symmetry is obviously broken. 

In the continuum theory the Polyakov loop can be most easily calculated in the Polyakov gauge
\beq
\partial_4 A_4 = 0 , \qquad A_4 \text{ color diagonal.} \label{G66}
\eeq
In this gauge one finds, for example, for the SU(2) gauge group that the Polyakov loop
\beq
P [A_4] (\vx) = \cos \lk \frac{1}{2} g A_4 (\vx) L \rkx \, \label{G67}
\eeq
is a unique
function of the gauge field, at least in the fundamental modular region of this gauge. It can be shown, 
see Refs.~\cite{Braun:2007bx, Marhauser:2008fz}, that instead of the expectation value of the Polyakov loop 
$\langle P[A_4] \rangle$ one may alternatively use the Polyakov loop of the expectation value, $P [\langle A_4
\rangle ]$, or the expectation value of the temporal gauge field itself, $\langle A_4 \rangle$, 
as order parameter of confinement in the gauge (\ref{G66}). This analysis also shows that the most efficient way to obtain the Polyakov 
loop is to carry out a so-called background field calculation with a temporal background field $a_4 (\vx) = \langle A_4 (\vx) \rangle$ 
chosen in the Polyakov gauge, and then calculate the effective potential $e [a_4]$ 
of that background field. From the minimum $\bar{a}_4$ of this potential one evaluates the Polyakov loop 
$P[\langle A_4\rangle] = P[\bar{a}_4]$, which can then serve as an order parameter for confinement.

\begin{figure}
\parbox{.48\linewidth}{\centering\includegraphics[width=\linewidth,clip]{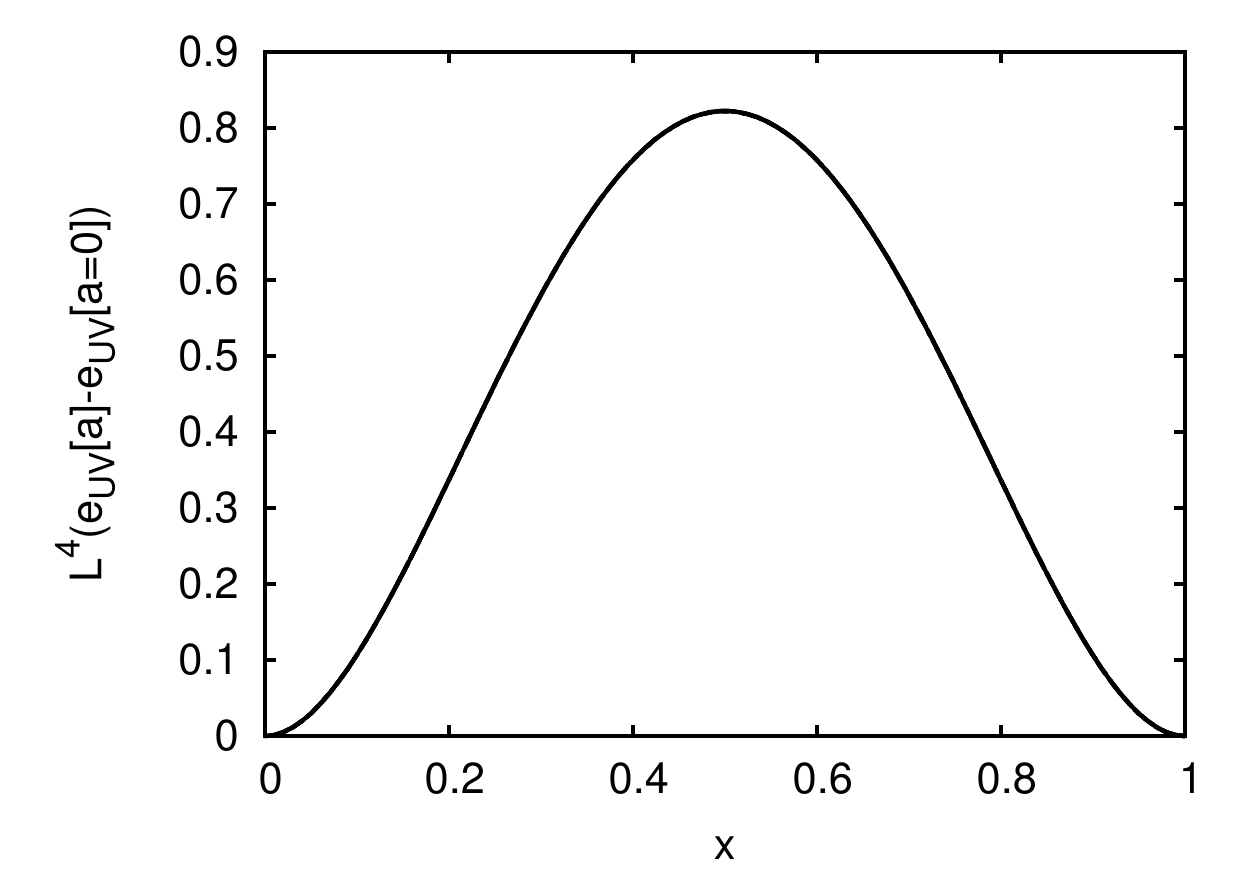}\\(a)}
\hfill
\parbox{.48\linewidth}{\centering\includegraphics[width=\linewidth,clip]{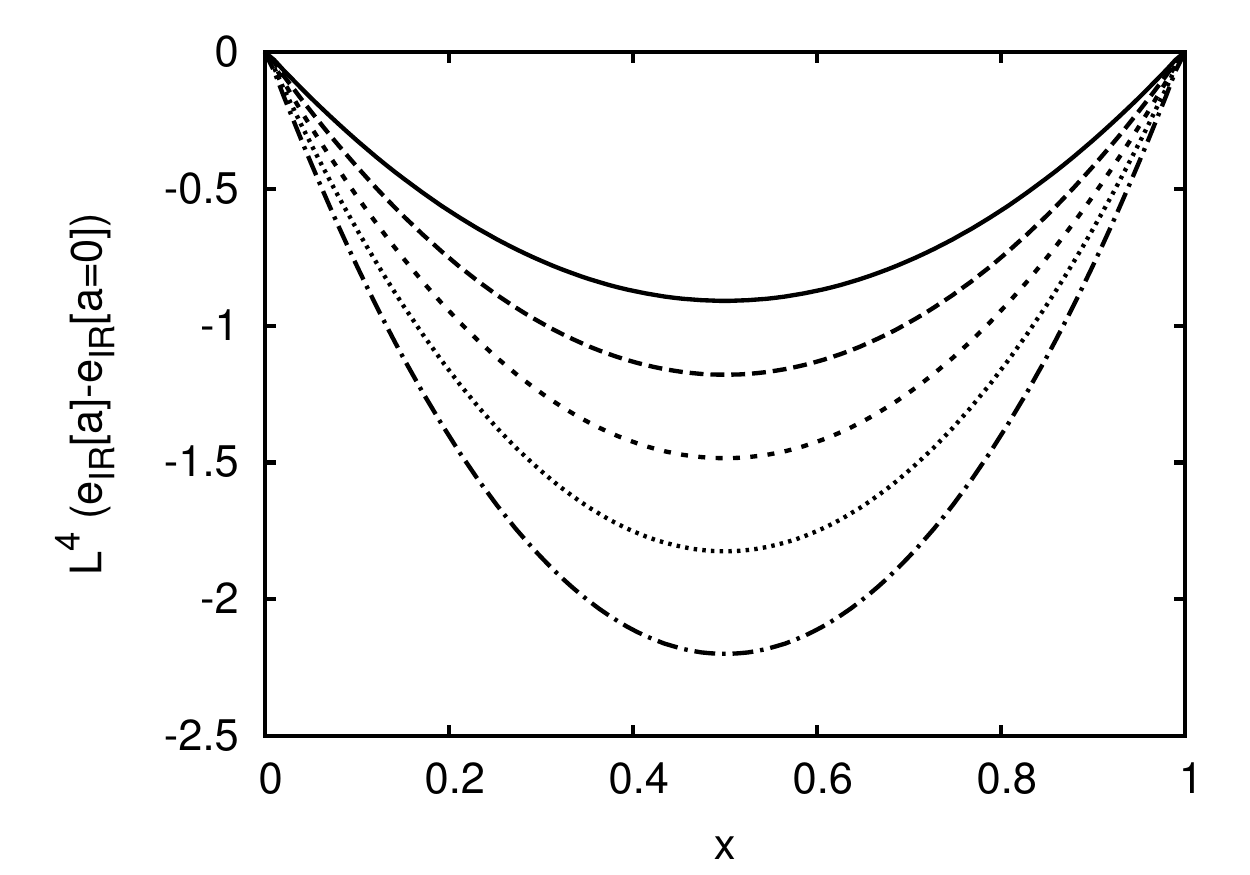}\\(b)}
\caption{The effective potential of the Polyakov loop $e(a, L)$ (\ref{G68}) as function 
of the background field $x = a_3 L / 2 \pi$. The curvature is neglected $(\chi = 0)$ and the gluon energy assumed to be (a) 
$\omega (p) = p$ (UV-form) and (b) $\omega (p) = M^2 / p$ (IR-form), respectively, for various temperaturs. Note that the UV form of the potential is 
(up to a global factor of $T^4$) independent on the temperature, see fig. (a). This is different for IR form of the potential,
see fig. (b), which contains a mass scale. Here the temperature increases from bottom to top.}
\label{fig-11}
\end{figure}

Such a calculation was done a long time ago in Ref.~\cite{Weiss:1980rj,Gross:1980br}, where the effective potential $e[a_4]$ was calculated
in one-loop perturbation theory. The result is shown in Fig.~\ref{fig-11}a. The potential is periodic due to center symmetry. The minimum of 
the potential occurs at vanishing background field, which gives $P[a_4 = 0] = 1$ 
corresponding to the deconfined phase. This is, of course, expected due to the use of perturbation theory. Below, I present the results of 
a non-perturbative evaluation of $e[a_4]$ in the Hamiltonian approach in Coulomb gauge.

At first sight it seems that the Polyakov loop cannot be calculated in the Hamiltonian approach due to the use of the Weyl gauge
$A_4 = 0$. However, we can now use the alternative Hamiltonian approach to finite temperature introduced in sect. \ref{4.5.1}, where the 
temperature is introduced by compactifying a spatial dimension. Here, we compactify the $x_3$-axis and consequently put also the 
background field along this axis\footnote{A constant background field directed along an uncompactified axis is irrelevant since it can be eliminated by a simple change of the corresponding  momentum variable.}, $\va = a \ve_3$. 
The Polyakov loop is then given by\footnote{Since we  are considering here only stationary vacuum wave functionals the time-argument can be skipped.}  (c.f. eq. (\ref{Gl: Polyakovloop}))
\begin{equation}
P [A_3] (\vx_\perp) = \frac{1}{d_r} tr \cP \exp \left[ i g \il^L_0 d x^3 A_3 (\vx_\perp, x^3) \right] \, ,
\end{equation}
where $\vx_\perp$ denotes the coordinates of the two uncompactified spatial dimensions.  We are interested  here in the  quantity $P [\langle A_3 \rangle] (\vx_\perp)$. For this purpose we calculate  the effective potential of $\langle A_3 \rangle$ using the background field method. 

In the Hamiltonian approach the effective potential of a spatial background field 
$\va$ can be easily calculated by minimizing the expectation value of the Hamiltonian under the constraint 
$\langle \vA \rangle = \va$ \cite{Weinberg:1996kr}. 
The resulting energy $\langle H \rangle_{\va} = L^2 \ell e(\va)$ is then (up to the spatial volume factor) the effective potential. So the effective potential $e(\va)$ is
nothing but the pseudo energy density considered earlier, but now calculated 
with the constraint $\langle \vA \rangle = \va$, see ref. \cite{Reinhardt:2013iia} for more details.

\subsubsection{The effective potential of the Polyakov loop}

After lengthy calculations,
exploiting the gluon gap equation (\ref{eq:part_iii:gap_equation_general}) and neglecting the Coulomb term, 
one finds for the effective potential of the Polyakov loop (or more precisely of $\langle A_3 \rangle = a$) the following expression \cite{Reinhardt:2013iia}
\beq
e(a, L) = \sli_\sigma \frac{1}{L} \sli^\infty_{n = - \infty} \int \frac{\dd^2 p_\perp}{(2 \pi)^2} \lk \omega (\vp^\sigma) - \chi (\vp^\sigma) \rkx \, , \label{G68}
\eeq
where $\omega(p)$ is the gluon energy and $\chi(p)$ is the ghost loop. These quantities have to be taken with the momentum variable
\beq
\vp^\sigma = \vp_\perp + \lk p_n - \sigma \cdot a \rkx \ve_3 \, , \label{G69}
\eeq
where $\vp_\perp$  is the momentum corresponding to the two non-compactified space dimensions while $p_n = 2 \pi n / L$ is the Matsubara frequency resulting from the compactification of the third dimension. Furthermore, $\sigma \cdot a \equiv \sigma^b a^b$ denotes the product of the color background field with the 
root vectors $\sigma^b$ of the color
group. Equation~(\ref{G68}) includes also the summation over the roots $\sigma$ of the gauge group. In Refs.~\cite{Reinhardt:2012qe, Reinhardt:2013iia}, the effective potential (\ref{G68}) was explicitly calculated using for $\omega (p)$ and $\chi (p)$ the results from the variational calculation in Coulomb gauge at zero temperature \cite{Epple:2006hv}. This represents certainly an approximation since, in principle, one should use the finite-temperature solutions obtained in Ref.~\cite{Heffner:2015zna}.

\begin{figure}
\parbox{.48\linewidth}{\centering\includegraphics[width=\linewidth,clip]{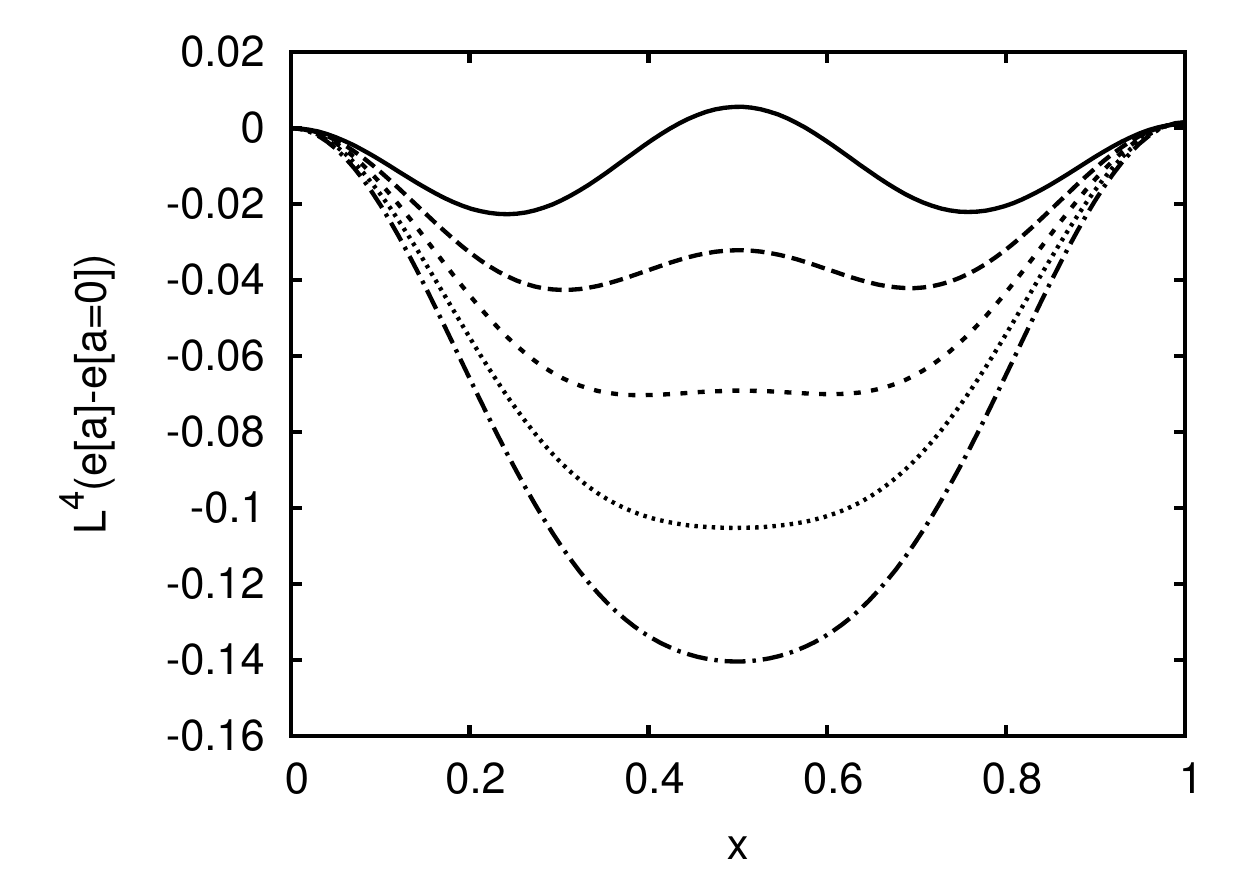}\\(a)}
\hfill
\parbox{.48\linewidth}{\centering\includegraphics[width=\linewidth,clip]{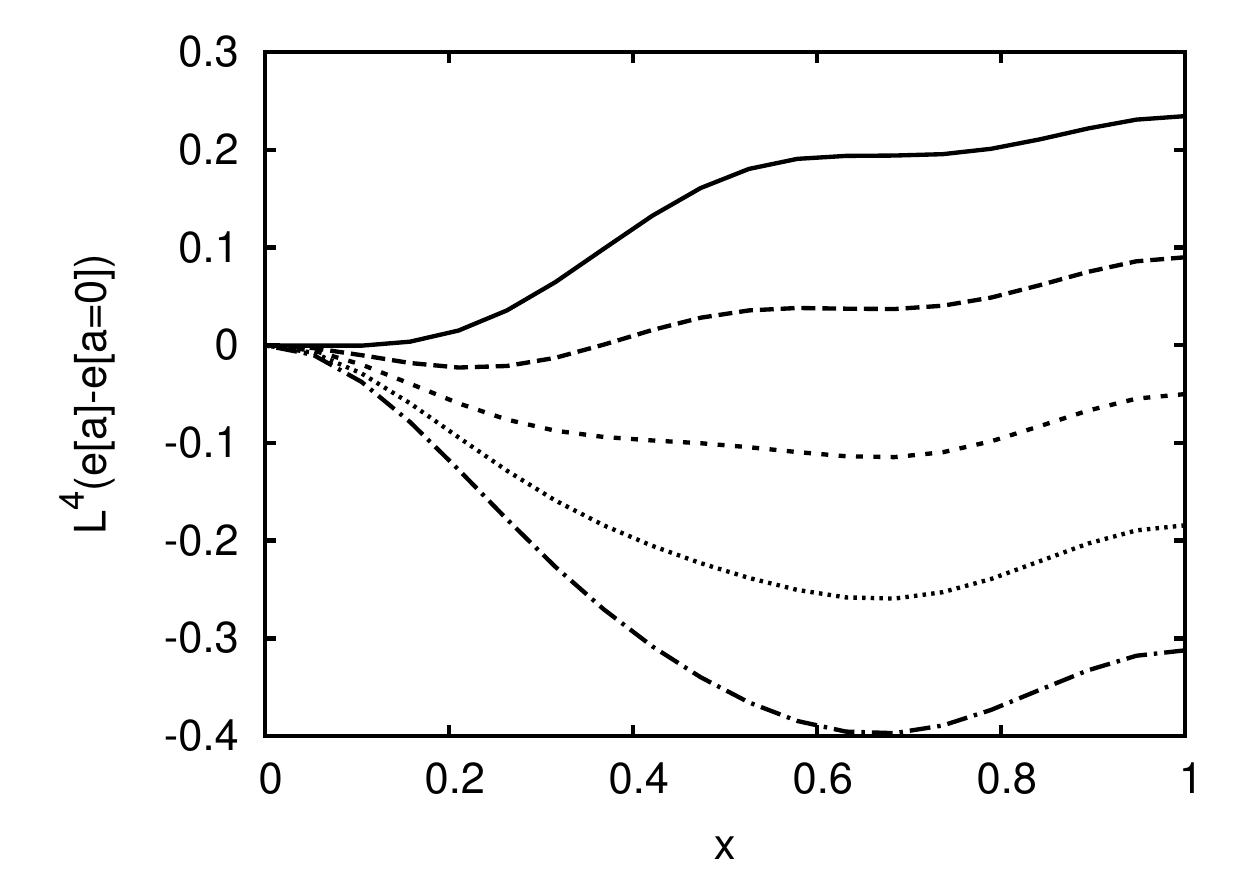}\\(b)}
\caption{Effective potential of the Polyakov loop (\ref{G68}) as function of the background field $x = a_3 L / 2 \pi$ at various temperatures, for the gauge group (a) SU(2) and (b) SU(3). The temperature increases from bottom to top. }
\label{fig-12}
\end{figure}

\begin{figure}
\parbox{.48\linewidth}{\centering\includegraphics[width=\linewidth,clip]{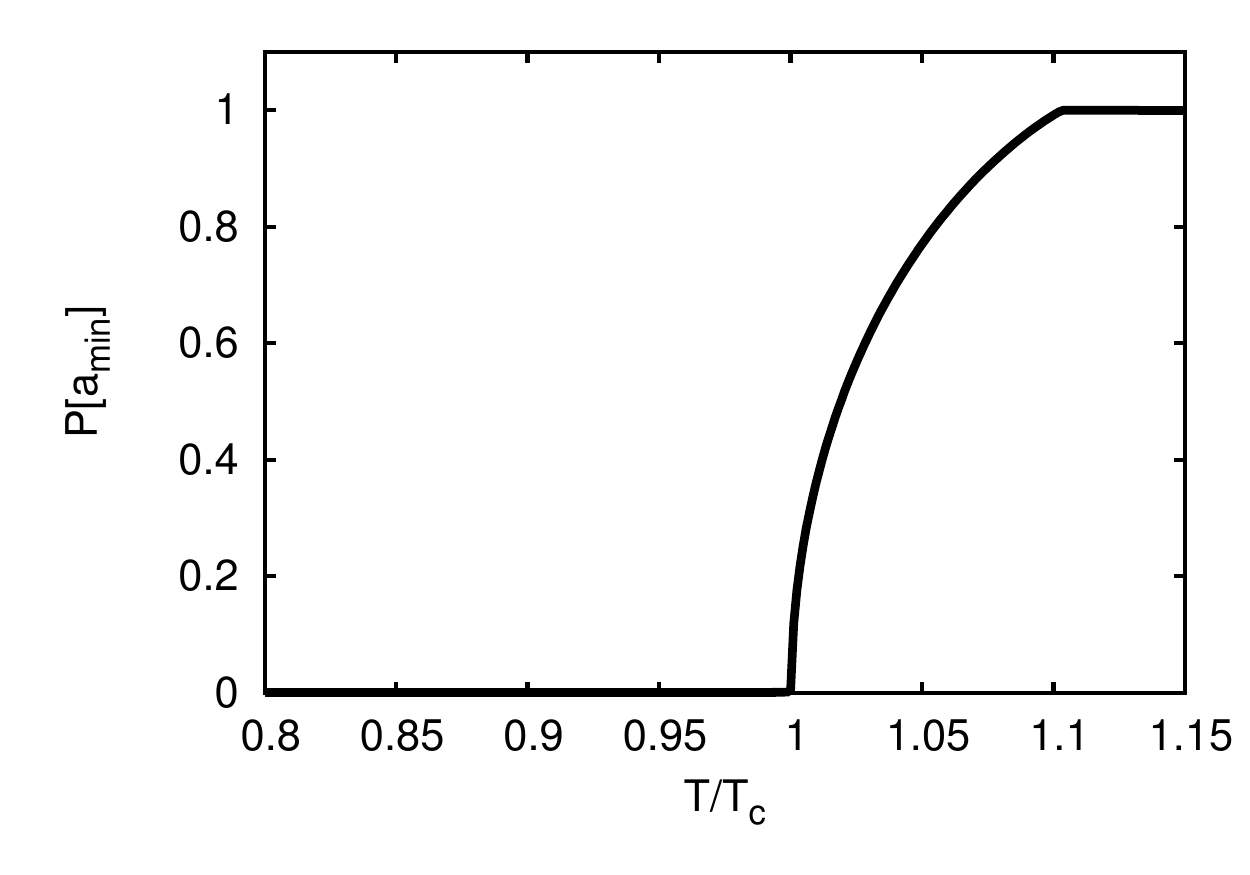}\\(a)}
\hfill
\parbox{.48\linewidth}{\centering\includegraphics[width=\linewidth,clip]{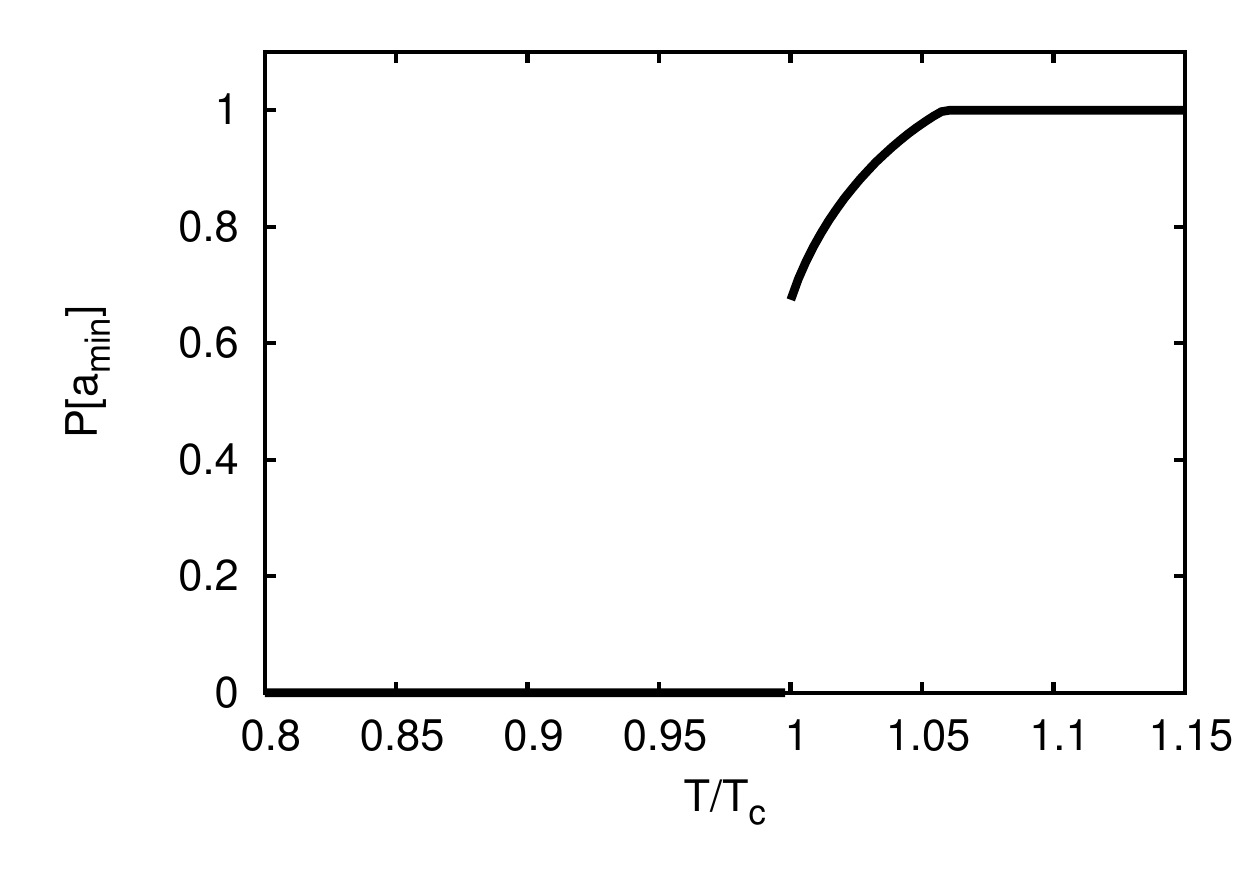}\\(b)}
\caption{The Polyakov loop as function of the temperature (a) for SU(2) and (b) for SU(3).}
\label{fig-13}
\end{figure}

Before I present the full results let me ignore the ghost loop $\chi(p)$ in Eq.~(\ref{G68}) and consider the ultraviolet and infrared limit of the gluon energy. If we choose the ultraviolet limit $\omega(p) = p$, we obtain from Eq.~(\ref{G68}) with $\chi(p) = 0$ precisely the Weiss potential, shown in Fig.~\ref{fig-11}a, which corresponds to the deconfined phase. Choosing for the gluon energy its infrared limit $\omega(p) = M^2 / p$, one finds from Eq.~(\ref{G68}) with $\chi(p) = 0$ the (center symmetric) potential shown in Fig.~\ref{fig-11}b. From its center symmetric minimum $\bar{a} = \pi / L$ one finds a vanishing Polyakov loop $P[\bar{a}] = 0$ corresponding to the confined phase. Obviously, the deconfining phase transition results from the interplay between the confining infrared 
and the deconfining ultraviolet dispersions. 
Choosing for the gluon energy the sum of its
UV- and IR-parts $\omega(p) = p + M^2/p$, which can be considered as an approximation to the Gribov formula
(\ref{eq:part_iii:gribov_fromula}), one has to add the UV and IR potentials and finds a phase transition at a critical temperature $T_{\mathrm{c}} = \sqrt{3} M / \pi$. With the Gribov mass $M \approx 880 \, \mathrm{MeV}$ this gives a critical value of $T_{\mathrm{c}} \approx 485 \, \mathrm{MeV}$ for the color group SU(2), which is much too high as compared to the lattice value of $312\,\mathrm{MeV}$ \cite{Lucini:2003zr}. One can show analytically \cite{Reinhardt:2012qe,Reinhardt:2013iia} that the neglect of the ghost loop $\chi(p) = 0$ shifts the critical temperature to higher values. If one uses the Gribov formula (\ref{eq:part_iii:gribov_fromula}) for the gluon energy $\omega(p)$ and includes the ghost loop $\chi(p)$, one finds the effective potential shown in Fig.~\ref{fig-12}a, which shows a second order phase transition and gives a transition temperature of $T_{\mathrm{c}} \approx 269 \, \mathrm{MeV}$ for the gauge group SU(2), which is in the right ballpark. The Polyakov loop $P[\bar{a}]$ calculated 
from the minimum $\bar{a}$ 
of the effective potential $e(a, L)$ (\ref{G68}) is plotted in Fig.~\ref{fig-13}a as function of the temperature.

\begin{figure}
\parbox{.48\linewidth}{\centering\includegraphics[width=\linewidth,clip]{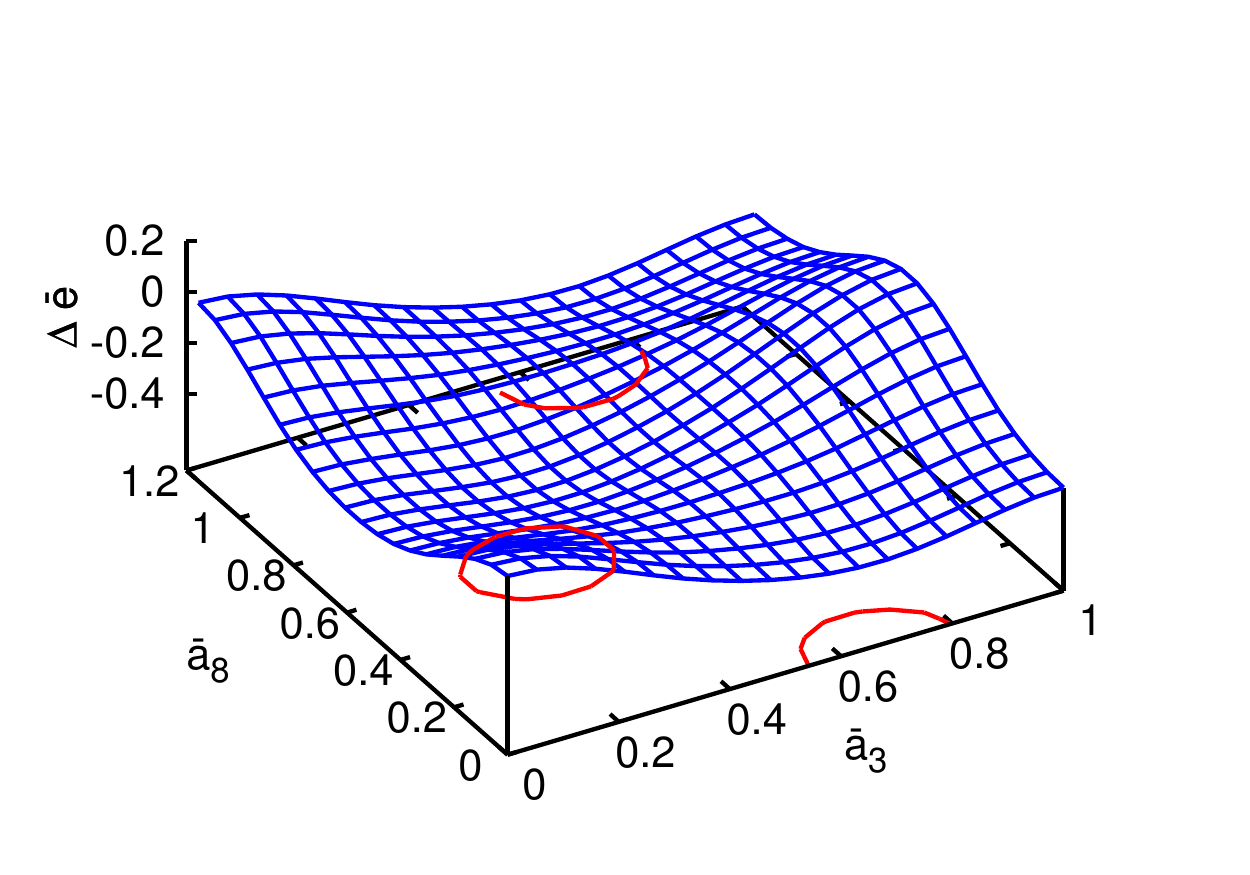}\\(a)}
\hfill
\parbox{.48\linewidth}{\centering\includegraphics[width=\linewidth,clip]{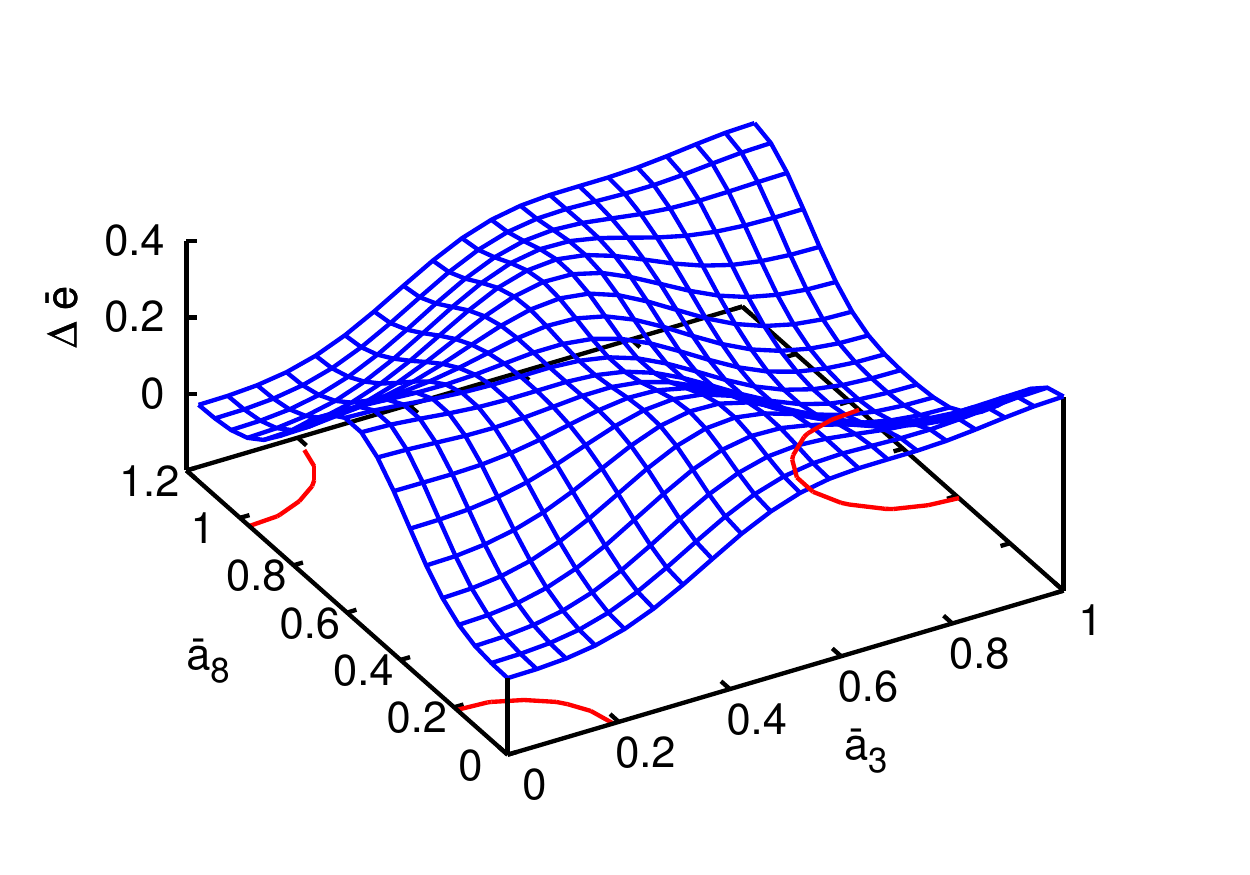}\\(b)}
\caption{The effective potential of the Polyakov loop for the gauge group SU(3) as function of the two Cartan components of the background field $x = a_3 L / 2 \pi$ and $y = a_8 L / 2 \pi$ for (a) $T < T_{\mathrm{c}}$ and (b) $T > T_{\mathrm{c}}$.}
\label{fig-14}
\end{figure}

The effective potential for the gauge group SU(3) can be reduced to that of the SU(2) group by noticing that the SU(3) algebra consists of three SU(2) subalgebras characterized by the three positive roots $\vec{\sigma} = (1, 0)$, $(1/2, \sqrt{3}/2)$, $(1/2 \, , - \sqrt{3} /2)$. One finds 
\beq
e_{\mathrm{SU(3)}} (a, L) = \sum_{\boldsymbol{\sigma} > 0} e_{\mathrm{SU(2)}}[\vec{\sigma}] (a, L) \, . \label{931-70}
\eeq
The resulting effective potential for SU(3) is shown in Fig.~\ref{fig-14} as function of the components of the background field in the Cartan algebra, $a_3$ and $a_8$. Above and below $T_{\mathrm{c}}$ the absolute minima of the potential occur in both cases for $a_8 = 0$. Cutting the two-dimensional potential surface at $a_8 = 0$, one finds the effective potential shown in Fig.~\ref{fig-12}b, which shows a first order phase transition with a critical temperature of $T_{\mathrm{c}} \approx 283 \, \mathrm{MeV}$. The first order nature of the SU(3) phase transition is also seen in Fig.~\ref{fig-13}b, where the Polyakov loop $P[\bar{a}]$ is shown as function of the temperature.

\subsubsection{The dual quark condensate}

\begin{figure}[ht]
\parbox{.48\linewidth}{\centering\includegraphics[width=\linewidth,clip]{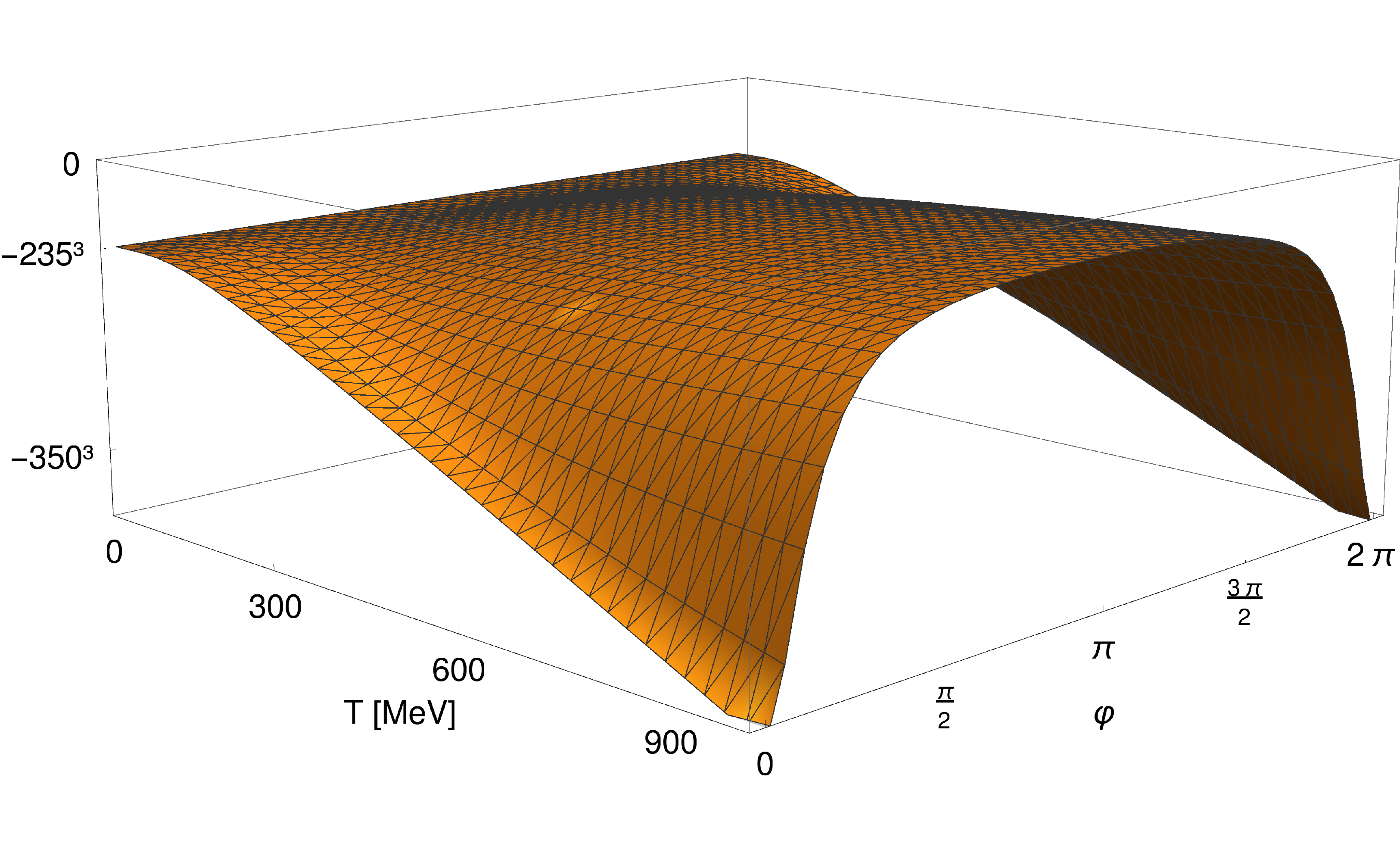}\\(a)}
\hfill
\parbox{.48\linewidth}{\centering\includegraphics[width=\linewidth,clip]{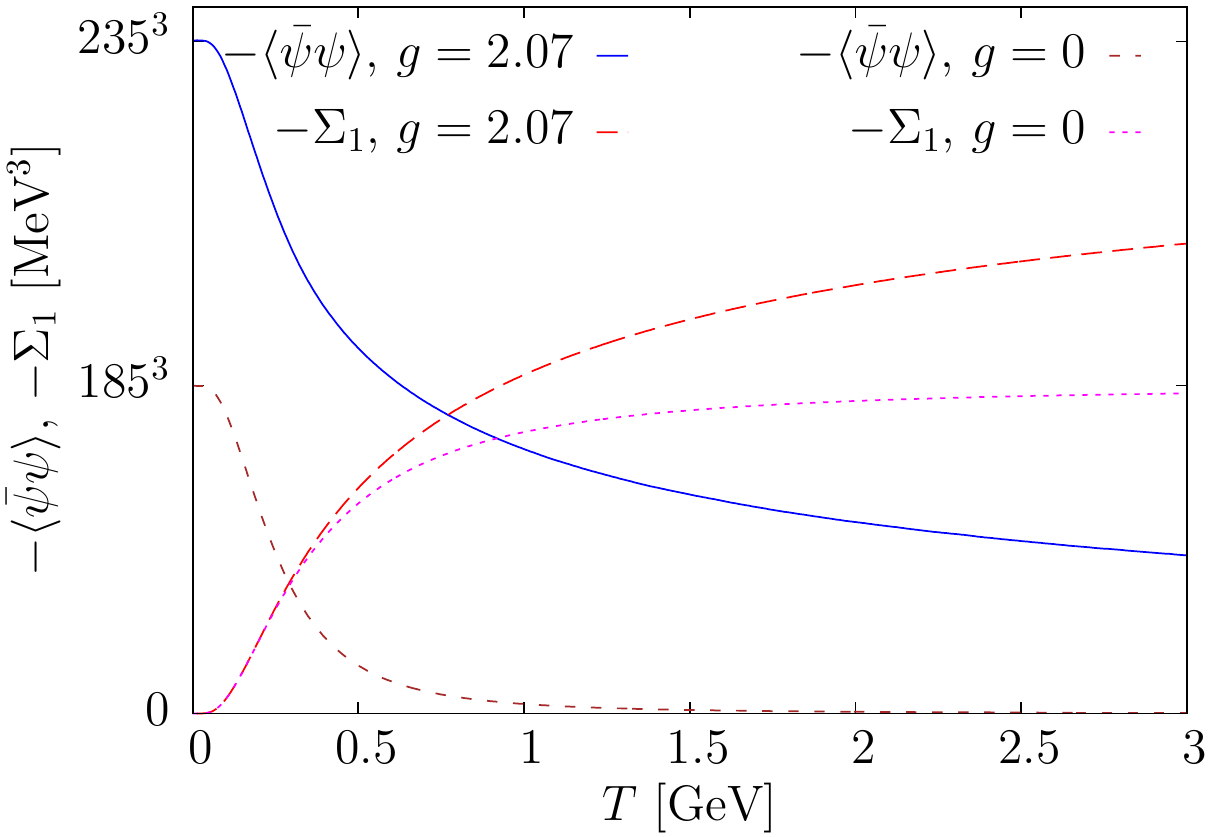}\\(b)}
\caption{(a) Chiral quark condensate $\langle \bar{\psi} \psi \rangle_\varphi$ as function of the temperature $T$ and the phase $\varphi$ of the boundary condition (\ref{28}). (b) Chiral and dual quark condensate as function of the temperature. Results are presented for both a coupling of $g \simeq 2.1$ and $g = 0$.}
\label{fig6}%
\end{figure}

The dual quark condensate was originally introduced in Ref.~\cite{Gattringer:2006ci} and was discussed in a more general context in Ref.~\cite{Synatschke:2007bz}. This quantity has been calculated on the lattice \cite{Bilgici:2008qy, Zhang:2010ui}, in the functional renormalization group approach \cite{Braun:2009gm} and in the Dyson--Schwinger approach \cite{Fischer:2010fx}. The dual condensate is defined by
\beq
\Sigma_n = \int\limits_0^{2 \pi} \frac{\dd \varphi}{2 \pi} \exp(-\ii n \varphi) \langle \bar{\psi} \psi \rangle_\varphi \, , \label{27}
\eeq
where $\langle \bar{\psi} \psi \rangle_\varphi$ is the quark condensate calculated with the $U(1)$-valued boundary condition
\beq
\psi(x^4 + L/2, \vx) = \mathrm{e}^{\ii \varphi} \psi(x^4 - L/2, \vx) \, . \label{28}
\eeq
For $\varphi = \pi$ these boundary conditions reduce to the usual finite-temperature boundary conditions of the quark field in the functional integral representation of the partition function, see Eq.~(\ref{656-32}). On the lattice it is not difficult to show that the quantity $\Sigma_n$ (\ref{27}) represents the vacuum expectation value of the sum of all closed Wilson loops winding precisely $n$-times around the compactified time axis. In particular, the quantity $\Sigma_1$ represents the expectation value of all closed loops winding precisely once around the compactified time axis and is therefore called the 
{\em dressed Polyakov loop}. The phase in the boundary condition (\ref{28}) can be absorbed into an imaginary chemical potential
\beq
\mu = \ii \frac{\pi - \varphi}{L} \label{29}
\eeq
for fermion fields satisfying the usual antisymmetric boundary condition $\psi(x^4 + L/2, \vx) = -\psi(x^4 - L/2, \vx)$. In the Hamiltonian approach to finite temperatures of Ref.~\cite{Reinhardt:2016xci}, where the compactified time axis has become the third spatial axis, the phase dependent boundary condition (\ref{28}) or equivalently the imaginary chemical potential (\ref{29}) manifests itself in the momentum variable along the (compactified) three-axis, which reads
\beq
p_3 = p_n + \ii \mu = \frac{2 \pi n + \varphi}{L} \, , \quad \quad p_n = \frac{2 n + 1}{L} \pi \, , \label{30}
\eeq
where $p_n$ is the usual fermionic Matsubara frequency [Eq.~(\ref{G57})]. Using the zero-temperature quark mass function $M(p)$ calculated in Ref.~\cite{Campagnari:2016wlt}, one finds in the Hamiltonian approach to QCD of Ref.~\cite{Vastag:2015qjd} for the dual quark condensate after Poisson resummation (\ref{2080-GX2})
the leading expression \cite{Reinhardt:2016pfe}
\beq
\Sigma_n = -\frac{N}{\pi^2} \int\limits_0^{\infty} \dd p \, \frac{p^2 M(p)}{\sqrt{p^2 + M^2(p)}} \left[\delta_{n 0} + 
\frac{\sin(n L p)}{n L p}\right] \, , \label{31}
\eeq
where $N$ denotes the number of colors. In the same way, one can compute the quark condensate $\langle \bar{\psi} \psi \rangle_\varphi$ shown in Fig.~\ref{fig6}a. For the dressed Polyakov loop $\Sigma_1$ one finds the temperature behavior shown in Fig.~\ref{fig6}b, where we also compare with the result obtained when the coupling to the transverse gauge field degrees of freedom is neglected ($g = 0$). As one observes there is no difference at small temperatures in accord with the fact that the mass function $M(p)$ has the same infrared behavior, whether the coupling to the transverse gluons is included or not. The slower UV decrease of the full mass function causes the dual condensate to reach its high-temperature limit
\beq
\lim_{L \to 0} \Sigma_1 = -\frac{N}{\pi^2} \int\limits_0^{\infty} \dd p \, \frac{p^2 M(p)}{\sqrt{p^2 + M^2(p)}} = \lim_{L \to \infty} \langle \bar{\psi} \psi \rangle_{\varphi = \pi} \label{32}
\eeq
only very slowly. However, we expect that this limit is reached faster when the finite-temperature solutions are used. This will presumably also convert the crossover obtained for the chiral condensate, see Fig.~\ref{fig6}b, into a true phase transition as expected for chiral quarks. From the inflexion points of the 
chiral and dual condensates one extracts the values of $T_{\chi}^{\mathrm{pc}} \simeq 170 \, \mathrm{MeV}$ and $T_{\mathrm{c}}^{\mathrm{pc}} 
\simeq 198 \, \mathrm{MeV}$ for the pseudo-critical temperatures of the chiral and deconfinement transition, respectively. 
For comparison, on the lattice one finds
for realistic quark masses $T_{\chi}^{\mathrm{pc}} \simeq 155 \, \mathrm{MeV}$ and $T_{\mathrm{c}}^{\mathrm{pc}} \simeq 165 \, \mathrm{MeV}$ \cite{Borsanyi:2010bp, Bazavov:2011nk}.

\subsection{Conclusions}

In this talk I have presented a survey of
results obtained within the Hamiltonian approach to QCD in Coulomb gauge. 
I have first considered the pure Yang-Mills sector and shown that the obtained gluon and ghost propagators reflect the confining 
properties of the theory and are in satisfactory agreement with the lattice data. The gluon energy and the ghost form-factor are
both IR divergent in accordance with the Griobov picture of confinement. I have also established the connection of this picture 
with the magnetic monopole (dual Mei\ss{}ner effect) and the center vortex picture. 

I have then studied the quark sector of QCD within the Hamiltonian approach 
in Coulomb gauge using a Slater determinant \emph{ansatz} for the quark wave functional, which includes in particular the quark-gluon 
coupling with two different Dirac structures. Our calculations show that there is no spontaneous breaking of chiral symmetry when the 
(linearly rising) infrared part of the Coulomb potential is excluded. Furthermore, choosing the Coulomb string tension from the lattice 
data we can reproduce the phenomenological value of the quark condensate when the coupling of the quarks to the transverse gluons is included.

I have then extended the Hamiltonian approach to QCD in Coulomb gauge to finite temperatures by compactifying a spatial dimension \cite{Reinhardt:2016xci}. Within this approach, I have calculated the effective potential 
of the Polyakov loop for pure Yang-Mills theory 
as well as the chiral and dual quark condensates for QCD as function of the temperature. 
Using our zero-temperature variational 
solution as input, from the Polyakov loop we predict a critical temperature for the 
deconfinement phase transition in pure Yang-Mills theory 
of about $T_{\mathrm{c}} \sim 275 \, \mathrm{MeV}$ for SU(2), and $T_{\mathrm{c}} \sim 280 \, \mathrm{MeV}$ for SU(3). Furthermore, 
the correct order of the phase transition was found for both
SU(2) and SU(3). For full QCD our calculations of the dual and chiral quark condensate
predict pseudo-critical temperatures of $T_{\chi}^{\mathrm{pc}} \simeq 170 \, \mathrm{MeV}$ for the chiral and $T_{\mathrm{c}}^{\mathrm{pc}} \simeq 198 \, \mathrm{MeV}$ for the deconfinement transition. In all these finite-temperature calculations the zero-temperature variational solutions were used as input, which is likely the reason that the critical temperatures currently obtained are too high 
as compared to the lattice data. The solution of the variational principle at finite temperature will be the next step in our investigation of the QCD phase diagram.

\section*{Acknowledgement}

This work was supported in part by DFG-RE856/9-2 and by DFG-RE856/10-1.
These lecture notes where prepared with the help of 
Maciej Lewicki (Lecture 1) and Michal Szymanski (Lectures 2, 3), whom I thank for their committed work.

%
%

\bibliography{biblio-spires}

\end{document}